\input harvmac.tex
\input epsf.tex

\def\abstract#1{\vskip .5in\vfil\centerline{\bf Abstract}\penalty1000{{\smallskip\ifx\answ\bigans\leftskip 2pc \rightskip 2pc\else\leftskip 5pc \rightskip 5pc\fi\noindent\abstractfont \baselineskip=12pt{#1} \smallskip}}\penalty-1000}

\def\abstractmod#1{\vskip.5in\vfil\centerline{\bf Abstract}\penalty1000{{\smallskip\leftskip0pc\rightskip0pc\noindent\abstractfont\baselineskip=12.5pt{#1}\smallskip}}\penalty-1000}

%

\lref\HIV{
  K.~Hori, A.~Iqbal and C.~Vafa,
  ``D-branes and mirror symmetry,''
  arXiv:hep-th/0005247.
}

\lref\GU{
  F.~Cachazo, B.~Fiol, K.~A.~Intriligator, S.~Katz and C.~Vafa,
  ``A geometric unification of dualities,''
  Nucl.\ Phys.\  B {\bf 628}, 3 (2002)
  [arXiv:hep-th/0110028].
}

\lref\hb{
  S.~Franco, A.~Hanany, K.~D.~Kennaway, D.~Vegh and B.~Wecht,
  ``Brane Dimers and Quiver Gauge Theories,''
  [arXiv:hep-th/0504110].
}

\lref\KS{
M.~Kontsevich and Y.~ Soibelman,
``Stability Structures, motivic Donaldson-Thomas invariants and cluster transformations,"
[arXiv:0811.2435v1]
}

\lref\Herzog{
  C.~P.~Herzog,
  ``Seiberg duality is an exceptional mutation,''
  JHEP {\bf 0408}, 064 (2004)
  [arXiv:hep-th/0405118].
}

\lref\aspb{
  P.~S.~Aspinwall and I.~V.~Melnikov,
  ``D-branes on vanishing del Pezzo surfaces,''
  JHEP {\bf 0412}, 042 (2004)
  [arXiv:hep-th/0405134].
}

\lref\Iqbal{
  A.~Hanany and A.~Iqbal,
  ``Quiver theories from D6-branes via mirror symmetry,''
  JHEP {\bf 0204}, 009 (2002)
  [arXiv:hep-th/0108137].
}

\lref\KV{
  B.~Feng, Y.~H.~He, K.~D.~Kennaway and C.~Vafa,
  ``Dimer models from mirror symmetry and quivering amoebae,''
  Adv.\ Theor.\ Math.\ Phys.\  {\bf 12}, 3 (2008)
  [arXiv:hep-th/0511287].
}

\lref\AK{
  M.~Aganagic and A.~Karch,
  ``Calabi-Yau mirror symmetry as a gauge theory duality,''
  Class.\ Quant.\ Grav.\  {\bf 17}, 919 (2000)
  [arXiv:hep-th/9910184].
}

\lref\JW{
  W.~y.~Chuang and D.~L.~Jafferis,
  ``Wall Crossing of BPS States on the Conifold from Seiberg Duality and
  Pyramid Partitions,''
  Commun.\ Math.\ Phys.\  {\bf 292}, 285 (2009)
  [arXiv:0810.5072 [hep-th]].
}

\lref\OY{
  H.~Ooguri and M.~Yamazaki,
  ``Crystal Melting and Toric Calabi-Yau Manifolds,''
  Commun.\ Math.\ Phys.\  {\bf 292}, 179 (2009)
  [arXiv:0811.2801 [hep-th]].
}

\lref\RM{
S.~Mozgovoy and M.~Reineke,
``On the noncommutative Donaldson-Thomas invariants arising from brane tilings,"
[arXiv:0809.0117].
}

\lref\DA{
R.~Kenyon, A.~Okounkov and S.~Sheffield,
``Dimers and Amoeba,"
[arXiv:math-ph/0311005].
}

\lref\ORV{
  A.~Okounkov, N.~Reshetikhin and C.~Vafa,
  ``Quantum Calabi-Yau and classical crystals,''
  arXiv:hep-th/0309208.
}

\lref\SK{
  S.~H.~Katz and E.~Sharpe,
  ``D-branes, open string vertex operators, and Ext groups,''
  Adv.\ Theor.\ Math.\ Phys.\  {\bf 6}, 979 (2003)
  [arXiv:hep-th/0208104].
}

\lref\Asp{
  P.~S.~Aspinwall,
  ``D-branes on Calabi-Yau manifolds,''
  arXiv:hep-th/0403166.
}

\lref\AspKatz{
  P.~S.~Aspinwall and S.~H.~Katz,
  ``Computation of superpotentials for D-Branes,''
  Commun.\ Math.\ Phys.\  {\bf 264}, 227 (2006)
  [arXiv:hep-th/0412209].
}

\lref\MWI{
  M.~Wijnholt,
  ``Large volume perspective on branes at singularities,''
  Adv.\ Theor.\ Math.\ Phys.\  {\bf 7}, 1117 (2004)
  [arXiv:hep-th/0212021].
}

\lref\sz{
B.~Szendroi,
``Non-commutative Donaldson-Thomas theory and the conifold,"
Geom.\ Topol. \ {\bf 12}, 1171 (2008)
[arXiv:0705.3419]
}

\lref\MNOPI{
D.~Maulik, N.~Nekrasov, A.~Okounkov and R.~Pandharipande,
``Gromov-Witten theory and Donaldson-Thomas theory, I,"
[arXiv:math/0312059]
}

\lref\MNOPII{
D.~Maulik, N.~Nekrasov, A.~Okounkov and R.~Pandharipande,
``Gromov-Witten theory and Donaldson-Thomas theory, II,"
[arXiv:math/0406092]
}

\lref\HVH{
  A.~Hanany, C.~P.~Herzog and D.~Vegh,
  ``Brane tilings and exceptional collections,''
  JHEP {\bf 0607}, 001 (2006)
  [arXiv:hep-th/0602041].
}

\lref\Hananythree{
  S.~Benvenuti, B.~Feng, A.~Hanany and Y.~H.~He,
  ``Counting BPS operators in gauge theories: Quivers, syzygies and
  plethystics,''
  JHEP {\bf 0711}, 050 (2007)
  [arXiv:hep-th/0608050].
}

\lref\Kennaway{
  K.~D.~Kennaway,
  ``Brane Tilings,''
  Int.\ J.\ Mod.\ Phys.\  A {\bf 22}, 2977 (2007)
  [arXiv:0706.1660 [hep-th]].
}

\lref\QF{
  A.~Iqbal, N.~Nekrasov, A.~Okounkov and C.~Vafa,
  ``Quantum foam and topological strings,''
  JHEP {\bf 0804}, 011 (2008)
  [arXiv:hep-th/0312022].
}

\lref\Wij{
  M.~Wijnholt,
  ``Parameter Space of Quiver Gauge Theories,''
  arXiv:hep-th/0512122.
}

\lref\HeckV{
  J.~J.~Heckman and C.~Vafa,
  ``Crystal melting and black holes,''
  JHEP {\bf 0709}, 011 (2007)
  [arXiv:hep-th/0610005].
}

\lref\DM{
  F.~Denef and G.~W.~Moore,
  ``Split states, entropy enigmas, holes and halos,''
  arXiv:hep-th/0702146.
}

\lref\DVV{
  R.~Dijkgraaf, C.~Vafa and E.~Verlinde,
  ``M-theory and a topological string duality,''
  arXiv:hep-th/0602087.
}

\lref\HV{
  K.~Hori and C.~Vafa,
  ``Mirror symmetry,''
  arXiv:hep-th/0002222.
}

\lref\CheungAZ{
  Y.~K.~Cheung and Z.~Yin,
  ``Anomalies, branes, and currents,''
  Nucl.\ Phys.\  B {\bf 517}, 69 (1998)
  [arXiv:hep-th/9710206].
}

\lref\GMNI{
  D.~Gaiotto, G.~W.~Moore and A.~Neitzke,
  ``Four-dimensional wall-crossing via three-dimensional field theory,''
  arXiv:0807.4723 [hep-th].
}

\lref\TV{
  M.~Aganagic, A.~Klemm, M.~Marino and C.~Vafa,
  ``The topological vertex,''
  Commun.\ Math.\ Phys.\  {\bf 254}, 425 (2005)
  [arXiv:hep-th/0305132].
}

\lref\GMNII{
  D.~Gaiotto, G.~W.~Moore and A.~Neitzke,
  ``Wall-crossing, Hitchin Systems, and the WKB Approximation,''
  arXiv:0907.3987 [hep-th].
}

\lref\GMNIII{
  D.~Gaiotto, G.~W.~Moore and A.~Neitzke,
  ``Framed BPS States,''
  arXiv:1006.0146 [hep-th].
}

\lref\VC{
  S.~Cecotti and C.~Vafa,
  ``BPS Wall Crossing and Topological Strings,''
  arXiv:0910.2615 [hep-th].
}

\lref\Gukov{
  T.~Dimofte, S.~Gukov and Y.~Soibelman,
  ``Quantum Wall Crossing in N=2 Gauge Theories,''
  arXiv:0912.1346 [hep-th].
}

\lref\OoguriWC{
  H.~Ooguri, P.~Sulkowski and M.~Yamazaki,
  ``Wall Crossing As Seen By Matrix Models,''
  arXiv:1005.1293 [hep-th].
}

\lref\Sulkowski{
  P.~Sulkowski,
  ``Wall-crossing, free fermions and crystal melting,''
  arXiv:0910.5485 [hep-th].
}

\lref\MJ{
  D.~L.~Jafferis and G.~W.~Moore,
  ``Wall crossing in local Calabi Yau manifolds,''
  arXiv:0810.4909 [hep-th].
}

\lref\AOVY{
  M.~Aganagic, H.~Ooguri, C.~Vafa and M.~Yamazaki,
  ``Wall Crossing and M-theory,''
  arXiv:0908.1194 [hep-th].
}

\lref\DB{
  D.~Berenstein and M.~R.~Douglas,
  ``Seiberg duality for quiver gauge theories,''
  arXiv:hep-th/0207027.
}

\lref\OYII{
  H.~Ooguri and M.~Yamazaki,
  ``Emergent Calabi-Yau Geometry,''
  Phys.\ Rev.\ Lett.\  {\bf 102}, 161601 (2009)
  [arXiv:0902.3996 [hep-th]].
}

\lref\FengBN{
  B.~Feng, A.~Hanany, Y.~H.~He and A.~M.~Uranga,
  ``Toric duality as Seiberg duality and brane diamonds,''
  JHEP {\bf 0112}, 035 (2001)
  [arXiv:hep-th/0109063].
}

\lref\Beasley{
  C.~E.~Beasley and M.~R.~Plesser,
  ``Toric duality is Seiberg duality,''
  JHEP {\bf 0112}, 001 (2001)
  [arXiv:hep-th/0109053].
}

\lref\benyoung{
B.~Young
``Computing a pyramid partition generating function with dimer shuffling,''
[arXiv:0709.3079]
}

\lref\Seiberg{
  N.~Seiberg,
  ``Electric - magnetic duality in supersymmetric nonAbelian gauge theories,''
  Nucl.\ Phys.\  B {\bf 435}, 129 (1995)
  [arXiv:hep-th/9411149].
}

\lref\Denef{
  F.~Denef,
  ``Supergravity flows and D-brane stability,''
  JHEP {\bf 0008}, 050 (2000)
  [arXiv:hep-th/0005049].
}

\lref\Mayr{
  P.~Mayr,
  ``Phases of supersymmetric D-branes on Kaehler manifolds and the McKay
  correspondence,''
  JHEP {\bf 0101}, 018 (2001)
  [arXiv:hep-th/0010223].
}

\lref\ADT{
  P.~S.~Aspinwall,
  ``D-Branes on Toric Calabi-Yau Varieties,''
  arXiv:0806.2612 [hep-th].
}

\lref\msw{
  J.~M.~Maldacena, A.~Strominger and E.~Witten,
  ``Black hole entropy in M-theory,''
  JHEP {\bf 9712}, 002 (1997)
  [arXiv:hep-th/9711053].
}

%

\lref\HerzogDP{
  C.~P.~Herzog,
  ``Exceptional collections and del Pezzo gauge theories,''
  JHEP {\bf 0404}, 069 (2004)
  [arXiv:hep-th/0310262].
}

\lref\HerzogDIB{
  C.~P.~Herzog and J.~Walcher,
  ``Dibaryons from exceptional collections,''
  JHEP {\bf 0309}, 060 (2003)
  [arXiv:hep-th/0306298].
}

\lref\HK{
  C.~P.~Herzog and R.~L.~Karp,
  ``Exceptional collections and D-branes probing toric singularities,''
  JHEP {\bf 0602}, 061 (2006)
  [arXiv:hep-th/0507175].
}

\lref\HKII{
  C.~P.~Herzog and R.~L.~Karp,
  ``On the geometry of quiver gauge theories: Stacking exceptional
  collections,''
  arXiv:hep-th/0605177.
}

\lref\Sharpe{
E.~Sharpe,
``Lectures on D-branes and sheaves,''
arXiv:hep-th/0307245.
}

\lref\Morrison{
  M.~R.~Douglas, B.~R.~Greene and D.~R.~Morrison,
  ``Orbifold resolution by D-branes,''
  Nucl.\ Phys.\  B {\bf 506}, 84 (1997)
  [arXiv:hep-th/9704151].
}

\lref\Okoun{
A.~Okounkov,
``Random surfaces enumerating algebraic curves,''
arXiv:math-ph/0412008
}

\lref\MOOP{
D.~Maulik, A.~Oblomkov, A.~Okounkov, and R.~Pandharipande,
``Gromov-Witten/Donaldson-Thomas correspondence for toric 3-folds,''
arXiv:0809.3976
}

\lref\Diaconescu{
  D.~E.~Diaconescu and J.~Gomis,
  ``Fractional branes and boundary states in orbifold theories,''
  JHEP {\bf 0010}, 001 (2000)
  [arXiv:hep-th/9906242].
}

\lref\DiacWC{
W.~Chuang, D.~Diaconescu, G.~Pan,
``Rank Two ADHM Invariants and Wallcrossing,''
arXiv:1002.0579
}

\lref\Stoppa{
J.~Stoppa,
``D0-D6 states counting and GW invariants,''
arXiv:0912.2923
}

\lref\Toda{
Y.~Toda,
``On a computation of rank two Donaldson-Thomas invariants,''
arXiv:0912.2507
}

{\Title{\vbox{\hbox{}}}
{\vbox{
\centerline{Wall Crossing, Quivers and Crystals}}}
\vskip 0.7 cm
\centerline{\bf Mina Aganagic$^{1,2,3}$ and Kevin Schaeffer$^{1}$}
}
\vskip.2cm
\centerline{\it ${}^1$Center for Theoretical Physics, Department of Physics,}
 \centerline{\it University of California, Berkeley, CA 94720, USA}
\centerline{\it ${}^2$Department of Mathematics, }
\centerline{\it University of California, Berkeley, CA 94720, USA }
\centerline{\it ${}^3$Institute for the Physics and Mathematics of the Universe, }
\centerline{\it University of Tokyo, Kashiwa City, Chiba 277-8568, Japan}

\abstractmod{We study the spectrum of BPS D-branes on a Calabi-Yau manifold using the $0+1$ dimensional quiver gauge theory that describes the dynamics of the branes at low energies.
The results of Kontsevich and Soibelman \KS\ predict how the degeneracies change. We argue that Seiberg dualities of the quiver gauge theories, which change the basis of BPS states, correspond to crossing the ``walls of the second kind" in \KS. There is a large class of examples, including local del Pezzo surfaces, where the BPS degeneracies of quivers corresponding to one D6 brane bound to arbitrary numbers of D4, D2 and D0 branes are counted by melting crystal configurations. We show that the melting crystals that arise are a discretization of the Calabi-Yau geometry. The shape of the crystal is determined by the Calabi-Yau geometry and the background B-field, and its microscopic structure by the quiver $Q$. 
We prove that the BPS degeneracies computed from $Q$ and $Q'$ are related by the Kontsevich Soibelman formula, using a geometric realization of the Seiberg duality in the crystal.  We also show that, in the limit of infinite B-field, the combinatorics of crystals arising from the quivers becomes that of the topological vertex. We thus re-derive the Gromov-Witten/Donaldson-Thomas correspondence.
} 

\goodbreak
\vfill
\eject


\newsec{Introduction}

There has been remarkable recent progress in understanding the spectra of BPS states of ${\cal N}=2$ theories in 4 dimensions,
driven in part by the mathematical conjectures of Kontsevich and Soibelman
 \KS . KS conjectured how the degeneracies of BPS states change as we cross walls of marginal stability. In some cases, we have a physical understanding of why the conjectures of \KS\ are true. In particular, for BPS states in a gauge theory, the results of \KS\
have been explained in \refs{\GMNI, \GMNII}, and from a different perspective in \VC\ (see also the very recent \GMNIII)\foot{See \OoguriWC\ for a matrix model perspective, and \refs{\DiacWC, \Stoppa, \Toda} for more mathematical progress.}.

In this paper we do three things. First, we give the physical explanation for the ``walls of the second kind" in \KS.
Second, we provide further evidence that the conjectures of \KS\ are true -- by proving that the spectrum of BPS bound states of a D6 brane with D4, D2 and D0 branes wrapping toric Calabi-Yau manifolds satisfies them, for certain walls. The spectrum here is in general very complicated (far more so than in the examples studied so far, corresponding either to four dimensional gauge theories \Gukov, or Calabi-Yau manifolds without compact four cycles, where the generating function of the spectrum is computable in closed form \refs{\MJ, \JW, \AOVY, \OY}.). 
Finally, we use this to shed new light on a relation between familiar objects: topological strings, Calabi-Yau crystals, and BPS D-branes. These were studied previously in \refs{\ORV, \QF,\MNOPI,\MNOPII,\OYII}.

\subsec{Walls of the Second Kind and Seiberg Dualities}

There are two distinct phenomena that affect the BPS spectrum. One, which was most studied in the literature, e.g. in \DM , corresponds to crossing a wall of marginal stability where central charges of a pair of states align. There, the degeneracies of bound states of the pair change. When the BPS states are described in a quiver gauge theory, we cross the wall by varying the Fayet-Iliopoulos terms. We interpret the ``walls of the second kind"  of \KS , as a kind of Seiberg duality, that changes the basis of BPS branes and the split of the spectrum into the branes and the anti-branes. In the quiver gauge theory, this can be made precise. The basis of the BPS branes is provided by the nodes of the quiver. The branes are described as linear combinations of these with positive coefficients which are the ranks of the quiver gauge groups. The walls of the second kind correspond to varying the gauge couplings $g^{-2}$ of the nodes. When one of them passes through zero, we need to change the description to a new quiver, related to the original one by Seiberg duality \refs{\DB, \FengBN, \Beasley, \GU}.  This replaces the brane of the corresponding node with its anti-brane. The action of Seiberg duality on the nodes of the quiver is exactly the change of basis that appears in \KS.  Once the basis vectors change, the possible bound states they can form change as well. So, crossing the walls of the second kind also affects the spectrum of BPS states. We will work out examples of this,
starting from a simple one, with an acyclic quiver, in section 2. In section 3, we set this up in more detail, in the context of D3 branes wrapping three-cycles of a local Calabi-Yau $Y$. $Y$ is related by mirror symmetry to a toric Calabi-Yau $X$, which maps the D3 branes to D6, D4, D2 and D0 branes, as we review in section 5.

\subsec{Dimer models, Seiberg Dualities, and BPS Degeneracies} 
Recently, \RM\ gave a remarkably simple way of computing degeneracies of BPS bound states of one D6 brane on a toric Calabi-Yau $X$, with D4 branes, D2 branes and D0 branes, generalizing the earlier work of \sz\ to essentially arbitrary toric Calabi-Yau singularities. Adding a D6 brane corresponds to extending the D4-D2-D0 quivers\foot{These were studied extensively in \refs{\Morrison,\Diaconescu, \hb, \DB, \aspb, \ADT, \FengBN, \Beasley, \GU, \HerzogDP, \HerzogDIB, \HK, \HKII, \Wij, \MWI, \HVH}.}  by a node of rank $1$ \OY, in a manner which we make precise.
While in principle BPS degeneracies of a quiver are computable for any given choice of ranks as an Euler characteristic of the moduli space, the direct computations become cumbersome as ranks increase. Instead, as we review in section 4, \RM\ give a combinatorial way to compute the degeneracies for any ranks, by counting perfect matchings of certain dimer models on a plane. (This can be rephrased in terms of counting melting crystals, as we will explain shortly.) The relevant dimers are the lift to ${\bf R^2}$ of dimers on ${\bf T}^2$ that correspond to D4-D2-D0 quivers in \refs{\hb, \KV}.

We show that crossing the walls of the second kind that take a quiver $Q$ to its Seiberg dual $Q'$  corresponds to a simple geometric transition in the dimer. In the case of the dimer on ${\bf T}^2$, this was known from \hb , and what we have here is a simple lift of this from the dimers on ${\bf T}^2$ to the dimers on the plane. Since perfect matchings of the planar dimer count BPS states, this gives a geometric description for how the spectrum jumps. We show (in section 4) that this can be used to prove the KS wall crossing formula in the context of the quivers of \RM .  

In section 6 we give another example of wall crossing of the second kind corresponding to varying the B-field on $X$. Since shifting the B-field brings us back to the same point in the moduli space of $X$, this should leave the spectrum of BPS states invariant.  Indeed, we show that if one turns on compact $B$-field, $B\in H^2_{cmpct}(X, {\bf Z})$, this corresponds to a sequence of Seiberg dualities of the quiver, but in the end the quiver comes back to itself. Shifting by a $B$-field in $B\in H^2(X, {\bf Z})/H^2_{cmpct}(X, {\bf Z})$ also corresponds to a change of basis of branes and Seiberg duality, but this time the quiver does not come back to itself. Instead, this change of basis permutes the quivers describing D6 branes with different amounts of noncompact D4 brane charge.

\subsec{Calabi-Yau Crystals, Quivers and Topological Strings}
There is an intriguing connection between the dimers that appear in \RM\ and an earlier appearance of dimers in the context of the closed topological string, in \ORV.
As explained in \refs{\ORV, \DA,\RM, \OY}\  there is a close relation between dimer models in the plane, with suitable boundary conditions, and three dimensional melting crystals.
We show that the melting crystals of \RM\ have a beautifully simple geometric description: the crystal sites are a discretization of the Calabi-Yau geometry. The shape of the crystal is determined by the geometry of the Calabi-Yau base, which is a singular cone at the point in the moduli space where the quiver is defined.  The crystal sites are the integral points of the Calabi-Yau.
The precise microscopic structure of the crystal depends on the quiver (and changes under Seiberg dualities). The refinement comes from the fact that we are not counting bound states of the D6 brane with D0 branes, but the more general bound states with D4, D2 and D0 branes, corresponding to splitting of the D0 branes into fractional branes.
We show that, increasing the B-field by $D\in H^2(X,{\bf  Z})$,  D6 brane bound states are counted by the crystal that takes the shape of the Calabi-Yau $X$ with Kahler class $D$, while the microscopic structure of the crystal does not change. 

The relation of topological string amplitudes on $X$ with certain melting crystals was observed in \ORV. The Calabi-Yau crystals that arise in our paper are the same as those in \refs{\ORV, \QF}, but only in the limit of infinite $D$, where the microscopic structure of the crystal is lost. The crystals in \ORV\ were interpreted in \QF\ as counting bound states of a D6 brane on $X$, formulated as the Witten index of a non-commutative ${\cal N} = 2$ SYM on $X$. The observations of \QF\ are known as the Gromov-Witten/Donaldson-Thomas correspondence \refs{\MNOPI,\MNOPII}. 
We thus provide a new derivation of the Gromov-Witten/Donaldson-Thomas correspondence from the perspective of the $0+1$ dimensional quiver quantum mechanics, but only in the limit of large $D$.
This is in agreement with \DM, which pointed out that the correspondence of  \QF\ can hold only in the limit of infinite B-field (this was also verified in \AOVY\ when $X$ has no compact 4-cycles). This is described in section 7.

\newsec{Walls of the Second Kind and Seiberg Duality}

Consider BPS states from D-branes wrapping cycles in a Calabi-Yau $Y$. For definiteness, take IIB string theory, so that the BPS particles are labeled by charge
$$\Delta \in H_{3}(Y, {\bf Z}).
$$ 
The mass of the BPS state and the supersymmetry it preserves are determined by the central charge,
$$ Z(\Delta) = \int_{\Delta} \Omega,$$
where $\Omega$ is the $(3,0)$ form on $Y$.
Near the singularities in $Y$, the D-branes can be described by quiver gauge theories in $0+1$ dimensions. The nodes of the quiver correspond to a basis of $H_3(Y,{\bf Z})$. Any bound state of the branes with charge $\Delta$ can be written as
\eqn\wedge{
\Delta = \sum_{\alpha} N_{\alpha} \Delta_{\alpha}, \qquad N_{\alpha} \geq 0.
}
The corresponding quiver quantum mechanics is a 
$$
G_{\Delta} = \prod_{\alpha} U(N_{\alpha}) 
$$
gauge theory. The quiver is a good description when all the central charges of the nodes are nearly aligned. 
By choosing an overall phase of $\Omega$, we can write
$$
Z(\Delta_{\alpha} ) = {i \over g_\alpha^2} +  \theta_{\alpha},
$$
corresponding to all $Z$'s being close to imaginary. Above, $\theta_{\alpha}$ is the Fayet-Iliopoulos parameter, and $g_{\alpha}$ is the gauge coupling in the quantum mechanics. The BPS degeneracy of a state $\Delta$ is determined by the Witten index %
\eqn\ind{
\Omega_Q(\Delta) = {\rm Tr}_{\Delta, Q}(-1)^F,
}
of the quiver with gauge group $G_{\Delta}$.\foot{From the space time perspective, we are computing $-2 {\rm Tr} F^2 (-1)^F$. The $F^2$ factor serves to absorb the contribution of zero modes on $R^{3,1}$. This reduces then to ${\rm Tr}(-1)^F$ where one traces internal degrees of freedom only \DM .}
Thus, the quiver gauge theory provides both a basis of D-branes, and a means to compute the degeneracies.

At walls in the moduli space, the spectrum of BPS states can change. One kind of wall is a wall of marginal stability,  where central charges of two states, for example of two nodes, align. At the wall, even though \ind\ is an index, it can jump, since the one particle states it is counting can split, or pair up. In the quiver, since we are near the intersection of walls anyhow, crossing this wall corresponds to some combination of FI parameters passing through zero.

Before we go on, note that the quiver only describes the states with non-negative $N_{\alpha}$. This is just as well, since the central charges of anti-branes $-\Delta_{\alpha}$ would be anti-parallel to the rest of the states in the quiver. The states with some of the $N_{\alpha}$'s negative and some positive do not form bound states, and the quiver does not miss anything\foot{There are of course the bound states with all $N_{\alpha}$ negative, but these are just CPT conjugates of the states at hand.}. Moreover, this remains true even when the central charges of $\Delta_{\alpha}$ are {\it not} nearly aligned, as long as they all remain in the upper half of the complex plane -- simply because there are no walls where $\Delta_\beta$ and $-\Delta_{\alpha}$ can align. 
The fact that only the states in the upper half of the complex $Z-$plane bind, implies that the spectrum can change as a node $\Delta_*$ leaves the upper half of the complex plane, and correspondingly $-\Delta_*$ enters it. 
The node $\Delta_*$ leaves the upper half of the complex plane when $g_*^{-2}$ passes through zero. This wall, in real codimension one of the moduli space, was called the ``wall of the second kind'' in \KS . %

On the wall, $g_*^{-2}=0$, the gauge coupling of the node is infinitely strong. Even though in $0+1$ dimensions the gauge fields have no local dynamics, the coupling enters the action as the coefficient of the kinetic terms for the fields on the brane, e.g.,
${1\over g_*^2} \int |\del \phi |^2$. We can continue past infinite coupling since clearly nothing special happens to the D-branes, as long as we stay away from $Z(\Delta_*)=0$, but we need to change description. To make $g_{*}^2$ positive, we need to flip 
\eqn\flop{
\Delta_{*} \rightarrow \Delta_*' = - \Delta_{*}.
}
The existence of the two dual descriptions of the same state, related by a continuation past infinite coupling \flop , is a Seiberg duality \refs{\GU, \DB, \Seiberg, \FengBN, \Beasley}. In fact, $\Delta_{*}$ is not the only node that changes. Were we to complete the circle around $Z(\Delta_*)=0$, all of the nodes would have changed due to monodromy, which maps $\Delta$ to $\Delta\pm(\Delta \circ \Delta_*) \Delta_*$, for any $\Delta$, with the sign depending on the path. Going halfway around the circle, there are partial monodromies, which are Seiberg dualities \DB.
On the other side of the wall, the theory is described by a dual gauge theory, based on
$\{\Delta_{\alpha}'\}$ and a new quiver $Q'$. 

A state $\Delta$ in \wedge , can have descriptions both in terms of $Q$ and $Q'$. In addition to the one based on $Q$ and $G_{\Delta}$, it may have another one, in terms of $Q'$ with gauge group
$$
G'_{\Delta} = \prod_{\alpha} U(N'_{\alpha}) 
$$
where
\eqn\wedgetwo{
\Delta = \sum_{\alpha} N'_{\alpha} \Delta'_{\alpha},
}
if $N_{\alpha}'$ are non-negative.  Crossing the ``wall of the second kind,'' the spectrum can change because the spaces of BPS bound states one can obtain 
 from $Q$ and $Q'$  (by varying FI parameters, while staying in the upper half of the complex plane) are different. In the terminology of KS, this is a change of $t$-structure.

\subsec{KS conjecture and Seiberg duality}
Kontsevich and Soibelman \KS\ predict how the degeneracies $\Omega_Q(\Delta)$ change as the central charges $Z$ are varied in some way.
For each charge $\Delta \in H_3(Y, {\bf Z})$, associate an operator $e_{\Delta}$, satisfying
\eqn\alg{
[e_{\Delta}, e_{\Delta'}] = \hbar (-1)^{\Delta\circ \Delta'}  \Delta \circ \Delta' \;e_{\Delta+\Delta'} 
}
and
$$
e_{\Delta} e_{\Delta'} = (-1)^{\Delta\circ \Delta'} e_{\Delta+\Delta'},
$$
where $\Delta \circ \Delta'$ is the intersection product on $H_3(Y)$, and an operator%
$$
A_{\Delta} = \exp({1\over \hbar}\sum_{n =1}^{\infty} {e_{n \Delta}\over n^2 }).
$$
Consider the product \foot{In general, the degeneracies $\Omega_Q(\Delta)$ in the KS formula may differ from the physical BPS degeneracies by a $\Delta$-dependent sign.  This additional sign factor will be important in section 4, when we use the KS formula to compute wall crossing for quiver gauge theories.}

\eqn\KSquiver{A_Q = 
\prod_{\Delta}^{\rightarrow} A_\Delta^{\Omega_Q(\Delta)}
}
taken over all states $\Delta$ with central charges in the upper half of the complex plane, and in the order of increasing phase of the central charges $Z(\Delta).$ 
The KS conjecture states that, if we change the central charges in any way, the BPS degeneracies adjust so that the product, taken over the states with central charges in the upper half plane, is invariant\foot{More precisely, \KS\ restrict to a ``strict" wedge, meaning one that subtends an angle less than $180^\circ$. This simply tells us how to define $A_Q$ exactly on the wall, when states $\Delta_{\alpha}$ and $-\Delta_{\alpha}$ both have central charges on the real line.}. 

Consider what happens as we cross the wall 
where the coupling $g_{*}^{2}$ of node $\Delta_{*}$
flips sign. Near the wall, the chamber corresponding to the quiver is as in the figure below. As we cross the wall, the state $\Delta_{*}$ leaves the upper half 
plane from the left, and the state $-\Delta_{*}$ enters it from the right, so that the new degeneracies should satisfy
\eqn\quiverwc{
A_Q' = A_{\Delta_{*}}^{-1} {A_Q} A_{-\Delta_{*}}.
}
%
\bigskip
\centerline{\epsfxsize 6.3truein\epsfbox{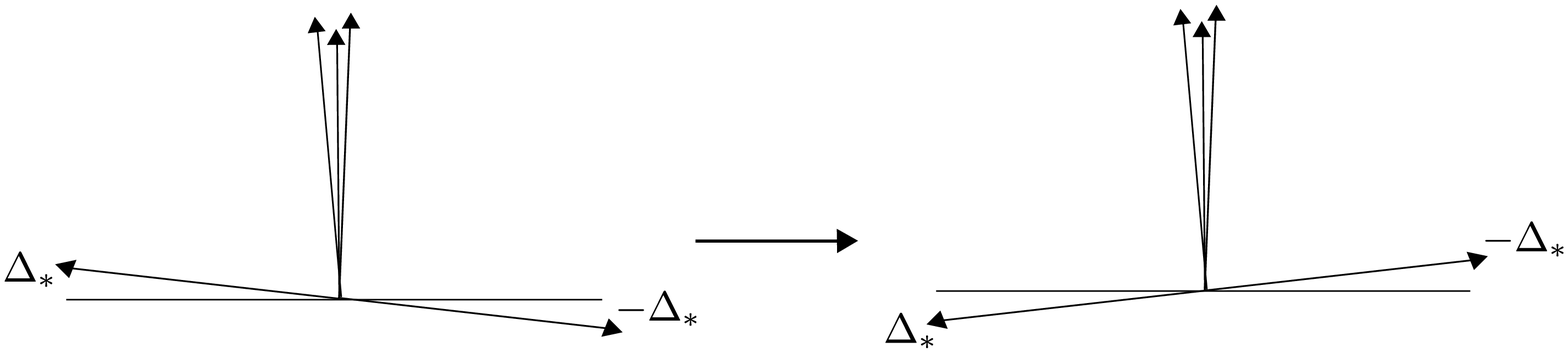}}
\noindent{\ninepoint \baselineskip=2pt {\bf Fig. 1.} {Wall of the Second Kind.  As we cross this wall, the central charge of $\Delta_{*}$ leaves the upper half plane.     }}
\bigskip

Crossing the wall as above, we generally do not come back to the same point in the moduli space, so $A_{Q}$ and $A_Q'$ are not themselves equal, not even after a change of basis. However, the spectrum should be determined uniquely by the point in the moduli space we are at. Going around a closed loop in the moduli space, the spectrum of BPS states must come back to itself, up to a monodromy that relabels the branes.  In particular, it should not matter which way we go around the singularity. 

\subsec{A simple example}
As a simple example, consider the quiver $Q$ with two nodes, $\Delta_1$ and $\Delta_*$, and one arrow, corresponding to the intersection number $\Delta_1 \circ \Delta_* = 1$.  With the central charges as in figure 2, there are only three BPS states, and the operator $A_Q$ is simply \KS\
\eqn\begn{
A_Q = A_{\Delta_1}A_{\Delta_1+\Delta_*}A_{\Delta_*}.
}
%
%
%
\bigskip
\centerline{\epsfxsize 5.0truein\epsfbox{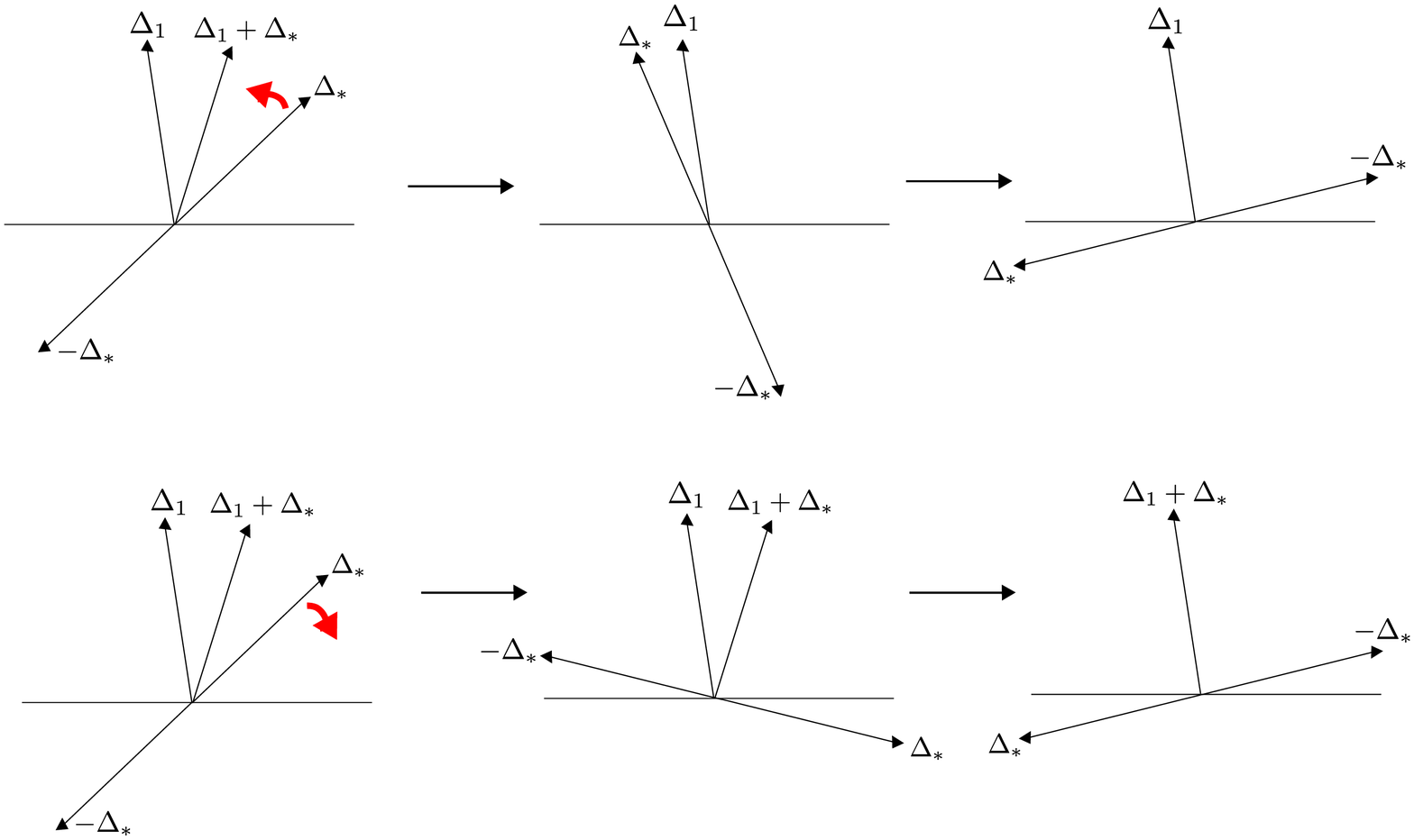}}
\noindent{\ninepoint \baselineskip=2pt {\bf Fig. 2.} {Rotating the central charge of $\Delta_{*}$ counterclockwise (above) and clockwise (below).  Note that in both cases we cross a wall of the first kind and a wall of the second kind.}}
\bigskip

Rotating the central charge of $\Delta_*$ counterclockwise, eventually $Z(\Delta)$ and $Z(\Delta_*)$ align, and we cross a wall of the first kind.
The bound state $\Delta_1+\Delta_*$ decays, and the product becomes\foot{We used the pentagon identity $A_{\Delta_1}A_{\Delta_1+\Delta_*}A_{\Delta_*}=A_{\Delta_*}A_{\Delta_1}$ which holds for any two states $\Delta_1$, $\Delta_*$ with intersection number $+1$\ \KS.}
$$A_Q=A_{\Delta_*}A_{\Delta_1}.$$
Continuing past this, eventually the gauge coupling of $\Delta_*$ becomes negative, so $\Delta_*$ leaves the upper half of the complex plane, and $-\Delta_*$ enters it. To get a good description, we need to change the quiver from $Q$, to $Q'$ with nodes 
$\Delta_1$, $ -\Delta_*.$ 
This corresponds to Seiberg duality on the node $\Delta_*$ (This quiver and its Seiberg dualities were studied in detail in \DB.)
with
$$
A_{Q'} = A_{\Delta*}^{-1} A_QA_{-\Delta_*}= A_{\Delta_{1}}A_{-\Delta_{*}}.
$$
Now, we could have reached the same point in the moduli space by starting with \begn\ and rotating $Z(\Delta_*)$ clockwise instead. Then $\Delta_*$ leaves the upper half of the complex plane, $-\Delta_*$ enters it. We again need to dualize node $\Delta_*$, but now we get the basis
$\Delta_1+\Delta_*$, $ -\Delta_*$ 
corresponding to quiver $Q''$. Moreover,
$$
A_{Q''} = A_{-\Delta*}  A_{Q}  A_{\Delta*}^{-1}=   A_{-\Delta_*} A_{\Delta_1}  A_{\Delta_1+\Delta_*}.
$$
To get this to correspond to the same point in the moduli space as $Q'$ above, we need to keep rotating $Z(\Delta_*)$, until $A_{Q''}$ becomes 
$$
A_{Q''} = A_{\Delta_1+\Delta_*} A_{-\Delta_*} .
$$ 
Since $Q'$ and $Q''$ now correspond to two quivers describing physics at exactly the same point in the moduli space, they are of course equivalent. The non-trivial relation between $A_{Q''}$ and $A_{Q'}$ is a consequece of the fact that to relate them, we need to go once around the $Z(\Delta_*) = 0$. In doing so, there is a monodromy acting on the cycles that maps
$$
\Delta \rightarrow  \Delta + (\Delta\circ \Delta_*) \Delta_* ,
$$  
looping counter-clockwise, which is precisely how these are related. As an aside, note that, taking for example the state $\Delta_1+\Delta_*$ of quiver $Q$,  depending on which way we go around the singularity there
are two different interpretations for its fate. Along the path corresponding to $Q'$, the state decays into ${\Delta}_1$ and ${\Delta_*}$ on the wall of marginal stability where $\Delta_*$ and $\Delta_1$ align. From the perspective of the split attractor flows \Denef, the flow corresponding to $\Delta_1+\Delta_*$ splits on this wall into a flow corresponding to $\Delta_*$, which crashes at $Z(\Delta_*)=0$, and an honest black hole attractor corresponding to $\Delta_1$. Along the path corresponding to $Q''$, the state crosses no walls, but the monodromy 
changes $\Delta_1+\Delta_*$ to $\Delta_1$. Following either flow, the attractor point is the same, as expected from \Denef.\foot{The change of basis of BPS states was also recently discussed in \ref\GM{G. Moore, Talk at Strings 2010 Conference.}, from the attractor viewpoint.}

\newsec{Quivers from Calabi-Yau Threefolds}

In this section we will review (following \GU) the quiver gauge theories $Q$ that arise for certain choices of a Calabi-Yau $Y$, and its moduli. One reason we choose these theories is that one has a very direct, geometric understanding of what happens to $Q$ as the central charges are varied, and the theory undergoes Seiberg dualities. 
The second reason is that the choice we make implies that the quiver theory has extra symmetries --  $Q$ are the so called toric quiver gauge theories of \refs{\hb, \KV} (see \Kennaway\ for an excellent review). 
The torus symmetries will allow us to extract very precise information about the quantum BPS spectra of $Q$, in the next section. The presence of the extra symmetries is related to the fact, aspects of which we will review in the next section, that $Y$ is mirror 
to a toric Calabi-Yau $X$.

Consider a Calabi-Yau $Y$ given by 
\eqn\mirrtwo{
W(e^x,e^y)=w\qquad\qquad uv=w.}
We can view $Y$ as a fibration over the $w$ plane.
At a generic point in the $w$-plane, the fiber is a product of a cylinder $uv=w,$ and a Riemann surface $W(e^x,e^y)=w$. Over special points, the fiber degenerates. Over a point $q$ with $w(q)=0$,  the $S^1$ of the cylinder pinches. The 1-cycles of the Riemann surface degenerate over
critical points of $W(e^{x},e^{y})$,
\eqn\crit{
p_{\alpha}:   \;\; \del_x W = 0=\del_y W ,  \qquad \alpha=1, \ldots, r
}
where $w(p_{\alpha}) =w_{\alpha}$. For each $p_{\alpha}$, a path in the $w$-plane connecting it with $q$, together with an $S^1 \times S^1$ fiber over it,
gives an 
 $S^3$ which we will denote 
$$\Delta_{\alpha}, \qquad \alpha=1, \ldots, r
$$
Above, one of the $S^1$'s corresponds to the cylinder, while the other corresponds to the 1-cycle of the Riemann surface pinching at $p_{\alpha}$).  The three-cycles $\Delta_{\alpha}$ provide a  basis for the compact homology of $Y$.
We can associate a quiver to the above singularity by considering D3 branes wrapping the cycles $\Delta_{\alpha}$. Since the $\Delta_{\alpha}$'s are spheres, the
D-branes wrapping them have no massless adjoint matter. For a collection of 
\eqn\wedge{
\Delta = \sum_{\alpha} N_{\alpha} \Delta_{\alpha}, \qquad N_{\alpha} \geq 0.
}
branes, we get a 
\eqn\gg{
G = \prod_{\alpha} U(N_{\alpha}) 
}
quiver gauge theory, with $\Delta_{\alpha}$ corresponding to node $\alpha$ of the quiver.
Moreover, for every point of intersection of $\Delta_{\alpha}$ and $\Delta_\beta$ we get a massless
chiral bifundamental in either $(N_{\alpha}, \bar{N}_{\beta})$, or $(N_{\beta}, \bar{N}_{\alpha})$ representation of the gauge group.
Pairs of these can get mass, so the net number of chiral multiplets going {\it from} node  ${\alpha}$ to node ${\beta}$ is 
$$n_{\alpha \beta} = \Delta_{\alpha}\circ {\Delta}_{\beta}.
$$ 

As we vary the moduli of $Y$, the locations of critical points $p_{\alpha}$ in the $w$ plane change.  If the critical point $p_{\alpha}$ crosses the cycle $\Delta_{\beta}$, due to monodromy, the homology of the cycle $\Delta_{\beta}$ changes to  $\Delta_{\beta}'$
\eqn\PL{
\Delta_{\beta}' = \Delta_{\beta} {\pm}(\Delta_{\beta} \circ \Delta_{\alpha}) \Delta_{\alpha},
}
where in \PL , $\Delta_{\beta}$ stands for the original homology class of the cycle (see figure 3).
%
%
\bigskip
\centerline{\epsfxsize 3.0truein\epsfbox{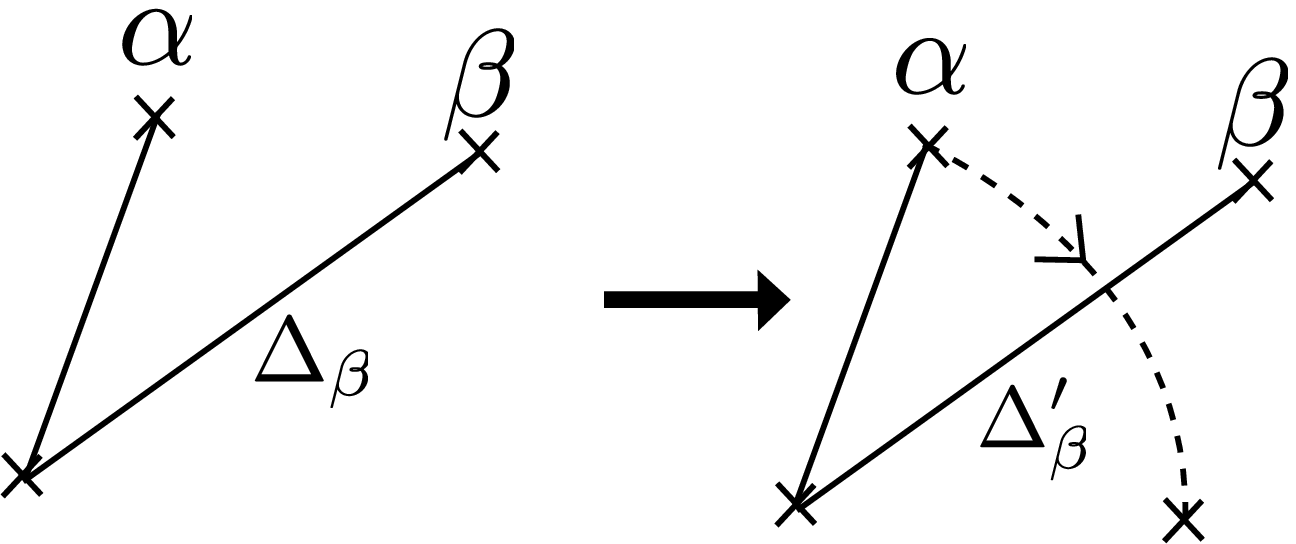}}
\noindent{\ninepoint \baselineskip=2pt {\bf Fig. 3.} {Picard lefschetz monodromy as $\alpha$ passes through the $\Delta_{\beta}$ cycle.}}
\bigskip

The fact that the homology classes of the cycles change implies that the quiver changes. For example, the intersection numbers $n_{\alpha\beta}$ change, and with them the number of arrows connecting the two nodes of the quiver. Consider varying the moduli so that the gauge coupling of the nodes $\Delta_*$ becomes negative. 
In the process, $p_*$ crosses the cycles $\Delta_{\beta_i}$, so that the basis of branes changes to
\eqn\plt{\eqalign{
\Delta_*'& = - \Delta_{*} \cr
\Delta_{\beta_j} '& = \Delta_{\beta_j} \pm n_{\beta_j *} \Delta_{*}\cr
\Delta_{\gamma_k}' &=  \Delta_{\gamma_k}}
}
%
%
\bigskip
\centerline{\epsfxsize 5.0truein\epsfbox{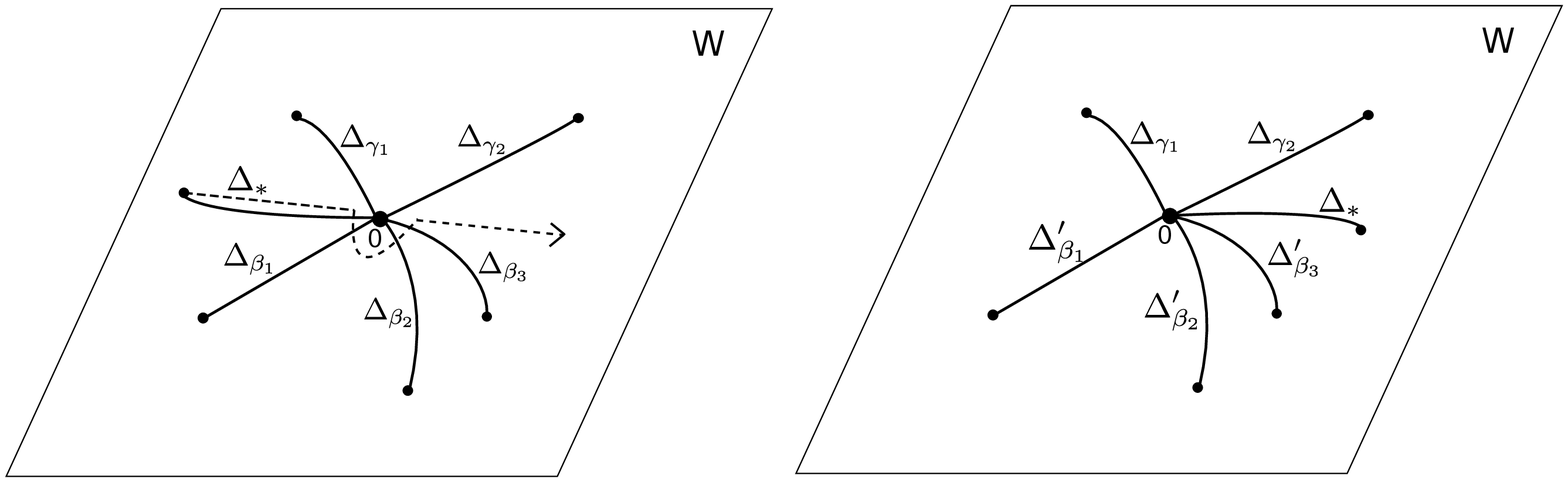}}
\noindent{\ninepoint \baselineskip=2pt {\bf Fig. 4.} {The $\Delta_{*}$ cycle is shown shrinking and then growing in the opposite direction.  In the process, $\Delta_{*}$ crosses the $\{\Delta_{\beta_{i}}\}$.}}
\bigskip
This implies that the intersection numbers change as 
\eqn\inter{\eqalign{
n'_{* \beta_{k}} & = - n_{* \beta_{k}} \cr
n'_{\beta_{k}\gamma_{j}}& = n_{\beta_{k}\gamma_{j}} \pm  n_{\beta_j *} n_{*\gamma_k}\cr
n'_{\gamma_j *} &= - n_{\gamma_{j} \alpha}}.
}

In addition, for the D3 brane charge to be conserved, the ranks of the quiver have to change, for the D-brane charge to be conserved. The numbers $N_{\alpha}$ of the branes on the node $\alpha$ have to change to $N_{\alpha'}$, so that 
$$
\Delta = \sum_{\alpha} N_{\alpha}\Delta_{\alpha} =  \sum_{\alpha} N'_{\alpha}\Delta'_{\alpha},
$$
in order to be consistent with \PL .

Note that \plt\inter\ are {\it exactly} the same changes of basis as on p. 134 of \KS . The transformation is a Seiberg duality of the quiver gauge theory \Herzog\aspb when either
$$
n_{* \beta_k } > 0, \qquad n_{* \gamma_j }\leq 0, 
$$
for all $\beta_k$, $\gamma_j$, or with the direction of inequalities reversed -- depending on the sign in \plt.\foot{When this is not the case, one should still get a dual description, though it may not correspond to the quiver gauge theory one would naively obtain. For example due to the presence of tachyons, the intersection numbers from the geometry need not count chiral multiplets. See, e.g. \aspb\Wij.} 
The choice of sign determines which way around the singularity we go. The new quiver $Q'$ is obtained from $Q$ by
(i) reversing the arrows beginning or ending on the node $*$ we dualized. The reversed arrows correspond to new chiral fields associated with the dual node, all of whose intersection numbers have flipped signs. (ii) The original bifundamentals transforming under the node 
$*$ are confined in the bifundamental mesons that no longer transform under gauge transformations on node $*$. (iii) There is an additional gauge invariant coupling of the mesons to the new bifundamentals charged under the dual node. These can however pair up with the existing bifundamentals of opposite orientations and disappear, since only the net intersection number, counted with signs is invariant. The net effect of (ii) and (iii) is to give a mass to all but $n_{\beta_{k}\gamma_{j}} \pm  n_{\beta_j *} n_{*\gamma_k}$  bifundamentals between the node $\beta_{k}$ and $\gamma_j$. 

 \subsec{The mirror of ${\bf P}^1\times {\bf P^1}$ example}

Consider for example,
\eqn\Wpp{
W(e^x, e^y) = e^x+z_t e^{-x} + e^{y} + z_s e^{-y} +1,
}
where $z_t = e^{-t}$, $z_s = e^{-s}$. There are four critical points in the $W$ plane, $w_{\alpha} = \pm 2{\sqrt z_t} \pm 2{\sqrt z_s}  +1$. 

\bigskip
\centerline{\epsfxsize 4.0truein\epsfbox{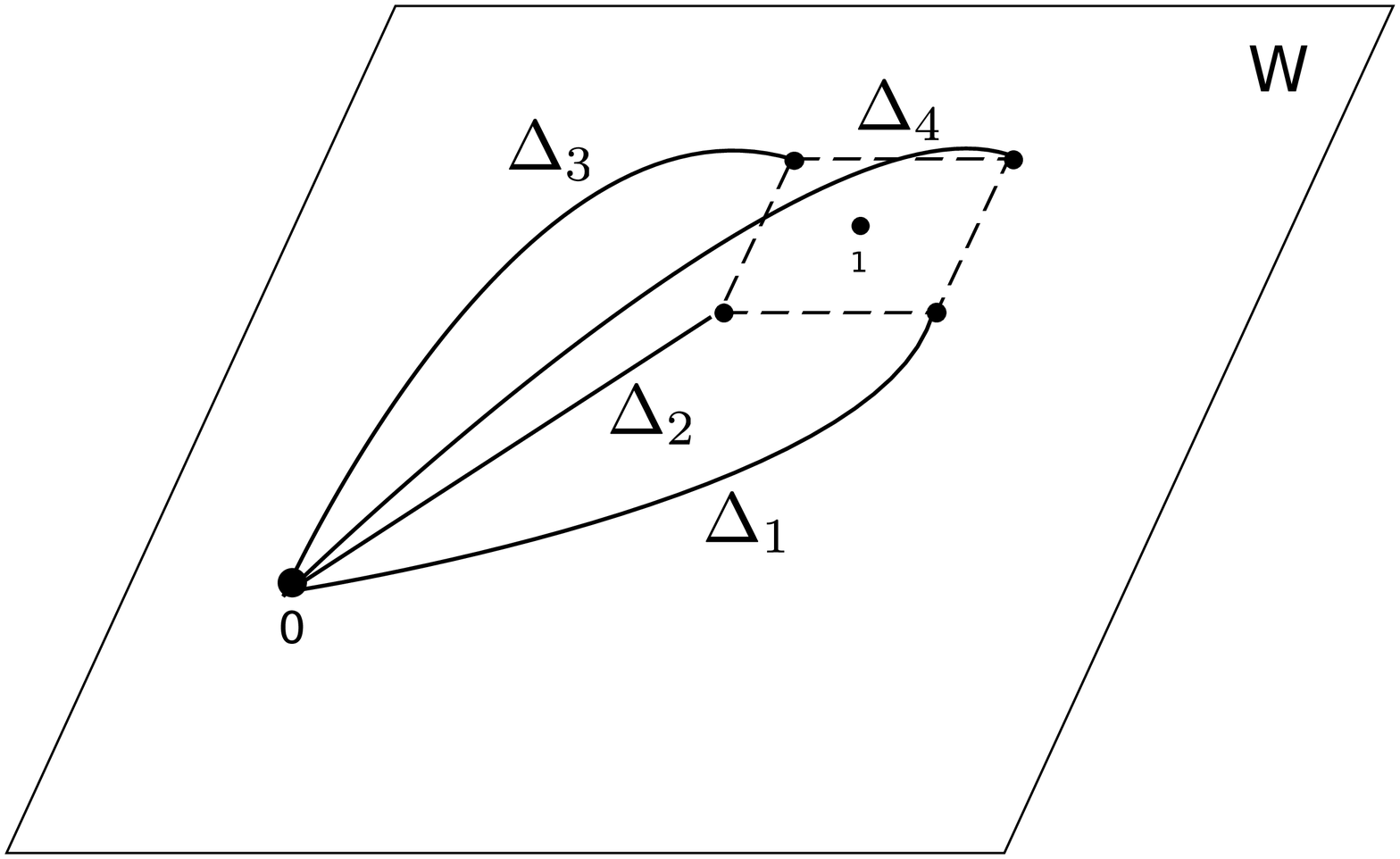}}
\noindent{\ninepoint \baselineskip=2pt {\bf Fig. 5.} {The W plane for the mirror of ${\bf P}^{1}\times {\bf P}^{1}$.}}
\bigskip

The four corresponding $S^3$'s are drawn in Figure 5, 
in the limit
$$z_s, z_t \rightarrow 0.
$$
The intersection numbers of the cycles were determined in \Iqbal . The quiver $Q$ that results has four nodes, connected by arrows
$$
n_{32} = n_{34} = n_{21}=n_{41} = 2, \qquad n_{13}= 4.
$$
The theory also has a superpotential, computed by the topological A-model on a disk, 
$$
{\cal W} = \sum_{i,j,a,b=1,2}\epsilon^{ij}\epsilon^{ab}\bigl(  {\rm Tr} B_{a} A_i D_{jb} + {\rm Tr} {\tilde A}_i {\tilde B}_a D_{jb}\bigr).
$$
where the fields are labeled in Figure 7.  As we vary the complex structure moduli of $Y$, the gauge coupling of one of the nodes, say node $2$ can become negative.
This can be achieved by sending 
$$z_{s}, z_{t} \rightarrow \infty,
$$ 
keeping the ratio $z_s/ z_t = e^{-T}$ fixed. As we vary $z$'s the cycles deform as in the Figure 6 \GU , corresponding to 
\eqn\chb{\eqalign{
\Delta_{1}'&= \Delta_1 + 2 \Delta_2 ,\cr
\Delta_2'& = -\Delta_2,\cr
\Delta_3'& = \Delta_3,\cr
 \Delta_4'& = \Delta_4.}
}
We chose the labeling of the new nodes in a manner that will be useful later.
%
\bigskip
\centerline{\epsfxsize 4.0truein\epsfbox{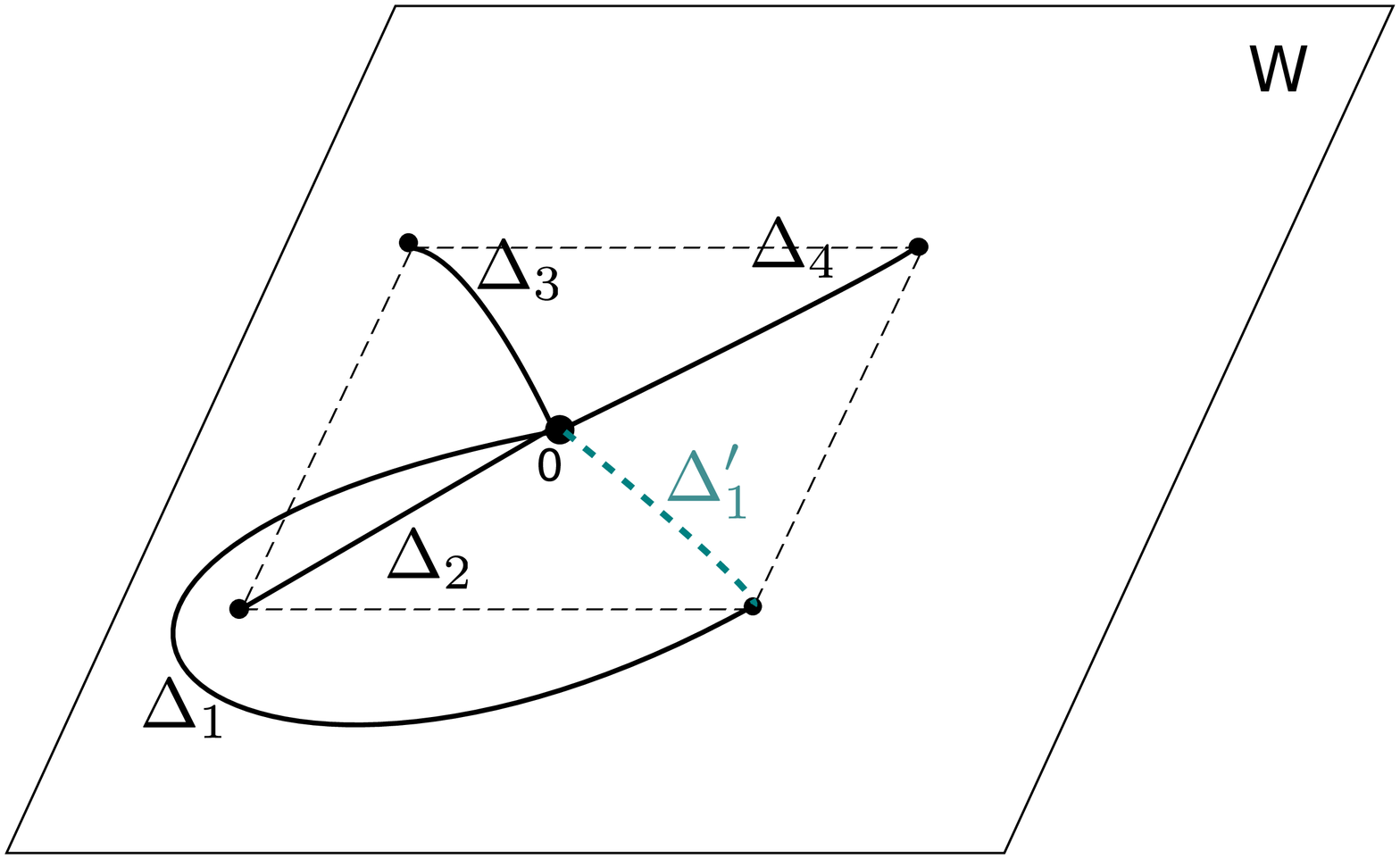}}
\noindent{\ninepoint \baselineskip=2pt {\bf Fig. 6.} {Deformation of the cycles as $z_{s}, z_{t} \to \infty$. Since $\Delta_{1}$ is deformed, the good cycle in this limit is now $\Delta_{1}'$.}}
\bigskip

%
This implies that the non-vanishing intersection numbers are now
$$
n'_{12} = n'_{23} = n'_{34}=n'_{41} = 2.
$$
This results in a new quiver, $Q'$, given in the Figure 7. The superpotential of the theory also changes, and becomes\foot{The dual theory is obtained by Seiberg duality --  instead of ${\tilde A}_i, {\tilde B}_\alpha$ we introduce two new pairs of fields $A_i', B_{\alpha}'$, with opposite orientation. The original fields are confined in mesons $M_{\alpha i} = {\tilde B}_{\alpha} {\tilde A}_i ,$ and the superpotential becomes 
$$W = \sum_{i,j, \alpha, \beta} \epsilon^{ij} \epsilon^{\alpha \beta}\bigl({\rm Tr} B_{\alpha} A_i D_{j \beta} +{\rm Tr}  M_{i \alpha} D_{j \beta} +  {\rm Tr} B'_{\alpha} A'_i M_{j \beta}\bigr). 
$$
The second term above makes both $M$ and $D$ massive, and they can be integrated out. This results in the effective superpotential we wrote.}
$$W = \sum_{i,j, \alpha, \beta} \epsilon^{ij} \epsilon^{\alpha \beta}{\rm Tr} B_{\alpha} A_i B'_{\beta} A'_{j}. 
$$
%
\bigskip
\centerline{\epsfxsize 4.0truein\epsfbox{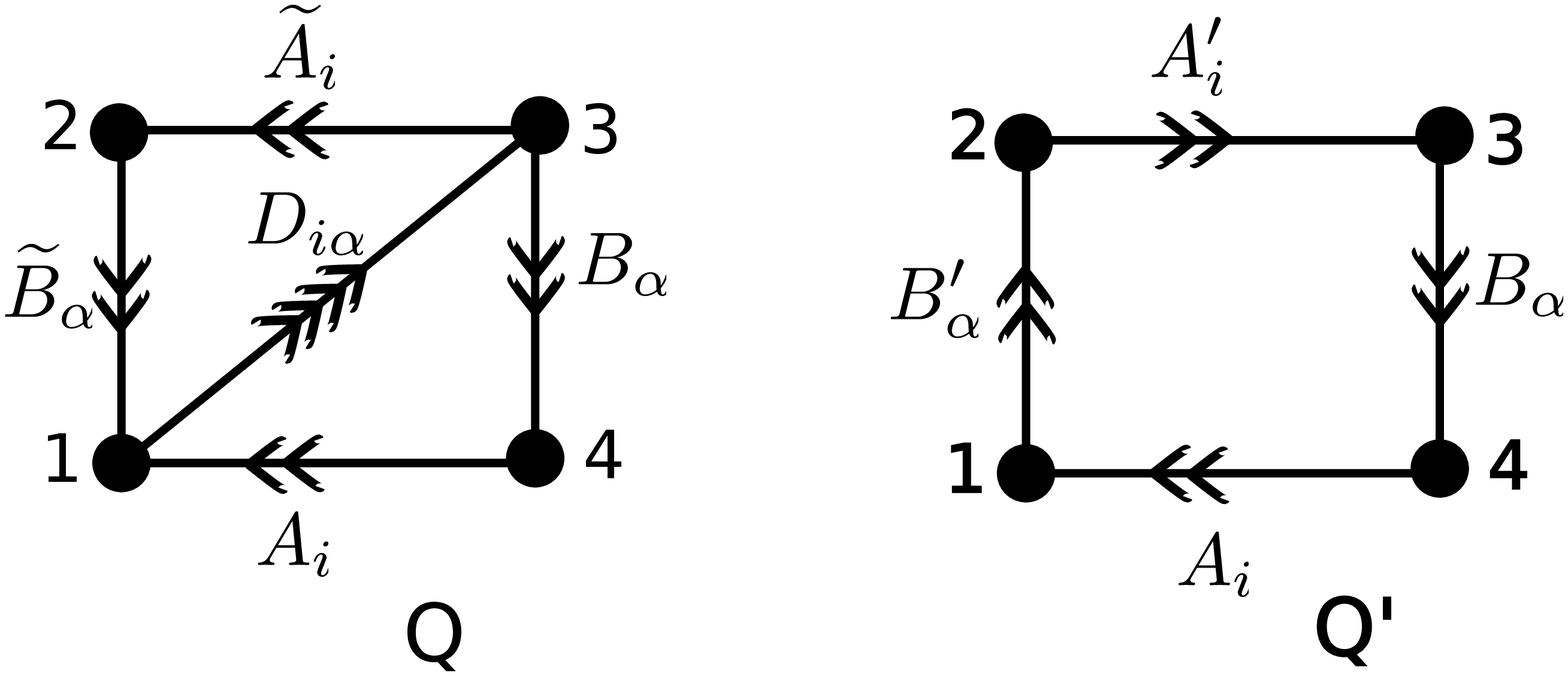}}
\noindent{\ninepoint \baselineskip=2pt {\bf Fig. 7.} {The Quivers for ${\bf P}^{1}\times {\bf P}^{1}$.  The $Q$ phase corresponds to $z_{s}, z_{t} \to 0$ while the $Q'$ phase corresponds to $z_{s},s_{t}\to\infty$.}}
\bigskip

\subsec{ Toric Quivers and Dimers on a Torus}

The two quiver gauge theories above are examples of the ``toric" gauge theories \hb (\Kennaway\ contains an excellent summary).
The quiver $Q$ of a toric gauge theory can be represented as a periodic quiver on a $T^2$ torus. The periodic quiver gives a tiling of the torus which turns out to encode not just the quiver, but also the superpotential $W$.  The terms in the superpotential correspond to the plaquettes on the torus defined by the quiver, whose boundaries are the bifundamental matter fields and where the orientation of the boundary of a plaquette determines the sign of the term. This implies, for example, that a given matter field enters precisely two superpotential terms, with opposite signs\foot{Any additional coefficients can be set to $1$ by a field redefinition.}. The dual graph, with faces and nodes exchanged, is per definition a bipartite graph on the torus. The bi-coloring of the nodes is determined by the sign of the corresponding superpotential term. Moreover, the edges of the dual graph connect pairs of nodes of different colors.

The structure is in part a consequence of mirror symmetry \KV, and the fact that $Y$ is fibered by three-tori $T^3$. If we view the $T^3$ as an $S^1$ fibration over $T^2$, where the $S^1$ fiber corresponds to the
$uv =w$ cylinder, and the $T^2$ is mapped out by the phases of $x$ and $y$ coordinates. Consider a D3 brane, wrapping a generic $T^3$ fiber (this is mirror to a D0 brane on $X$), and let it approach $w=0$ in the base. There, the $S^1$ fibration degenerates over a graph on the $T^2$ cut out by the Riemann surface $W(e^x,e^y)=0$. In the nice cases, corresponding to the quiver being toric, the fibration is such that
this sections the D3 brane out into plaquetes, which are the plaquettes of the bipartite graph \AK\KV . In particular, 
$$
{\rm T^3} = \sum_{\alpha} \Delta_{\alpha}
$$
this implies that the moduli space of
the $U(1)^r$ quiver gauge theory is the mirror manifold $X$, since mirror symmetry maps D3 brane on $T^3$ to D0 brane on $X$.

The toric quiver gauge theories have a global 
$$T = U(1)^2 \times U(1)_R$$ 
symmetry. The $U(1)^2$ symmetry is inherited from the $U(1)^2$ symmetry of the Calabi-Yau, and leaves the superpotential invariant. The $U(1)_R$ is an $R$-symmetry, 
under which the superpotential is homegenous, of degree $2$.

There is a simple geometric relation between bipartite graphs of a pair of dual quivers \hb.  
The transformation is local, acting only on the face of the bipartite graph we dualize.  We replace the face corresponding to the dualized node  of the quiver (in the present case, this is node 2), with a dual face. The dual face is the copy of the original, but with the colors of all the vertices reversed. For this to fit into the original bipartite graph consistently, we add a link connecting each original  vertex bonding the face to its dual of opposite color.  The new links that appear in this way correspond to mesons. Finally, we can erase links corresponding to massive fields. The result is the bipartite graph corresponding to the dual gauge theory.  An example of this, relating the dimers of quivers $Q$ and $Q'$ is in the figure 8. 

\bigskip
\centerline{\epsfxsize 3.0truein\epsfbox{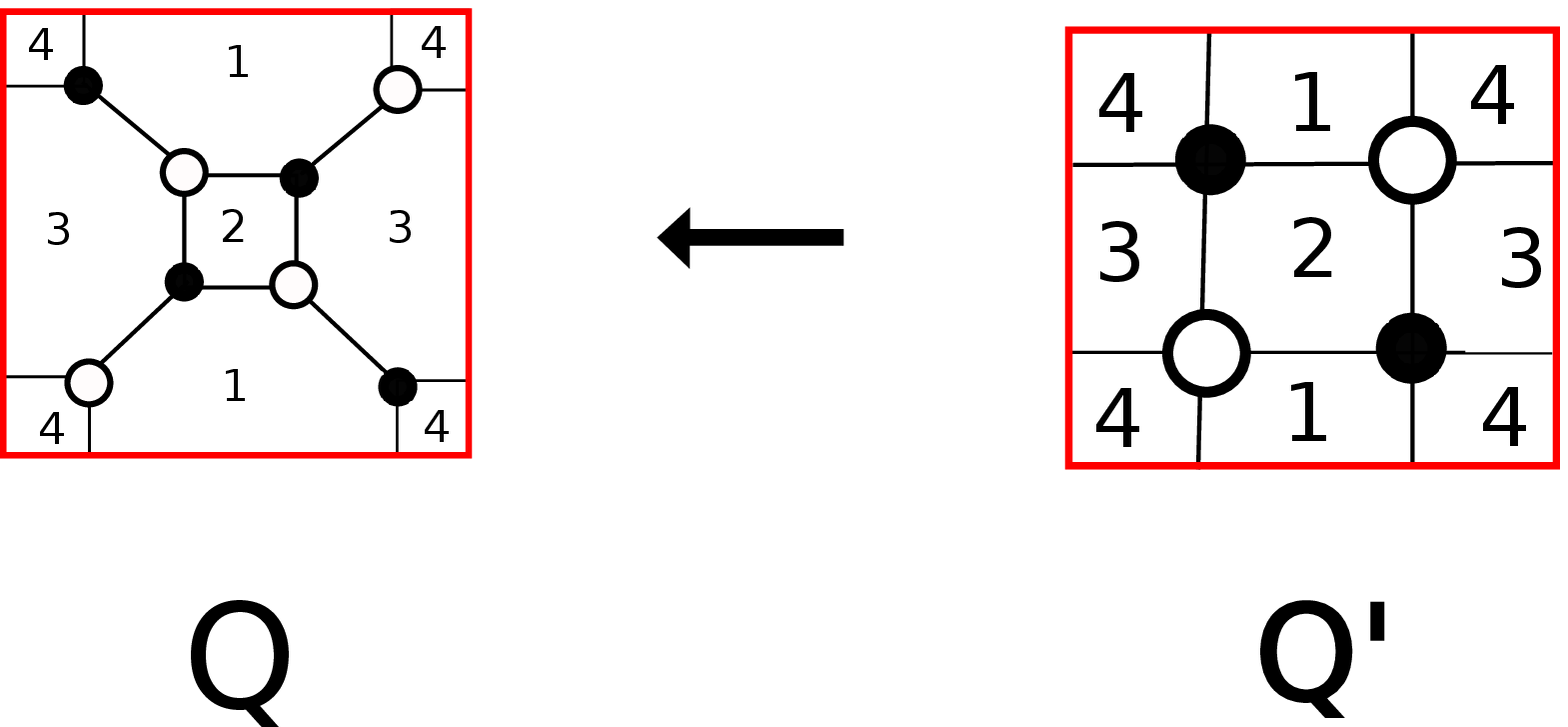}}
\noindent{\ninepoint \baselineskip=2pt {\bf Fig. 8.} {The dimer models on $T^{2}$ for $Q$ and $Q'$ are related by dualizing face 2.}}
\bigskip

\newsec{BPS Degeneracies and Wall Crossing from Crystals and Dimers}

There is a combinatorial way to compute the BPS degeneracies of toric quiver theories \refs{\RM, \OY, \Sulkowski}\ in terms of enumerating certain melting crystal configurations, or equivalently \refs{\ORV, \RM}, by counting dimer configurations. In the language of dimers, Seiberg duality becomes geometric. We will use this to prove that the BPS degeneracies of of two Seiberg dual quivers satisfy the relations \quiverwc\ for a certain infinite class of states. (As we will see in the next section, these will turn out to be bound states of a single D6 brane bound to D4, D2 and D0 brane wrapping cycles in a mirror toric Calabi-Yau $X$.)

\subsec{BPS states of Quivers and Melting Crystals}

The BPS degeneracy of a state $\Delta$ is the Witten index 
$$\Omega_Q(\Delta) = Tr_{Q, \Delta} (-1)^F$$
of the quiver $Q$ with gauge group $G_{\Delta}$ \gg . It can be computed as the Euler character of the moduli space of the quiver, defined by setting F- and D-terms to zero, and dividing by the gauge group. In practice, this is doable for any fixed $\Delta$, but it quickly gets cumbersome.

There is a combinatorial way to compute the degeneracies $\Omega_Q(\Delta)$, for any $\Delta,$ for a toric gauge theory, using the torus ${\bf T}$ symmetry and localization. 
The price to pay is that the results correspond to degeneracies not of the quiver $Q$ alone, but of its extension by adding one extra node
of rank $1$. In the figure below we show the extension of the two dual quivers we considered above by an node $\alpha=0$ and an arrow. 
\bigskip
\centerline{\epsfxsize 4.0truein\epsfbox{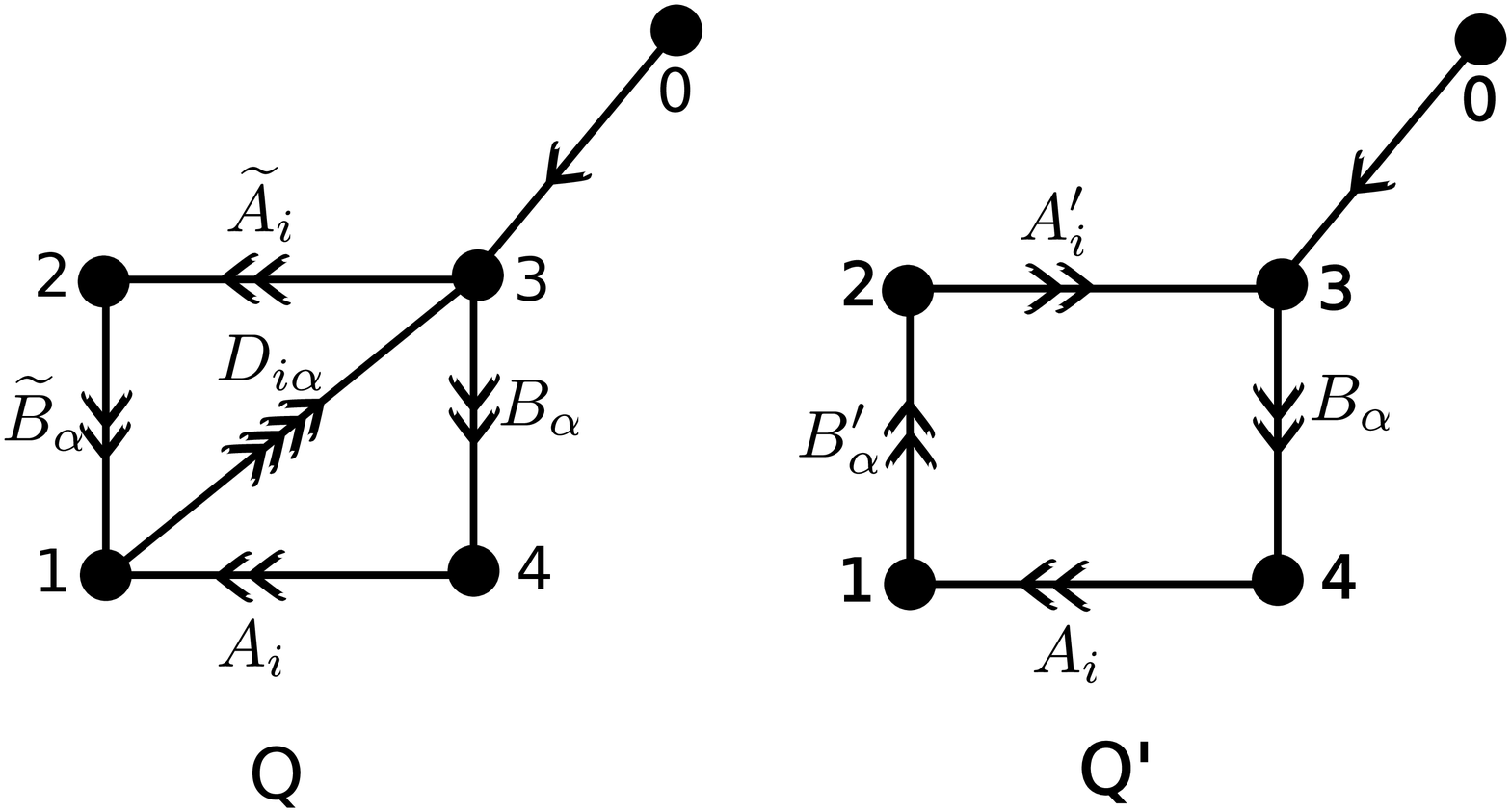}}
\noindent{\ninepoint \baselineskip=2pt {\bf Fig. 9.} {The extended quivers for ${\bf P}^{1}\times {\bf P}^{1}$.}}
\bigskip

The extended quiver has the same superpotential as before, since there are no gauge invariant operators we can add.
We will return to the physical meaning of this extension in a moment\foot{This was discussed in \JW\ in the case of the conifold, and in \OY\ for toric Calabi-Yau without compact 4 cycles.}, but for now we 
simply explain the statement \refs{\RM,\OY}.

Recall that a quiver defines a path algebra $A$, whose elements are all paths on the quiver obtained by joining arrows in the obvious way, where we consider equivalent two paths related by F-term constraints.  Since we allow paths on the quiver $Q$ that wind around the torus arbitrary numbers of times, consider the lift to a periodic quiver on the plane, $Q_{R^2}$. 
Let $A_0$ be the subspace of $A$, corresponding to paths starting on node $0$. The ${\bf T}$-charge of the arrows in the quiver assigns ${\bf T}$ charge to paths in $A_0$. 
A theorem of \refs{\Hananythree, \RM}\ states that any two elements of $A_{0}$ with the same ${\bf T}$ charge are equivalent modulo F-terms. The set of ${\bf T}$ charges of endpoints of $A_0$ is a three-dimensional crystal ${\cal C}$, which is a cone (see Figure 10).  Keeping track of only the $U(1)^2$ charge, ${\cal C}$ projects down to the two-dimensional planar quiver we started with. The $U(1)_R$ charge
defines height of the nodes, making the crystal three dimensional.\foot{It is crucial for the structure of the crystal that there exist a $U(1)_R$ symmetry so that every path has positive R-charge.  One choice is the $U(1)_R$ symmetry that can be geometrically realized in the dimer model as an ``isoradial embedding'' of the bipartite graph in the plane \Kennaway.}

\bigskip
\centerline{\epsfxsize 1.5truein\epsfbox{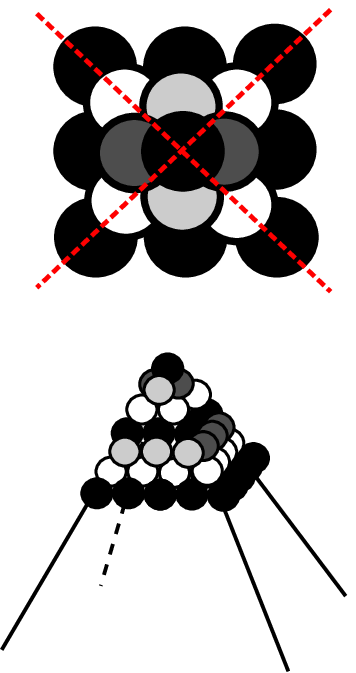}}
\noindent{\ninepoint \baselineskip=2pt {\bf Fig. 10} {The crystal for local ${\bf P^1}\times{\bf P}^1$.}}
\bigskip

Starting from ${\cal C}$ one can explicitly construct ${\bf T}$-invariant solutions to F-term equations.
A melting of the crystal is an ideal ${{\cal C}}_{\Delta}$ 
such that if a path $p$ is in ${\cal C}_{\Delta}$, than $pa$ is also in ${\cal C}_{\Delta}$ for any path $a$ in $A$. ${\cal C}_{\Delta}$
is obtained from ${\cal C}$ by removing $N_\alpha$ sites corresponding to node ${\alpha}$ where 
$\Delta= \sum_{\alpha}N_{\alpha}\Delta_{\alpha}$.
The melting crystal configurations ${\cal C}_{\Delta}$ are in one-to-one correspondence with ${\bf T}$-fixed solutions to F-term equations corresponding to quiver $Q$, with gauge group $G_{\Delta}$.\foot{To sketch this, 
consider the finite set of sites we removed to get ${\cal C}_\Delta$ from ${\cal C}$.  The
set of sites corresponding to node $\alpha$ give vector spaces $V_{\alpha}$ of Chan-Paton factors of rank
${\rm dim}(V_{\alpha}) =N_{\alpha}$. The algebra $A$ is represented on this by matrices, with non-zero entries corresponding to paths in the crystal. By construction, these satisfy $F$-term constraints. The vector spaces $V_{\alpha}$ come with the grading by the torus ${\bf T}$ charge assigned to them by paths in $A_{0}$. 
The solutions to the F-term equations we obtained above are fixed under the torus action that transforms the $V_{\alpha}$ and elements of $A$ by their corresponding weights. We still need to impose D-term constraints, and divide by the subgroup of the gauge group $G_{\Delta}$ that is preserved by the solution. Since the nodes of $V_{\alpha}$ all carry different ${\bf T}$ charge, $G_{\Delta}$ is broken to the maximal abelian subgroup. We thank D. Jafferis for discussions and explanations of this point.} We will choose the FI parameters $\theta_{\alpha}$ in such a way that every solution we constructed is stable -- they can solve the D-term constraints. This corresponds to $\theta_0>0$ and $\theta_{\alpha\neq 0}<0$. Physically, this means that the central charges of the nodes $\Delta_{\alpha}$ are all roughly aligned, and at an angle with the central charge of the node $\Delta_0$. Then, setting D-terms to zero and dividing by the gauge group is equivalent to dividing by the complexified gauge group. Doing so, the fixed points in the quiver moduli space are isolated, and counting them reduces to enumerating crystal configurations ${\cal C}_{\Delta}$.

Counting $T$-fixed points in the classical moduli space gives $Tr_{\Delta} (-1)^F$ up to a sign 
$(-1)^{d(\Delta)},$
corresponding to the fermion number of the fixed point (which determines whether the BPS multiplet is bosonic or a fermionic).
Thus, the BPS degeneracies of the quiver are obtained by counting melting crystal configurations, up to sign
$$
\Omega_Q(\Delta)  = \sum_{{{\cal C}}_{\Delta}} (-1)^{d(\Delta)} = (-1)^{d(\Delta)}\chi({\cal M}^{\Delta}_{Q}),
$$
where the sum is over crystals  ${\cal C}_{\Delta}$ with charge $\Delta$ nodes removed, and where we have also written the degeneracies in terms of the euler characteristic of the quiver moduli space ${\cal M}^{\Delta}_{Q}$ with fixed ranks given by $\Delta$.
The sign is computed by
\eqn\dsign{
d(\Delta) = \sum_{\alpha} N_{\alpha}^2 + \sum_{\alpha \rightarrow \beta} N_{\alpha} N_{\beta}
}
where the last sum is over all the arrows in the quiver $Q$ connecting nodes $\alpha$, and $\beta$, including node $0.$
This counts the dimension of the tangent space to the fixed point set.

Finally, let us define a generating function for the degeneracies%
\eqn\pp{
Z_Q(q) = \sum_{\Delta,{\cal C}_{\Delta}} (-1)^{d(\Delta)} q^{\Delta}
}
where $q^{\Delta}$ is the chemical potential,  induced by giving weight $q_{\alpha}$ to node $\alpha$, i.e.
$$
q^{\Delta} = \prod_{\alpha>0} q_{\alpha}^{N_{\alpha}}.
$$

\subsec{Wall Crossing and Crystals}

We have seen that the degeneracies of $N_0=1$ BPS states of a toric quiver $Q$ extended by a node, 
can be computed by counting crystal configurations. Consider now two toric quivers $Q$, $Q'$ related by 
dualizing a node $\Delta_*$. We will prove that the degeneracies computed by the corresponding crystals satisfy
the wall crossing formulas of \KS .

The wall crossing formula \quiverwc , predicts that degeneracies corresponding to
two quivers $Q$ and $Q'$ are related by
\eqn\algt{
A_{Q'} = A_{\Delta_{*}}^{-1} {A_Q} A_{-\Delta_{*}},
}
together with a change of basis following from \plt\
\eqn\plte{\eqalign{
e_{\Delta_{*}}' &= e_{-\Delta_{*}} \cr
e_{\Delta_{\beta_j}}' &=e_{(\Delta_{\beta_j} + n_{\beta_{j *}}\Delta_{*})} \cr
e_{\Delta_{\gamma_k}}' &=e_{\Delta_{\gamma_k}}}
}
where $n_{\beta_j *} >0$, and $n_{\gamma_k*}<0$. (In \algt\ we made a particular choice of the route around the singularity. The other choice is related to this by monodromy, which does not affect the degeneracies, but relabels the charges.)
In the above, $A_Q$ contains the information about the BPS states with any $N_0$,
and not just $N_0=1$ states that are counted by the crystal. To apply this to the present context, consider a truncation of the algebra \alg\  to operators  $e_{\Delta}$ with $N_0=0$. Denote by $A_{Q}^{(0)}$ the restriction of $A_Q$ operator product to states with vanishing $N_0$. This can be implemented by setting $e_{\Delta_0}=0$. Then, 
\algt\ reads the same, just restricted to $A_{Q}^{(0)}$.  
%
%

Next, restrict  to operators $e_{\Delta}$, with $N_{0} = 0,1$. (Note that this implies that any two operators with $N_0=1$ commute.) By our choice of the FI parameters,  the central charges of all the states with $N_0=0$ are approximately aligned,  and the central charges of all the states with $N_0=1$ are also aligned, but at an angle to the former.\foot{Note that this implies that 
$|Z(\Delta_{\alpha})| \ll |Z(\Delta_{0})|,$
an assumption that we will justify in the next section, when we identify node $0$ with a D6 brane wrapping $X$.}  To this order, $A_{Q}$ reduces to a product  
$
A_Q^{(0)} A_Q^{(1)}.
$
Then, \algt\ implies that $A_{Q}^{(1)}$ transforms by conjugation with $A_{-\Delta_{*}}$:
\eqn\prd{
A_{Q'}^{(1)} = A_{-\Delta_{*}}^{-1} A_Q^{(1)} A_{-\Delta_{*}}.
}
To compare $A_{Q'}^{(1)}$ and $A_Q^{(1)}$, we in addition need to redefine the variables using \plte .
Note that, in addition to crossing the wall of the second kind, corresponding to a Seiberg duality, we have 
adjusted the FI terms so that the computation of \RM\ applies. In other words, all the central charges of all the $\Delta_{\alpha\neq 0}'$ are aligned, and at an angle to $\Delta_0'$.

Since we are only interested in states with $N_0=1$, and using that conjugation by $A_{\Delta}$
acts as 
$$
A_{\Delta} e_{\Delta'} A_{\Delta}^{-1} \; =\; (1- e_{\Delta} )^{\Delta \circ \Delta'} e_{\Delta'}
$$
for any two $\Delta, \Delta'$,
we can equivalently rewrite \prd\  as
$$
\sum_{\Delta} \; \chi({\cal M}^{\Delta}_{Q'})  \;e_{\Delta}= \sum_{\Delta}\;\chi({\cal M}^{\Delta}_Q) \;(1- e_{-\Delta_*})^{\Delta_* \circ \Delta} \; e_{\Delta}.
$$
Note that we have written this KS formula in terms of the (unsigned) moduli space euler characteristics rather than the true (signed) BPS invariants.  As noted in section 2, this is because KS naturally counts the euler characteristic moduli spaces without the additional sign.  However, this sign is naturally restored in the change of variables below.
The sum is over all $\Delta$ with $N_0=1$. 
Finally, let us write how the partition function transforms. Writing 
$$
e_{\Delta} \rightarrow (-1)^{d(\Delta)} q^{\Delta},
$$
and noting that
$$
e_{\Delta} e_{\Delta'} = (-1)^{\Delta \circ \Delta'} e_{\Delta+\Delta'}
$$
we can write 
\eqn\wcq{
Z_{Q'}(q') = \sum_{\Delta} \; \Omega_{Q'}(\Delta)  \;q'^{\Delta}= \sum_{\Delta}\;\Omega_Q({\Delta}) \;(1- (-1)^{\Delta_{*} \circ \Delta}q_{*}^{-1})^{\Delta_{*} \circ \Delta} \; q^{\Delta},
}
where we change the variables consistent with \plte , in other words
\eqn\wcb{\eqalign{
q_{*}'& =(q_*)^{-1}\cr 
q_{\beta_{j}}' & = q_{\beta_j} (q_{*})^{n_{\beta_j *}}\cr
q_{\gamma_k}' & = q_{\gamma_k}}.
}
Note that this is the semi-primitive wall crossing formula of \DM , because always one of the products has $N_0=1$. When we consider passing through several walls of the second kind this will end up involving general decays, because the degeneracies of decay products at one wall can jump on the next. 
This gives us an explicit prediction for the degeneracies computed from one crystal in terms of the other, which
one can check term by term.
We can however do better, and {\it prove} that the BPS degeneracies of the two quivers are indeed related by \prd . 
The proof is elementary, using the dimer point of view on the crystals.

\subsec{Dimers and Wall Crossing}

Recall that Seiberg duality relating quivers $Q$ and $Q'$ has a simple geometric realization in terms of bipartite graphs on $T^2$.
This will allow us to give a geometric proof of the wall crossing formula \wcq\ in terms of dimers. But, for this we need to translate the counting of BPS states in the quiver from the language of crystals, which we used so far, to dimers.

Consider the bipartite graph dual to the periodic quiver $Q_{T^2}$ on the torus. The lift of this to the covering space is
a bipartite graph in the plane. The path set $A_{0}$ gives rise to a canonical perfect matching of the bipartite graph.
Consider the paths in $A_{0}$ that lie on the surface of the crystal. These correspond to short paths in the planar quiver $Q_{R^2}$, paths containing no loops. These short paths define a set of paths on the bipartite graph, where they pick out a subset of edges crossed by them. The edges in the complement of this define a perfect matching $m_0$ of the bipartite graph \refs{\RM,\OY}. 

The finite melting crystal configurations are in one to one correspondence with perfect matchings $m$ which agree with $m_0$ outside of a finite domain. Removing an atom in the crystal is rotating the dimers around the corresponding face in the bipartite graph. The difference $m-m_0$ of the two perfect matchings defines closed level sets on the bipartite graph. We can use them to define a height function whose value is $0$ at infinity, and increases by $\pm$ 1 each time we cross a level set, depending on the orientation. Defined this way, the height function lets us keeps track of the height of the the sites we melt from the crystal.  The figure 11 shows the planar dimers and their backgrounds $m_0$ corresponding to the two quivers $Q$ and $Q'$.
%
\bigskip
\centerline{\epsfxsize 5.0truein\epsfbox{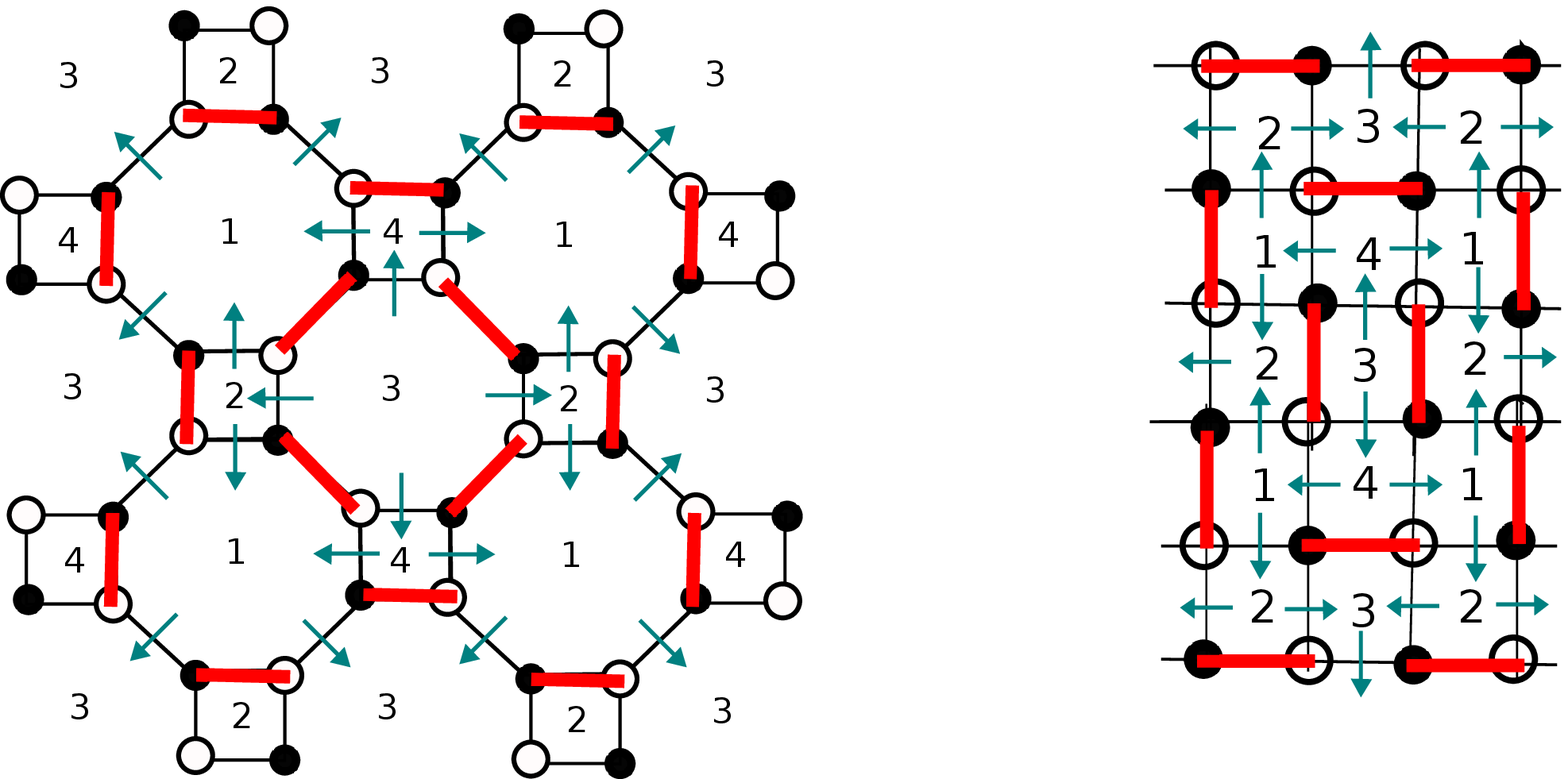}}
\noindent{\ninepoint \baselineskip=2pt {\bf Fig. 11.} {The vacuum perfect matchings, $m_{0}$, for the dimer model associated to the quivers $Q$ and $Q'$ of local  ${\bf P}^{1}\times {\bf P}^{1}$.}}
\bigskip
%
The partition function \pp\  can thus be written in terms of the dimer model as
\eqn\ppd{
Z_{Q}(q) = \sum_{m\in {\cal D}_Q} (-1)^{d(\Delta (m))} q^{\Delta(m)}
}
where $q^{\Delta(m)}$ is the weight of the dimer configuration, chosen to agree with the weights of the corresponding crystal \refs{\ORV,\DA}. We assign a fixed weight $w(e)$ to every dimer $e$ on the plane, 
in such a way that the product of weights of edges around a face, corresponding to node $\alpha$ of $Q_{R^2}$
equals $q_{\alpha}$. The edges have a natural orientation, from a white vertex say to black, and contribute to the product by $w(e)$ or $w(e)^{-1}$ depending on whether orientation of the edge agrees or disagrees with the orientation of the cycle. The weight of the dimer configuration $m$ is a product over the weights of the edges in the dimer,  $w(m) =\prod_{e\in m} w(e)$, normalized by  $w(m_0) = \prod_{e\in m_0} w(e)$,
and
$$
q^{\Delta(m)} = {w(m)\over w(m_0)}
$$ 
The sign in \ppd\ does not come from the weights (i.e. for a general $Q$ it cannot be absorbed into the weights), but is added in by hand.

The duality transformation relating the two quivers $Q$ and $Q'$ has a simple geometric interpretation in terms 
of the bipartite graph, corresponding to replacing a face of the node we dualize, with the dual face. 
This operation lifts to perfect matchings of the corresponding bipartite graphs as well -- from the perfect matching of one graph, we can obtain a perfect matching of the dual graph.
Were the operation one to one, the degeneracies computed from one dimer and its dual would have been the same. The operation is in fact almost one to one, except for an ambiguity in the mapping of certain dimers on the faces we dualized. 

Let ${\cal D}_Q$ and ${\cal  D}_{Q'}$ be the sets of the perfect matchings of the two dual bipartite graphs in the plane (with the suitable asymptotics). Any two perfect matchings in ${\cal D}_{Q'}$ differing by ``configurations of type $1$" in figure 12 come from the same configuration in ${{\cal D}}_{Q}$.  In addition, any two perfect matchings in ${\cal D}_{Q}$  containing a ``configuration of type $2$" correspond to the same dimer in ${\cal D}_{Q}'$. 

\bigskip
\centerline{\epsfxsize 4.0truein\epsfbox{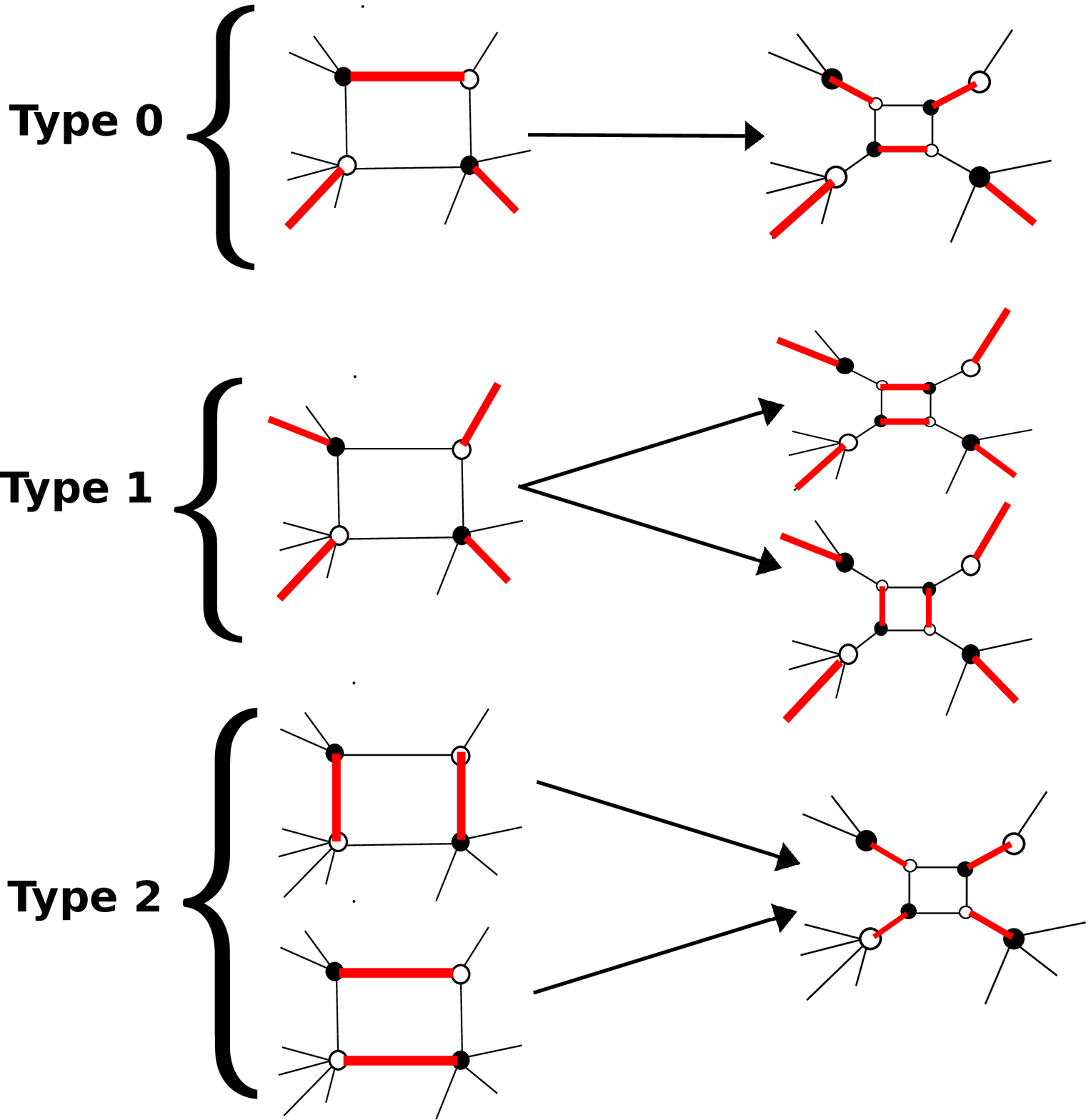}}
\noindent{\ninepoint \baselineskip=2pt {\bf Fig. 12.} {How dimer configurations transform under Seiberg Duality.}}
\bigskip

This means we can use perfect matchings of ${\cal D}_Q$ to enumerate perfect matchings of ${\cal D}_{Q'}$, provided we know the numbers $\#1(m)$, $\#2(m)$ of configurations of type $1$ and type $2$ in each perfect matching $m$ of ${\cal D}_Q$ .
To count the perfect matchings in ${\cal D}_Q'$, we need to sum over all the perfect matchings of
${\cal D}_Q$, and compensate for configurations over- or under-counted. Schematically, in terms of counting dimer configurations without signs (later, we will be precise about the weights), this gives
\eqn\wcd{
\sum_{m\in {\cal D}_{Q'}} (q')^{\Delta(m')}   = \sum_{m \in {\cal D}_Q} {(1+q_{*})^{I(m)}} q^{\Delta(m)} 
}
where
%
%
 %
%
$$I(m) = \#1(m)- \#2(m).$$
The factor 
$$(1+q_{*})^{I(m)}= {(1+q_{*})^{\#1(m)}\over (1+q_*)^{\#2(m)}}
$$
appears since in ${\cal D}_Q$ there are too many configurations of type $2$, and too few of type configurations of type $1$.

Notice that \wcq\ and \wcd\ are of the same form, provided $I(m)$,
which we will call the ``dimer intersection number", depends only on the charge $\Delta$ of the dimer configuration $m$, and not on $m$ itself, and equals the intersection number of $\Delta_{*}$ and $\Delta$,
$$
I(m)= \Delta_{*} \circ \Delta,
$$
where $\Delta_{*}$ is the node being dualized. In the next subsection, we will show that this is indeed the case. Using this, we will prove that the degeneracies of two Seiberg dual toric quivers indeed satisfy \wcq .

\subsec{The proof}

%


To be able to count the BPS states using dimers, both the original and the dual quiver need to be toric.
The conditions for this were studied in \hb. The Seiberg duality needs to preserve the fact that a D3 brane wrapping the $T^3$ fiber corresponds to all ranks of the quiver being one. (This is required by the stringy derivation of the relation of quivers and dimer models \KV , as we reviewed earlier.)  For this to be the case, as is easy to see from \plt\ and charge conservation of $T^3=\sum_{\alpha}\Delta_{\alpha}$, the node we dualize has to have two incoming and two outgoing arrows. Under this restriction, the most general face of the dimer model that can be dualized is shown in Figure 13.    

Now consider some arbitrary matching $m$ corresponding to charge $\Delta$, with
$$
\Delta = \sum_{\alpha}N_{\alpha}\Delta_{\alpha}.
$$
If we denote the face to be dualized by, $\Delta_{*}$, then it follows that its intersection with $\Delta$
is
\eqn\intn{
\Delta_{*}\circ \Delta = \sum_{\alpha}N_{\alpha} n_{* \alpha} = N_{2} + N_{4} - N_{1} - N_{3} - n_{0*}
}
where we used the fact that $n_{* \alpha}$ is non-zero only for faces that share an edge with $*$, and its value, including the sign can be read off from the bipartite graph. We are using here the conventions of figure 13. The  last term is not geometric, $n_{0*}$ is the number of framing nodes from $0$ to node $*$. In our setup so far, this is either 0 or 1.  

We will use induction to find the dimer intersection number $I(m)$, and show it equals the physical intersection number \intn .  
Starting with some arbitrary perfect matching, we consider the effect of melting an additional node.  We will show that the $I(m)$ and 
$\Delta_*\circ \Delta$ change in the same way. In the end, we will show that the cannonical perfect matching $m_0$ also satisfies
the relation, so in fact any perfect matching does so as well.

In the dimer model, melting a node simply corresponds to exchanging occupied and unoccupied bonds along the perimeter of that face.  Further, from the ingoing and outgoing arrows at face *, a bond on edge $a$ corresponds to face 1 being melted and bond $c$ corresponds to 3 melted, while $b$ corresponds to 2 unmelted, and $d$ corresponds to 4 unmelted.
Now we observe that removing a bond from the perimeter of $*$ always changes the dimer intersection number by $+1$, since it either removes a type 2 configuration or adds a type 1 configuration.  Thus we find that melting faces 1 or 3 both change the dimer intersection number by $-1$ and melting faces 2 or 4 change the dimer intersection number by $+1$.

To complete the inductive argument, we need to compute the dimer intersection number for the vacuum configuration.  If $n_{0*}=0$ then there are no ``removable'' $*$ faces in the initial dimer configuration so no type 2 configurations appear.  Since the vacuum configuration can be seen as the complement of those bonds that intersect arrows in the planar quiver, there are also no type 1 configurations.  Such a type 1 configuration would have both incoming and outgoing arrows present, but such a configuration cannot appear on the surface of the crystal. 
One way to understand this is to note that the vacuum dimer configuration breaks the rotational and translational symmetry of the bipartite graph by specifying the tip of the crystal.  For any face in the bipartite graph (which must correspond to an atom on the surface of the crystal), the direction toward the tip of the crystal is a preferred direction, and the dimer model reflects this preference.  However, a type 1 configuration has symmetric arrows with no preferred direction.  Thus, it cannot exist in the vacuum configuration.\foot{We can also see this by breaking the planar dimer into a tiling of $T^{2}$ dimers, so that the vacuum configuration corresponds to a set of $T^{2}$ dimer configurations.  These $T^{2}$ dimer matchings are in one-to-one correspondence with points on the toric diagram \hb.  It has been conjectured \OY\ that only matchings corresponding to external points on the toric diagram appear along the surfaces of $m_{0}$.  A general $T^{2}$ matching containing a type 1 or type 2 configuration will always be an internal point in the toric diagram, and thus can only appear at the apex of the crystal.}    If $n_{0*}=1$ then there is exactly one ``removable'' $*$ face in $m_{0}$, which corresponds to a type 2 configuration, giving a dimer intersection number of -1.  Combining these results, we find,
$$
I(m) =  -n_{0*} + N_{2} + N_{4} - N_{1} - N_{3}
$$


We will now tie this all together, and show that, explicitly mapping the dimer configurations of ${\cal D}_Q$ to configurations in ${\cal D}_{Q}'$
{\it together with their weights}, we reproduce the change of degeneracies \wcq , together with the change of basis \wcb.
If we denote the dimer weight for configuration $m$ by $w(m)$ and the partition function variables by $q^{\Delta_{m}}$, then by definition,
$$
q^{\Delta(m)} = {w(m) \over w(m_{0})}
$$

Now we must decide how our weights transform under the duality.  There is a large redundancy in the weight assignments, since the partition function (normalized by the weight of the canonical perfect matching) does not depend on the weights of the individual edges, but only on gauge invariant information, the products of weights around closed loops. A convenient choice (Figure 13) is one which does not change the weighting of Type 0 configurations, so that the weighting of the vacuum, $w(m_{0})$ remains the same.  We accomplish this by flipping the edge weights across the dualized face and assigning weight 1 to the meson edges.  If we denote the old variables by $\{q_{\alpha}\}$ and the new variables by $\{q'_{\alpha}\}$ and take into account the change in orientation of arrows, we find,
\eqn\dchb{\eqalign{q_{1}' \; &= \; q_{1}(w_{a}w_{c})^{-1}\cr
q_{2}' \; &= \; q_{2}(w_{b}w_{d})\cr
q_{3}' \; &= \; q_{3}(w_{a}w_{c})^{-1}\cr
q_{4}' \; &= \; q_{4}(w_{b}w_{d}) \cr
q_{*}' \; &= \;  q_{*}^{-1} = (w_{a}w_{c})(w_{b}w_{d})^{-1}\cr
q_{\gamma}' &=q_{\gamma} }}
%
where $q_{\gamma}$ are correspond to nodes with no intersection with node $*$. 
Now we can explicitly transform the dimer configuration, together with its weight,
$$ w(m)\to w'(m) = w(m){(w_{b}w_{d} + w_{a}w_{c})^{\#_{1}}\over (w_{b}w_{d} + w_{a}w_{c})^{\#_{2}}}.
$$

We still have leftover arbitrariness in the weights, since nothing depends on $w_a$, $w_b$,$w_c$,$w_d$ separately.
We can set for example 
$$w_b w_d = 1, \qquad w_a w_c = q_*^{-1},$$
and then we get the correct change of variables for the Seiberg duality corresponding to $+$ sign in \plt, and going around the singularity clockwise. This is because $n_{3*}, n_{1*}$ are positive,
$n_{2*}$, $n_{4*}$ negative, and $n_{\gamma *}$ vanishes. This is also the choice we have been making in this section. Setting $w_a w_c=1$, instead, the change of variables is correct for a Seiberg duality with a $-$ sign in \plt. The two choices are related by full monodromy around $Z(\Delta_*)=0$, which is just the change of variables at hand. Proceeding with the $+$ sign choice, \dchb\ is the change of basis in \wcb . In addition, the contribution of dimer configuration $m\in {\cal D}_Q$ to ${\cal D}_{Q'}$ is obtained by replacing 
and
$$ w(m)\;\;\to\;\; w'(m) = w(m){(1+ q_*^{-1})^{\#_{1}-\#_{2}}}.
$$
To find the contribution to the partition function, we still need to divide by the weight of the vacuum configuration, so
$$ w(m)/w(m_0) =q^{\Delta(m)}\;\;\rightarrow\;\;  w'(m)/w'(m_0)=
 q^{\Delta(m)} (1+ (q_{*})^{-1})^{\Delta_{*}\circ\Delta}.
$$
Finally, to compute BPS degeneracies from the dimer configurations, we need to reintroduce the sign twists, which means replacing 
$q^{\Delta}$ by $(-1)^{d(\Delta)}q^{\Delta} $, and sum over all matchings $m$. This precisely reproduces \wcq\ and \wcb, since $(-1)^{d(\Delta-\Delta_{*})} =- (-1)^{d(\Delta)}(-1)^{(\Delta_{*}\circ\Delta)}$. 
Thus we have derived the KS wall crossing formula purely from  how perfect matchings in the dimer model transform.

\bigskip
\centerline{\epsfxsize 5.0truein\epsfbox{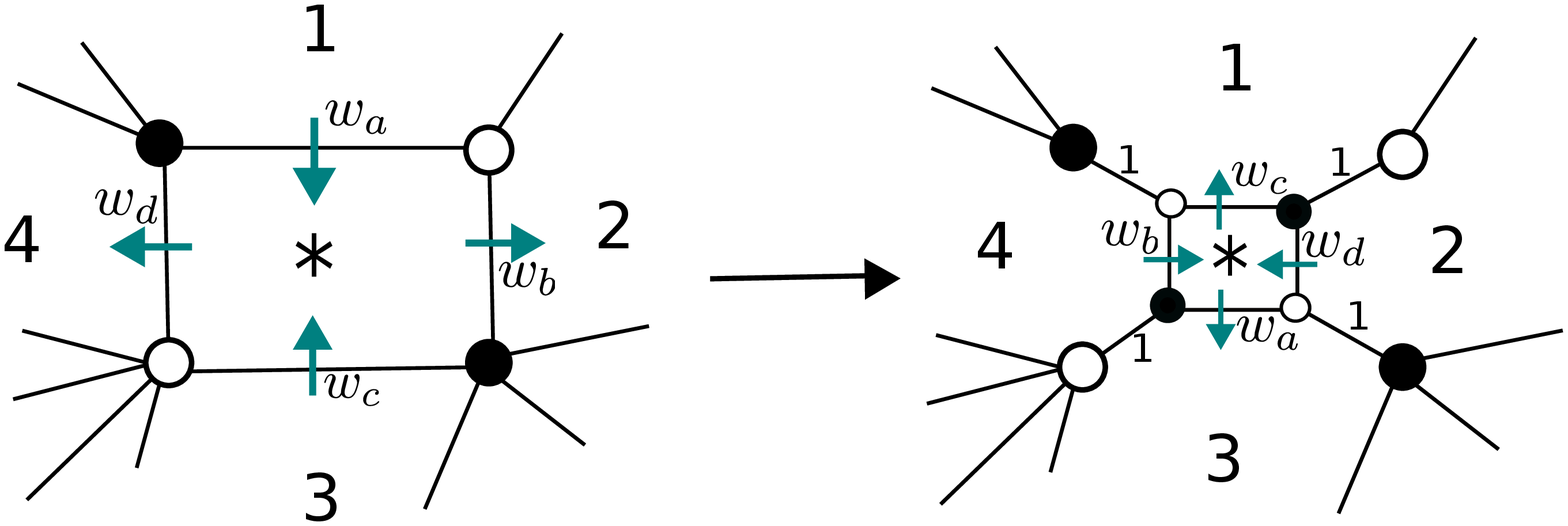}}
\noindent{\ninepoint \baselineskip=2pt {\bf Fig. 13.} {The effect of Seiberg Duality on face $*$ for a general brane tiling.  Note that the dimer weights, $w_{i}$ are flipped on the inner square and are 1 on the new legs.}}
\bigskip
%

\newsec{Mirror Symmetry and Quivers}

Mirror symmetry provides a powerful perspective on the quivers in section 3 and 4. The mirror of the manifold $Y$ of section 3 is a toric Calabi-Yau manifold $X$.
Mirror symmetry also maps  D3 branes wrapping three-cycles in IIB on $Y$ to D0-D2-D4-D6-branes wrapping holomorphic submanifolds in IIA string on $X$.
In the mirror, many aspects of the quiver construction become more transparent. In particular, computing the  quiver gauge theory on the branes becomes 
a question in the topological B-model on $X$, with the answers provided by a vast machinery of the derived category of coherent sheaves on $X$ (see \refs{\Asp,\Sharpe} for excellent reviews).
The goal of this section is to explain what are the BPS D-branes counted by the crystals in section 4. For the ``small quivers" of section 2, this is well understood. The
D3 branes on compact three cycles of $Y$, are mirror to D4,D2, and D0 branes wrapping compact submanifolds of $X$
\refs{\HIV, \GU}. The specific combinations of branes involved correspond to collections of spherical sheaves on the surface $S$ as we will review. 
The extended quivers of section 4 correspond to adding a D6 brane wrapping all of $X$ \OY .
We will show that, which node of the D4-D2-D0 quiver ends up extended, is determined by a choice of a suitable bundle on the D6 brane. These are essentially the tilting line bundles of 
\refs{\Mayr, \aspb, \ADT}.\foot{The fact that the D-brane in question is a D6 brane was proposed in \OY , however the specific choices of bundles are very important.}
Having understood this, the mirror perspective gives a simple interpretation to some of the Seiberg dualities, as turning on B-field on $X$. The effect on the quiver ends up depending on the class of B in $H^2(X, {\bf Z})/H^2_{cmpct}(X,{\bf Z})$, as we will explain in the next section. 

\subsec{Mirror Symmetry}

The Calabi-Yau manifold $Y$ given by \mirrtwo\
$$W(e^x,e^y)=w\qquad\qquad uv=w
$$
is mirror to IIA on $X$, where $X$ is a toric Calabi-Yau threefold. The monomials $w_i =e^{m_i x + n_i y}$ in $W$ 
satisfy relations
$$
\prod_{i} w_i^{Q_i ^a} = e^{-t_a}
$$
for some complex constants $t_a$, and integers $Q_i^a$, satisfying $\sum_i Q_i^a=0$, since $Y$ is Calabi-Yau. The mirror $X$ is given in terms of coordinates $z_i,$ one for each monomial $w_i$, 
satisfying 
\eqn\cons{
\sum_i Q_i^a |z_i|^2 = r_a,
}
and modulo gauge transformations
\eqn\gauge{
z_i \sim z_i e^{i \theta_a Q_i^a}.
}
Above, $r_a = {\rm Re}(t_a)$, and the imaginary part of $t_a$, gets related to the NS-NS B-field on $X$.
For each $a$, we get a curve class $[C_a]\in H_2(X, {\bf Z})$, whose volume is $r_a$. Dual to this are divisor classes $D_a$, corresponding to 4-cycles with $D_a \circ  C^b = \delta_{a}^b$. The toric divisors $D_i$, obtained by setting $z_i=0$, are given in terms of these by
$$  D_i = \sum_i Q_{i}^a D_a
$$
By Poincare duality, we can think of $D_a$ as spanning $H^2(X, {\bf Z})$. Among the toric divisors are compact ones, $D_S$,
which restrict to compact surfaces $S$ in $H_4(X,{\bf Z} ) \approx H^2_{cmpct}(X, {\bf Z})$. For simplicity, we will restrict to the case when $X$ is a local del Pezzo surface, i.e. when there is only one compact $S$.

\subsec{Mirror Symmetry, D-branes and Quivers}

Mirror symmetry maps D3 branes on three-spheres $\Delta_{\alpha}$ in $Y$ to a
collection of sheaves\foot{By $E_{\alpha}$ we will really mean a sheaf $i_* E_{\alpha}$ induced on  $X$ from a sheaf $E_{\alpha}$ on $S$ by the embedding  $i: S \rightarrow X$ of $S$ in $X$.}  $E_{\alpha}$, supported on $S$, for $\alpha=1, \ldots, r$.
Physically, $E_{\alpha}$ are D4 branes wrapping $S$, with some specific bundles turned on, giving the branes specific D2 and D0 brane charges. (More precisely, we need to include both the D4 branes and the anti-D4 branes.) 
The specific bundles turned on correspond to $E_{\alpha}$ being an exceptional collection of spherical sheaves. 
Spherical means that the sheaf cohomology group $Ext_X^k(E_{\alpha},E_{\alpha})$ is the same as $H^k(S^3)$, so both sets of D-branes have no massless adjoint-valued excitations beyond the gauge fields. The collection of sheaves is in addition such that, for $\alpha$ and $\beta$ distinct,  $Ext^{k}(E_{\alpha}, E_{\beta})$ is non-zero only for $k=1,2$, corresponding to chiral bifundamental matter.\foot{Non-zero $Ext^{0,3}(E_{\alpha},E_{\beta})$ would have corresponded to the presence of ghosts.} 

The net number of chiral minus anti-chiral multiplets is an index
$$
n_{\alpha \beta} = \sum_{k=0}^3 (-1)^k {\rm dim}\,Ext_X^k(E_{\alpha}, E_{\beta}) 
$$
On $Y$, this corresponded to the intersection numbers of cycles $n_{\alpha \beta} =\Delta_\alpha \circ \Delta_{\beta}$.
In fact, $n_{\alpha \beta}$ also has geometric interpretation on $X$.
A D-brane corresponding to $E_{\alpha}$ has charge
$$
\Delta_{\alpha} = {\rm ch}(E_{\alpha}) \sqrt{td(X)},
$$
where we made use of Poincare duality to express the Chern class in terms of the dual homology class. Using an index theorem, we can relate the $n_{\alpha\beta}$ to computing intersections on $X$,
$$
n_{\alpha \beta} = \int_X {\rm ch}(E_{\alpha})^v {\rm ch}(E_{\beta})td(X),
$$
where $\omega^v$ denotes $(-1)^n \omega$ for any $2n$-form $\omega$. 

In this case, we have an additional simplification\foot{As explained in \SK\Asp\ even though the sheaf $E_{\alpha}$ on $X$
comes from a vector bundle $E_{\alpha}$ on $S$ (see footnote 17), it does not correspond to a D-brane on $S$ with a vector bundle $E_{\alpha}$.  It corresponds to a D-brane with a vector bundle $V_{\alpha}$, where $V_{\alpha} = E_{\alpha} \otimes K_S^{-{1\over 2}}$, where $K_S$ is the canonical line bundle of $S$.} that
the sheaves $E_{\alpha}$ on $X$ correspond to line bundles ${V}_{\alpha}$ on $S$. This implies \GU\
\eqn\exts{
Ext_X^k(E_{\alpha}, E_{\beta}) = Ext_S^k({V}_{\alpha}, {V}_{\beta}) \oplus Ext_{S}^{3-k}(V_{\alpha}, V_{\beta}).
}
Relative to the naive expectations, the relevant vector bundles on $S$ are twisted \SK\ by $K_S^{-1/ 2}$. This accounts for the fact that the theory on the brane is naturally twisted if $S$ is curved.  The relation to bundles on $S$ simplifies things, since for a holomorphic vector bundle $Ext_S^k(V_{\alpha}, V_{\beta}) = H^{k}(S, V_{\alpha}^{{-1}} \otimes V_{\beta})$.The superpotential can also be computed by fairly elementary means, at least if it is cubic \refs{\GU, \MWI}. For the more general case, see \AspKatz. We can then write (see for example \Asp )
\eqn\chd{
\Delta_{\alpha} = ch({V}_{\alpha})\sqrt{td(S)\over td(N) } = ch({V}_{\alpha})  {\Big(1+{e(S)\over 24}\Big)}.
}
where $N$ is the normal bundle to $S$ in $X$. We are implicitly using mirror symmetry and Poincare Duality to identify a Chern class on $X$ with a cycle on $Y$.

When we add a D6 brane wrapping all of $X$, the quiver gets extended by one node corresponding to $E_0$, a sheaf supported on all of $X$. The charges of the brane are
$$
\Delta_0 = ch(E_0) {\sqrt{td(X)}} = ch(E_0)\Big(1+ {c_2(X)\over 24}\Big)
$$
%
%
We want to choose $E_0$ so that $Ext^k(E_0,E_0)$ is non-vanishing only for $k=0,3$, and extending the quiver by this node does not introduce exotic matter in the quiver, which means $Ext^k(E_0,E_{\alpha})$ is non-vanishing only for $k=1,2$. For the quivers in section $4$ where $n_{0\alpha}=1$ for only one node $\alpha$ and zero for all the others, there is a natural construction of $E_0$. There is a ``duality" that pairs up sheaves $E_{\alpha}$ supported on $S$, with the dual line bundles $L_{\beta}$ on $X$ \refs{\ADT, \Mayr, \aspb}, such that $\chi(E_{\alpha}, L^{\beta}) = \delta_{\alpha}^{\beta}$. Thus, the requisite extension corresponds simply to choosing $E_0 = L_\alpha$.

\bigskip
\centerline{\epsfxsize 1.8truein\epsfbox{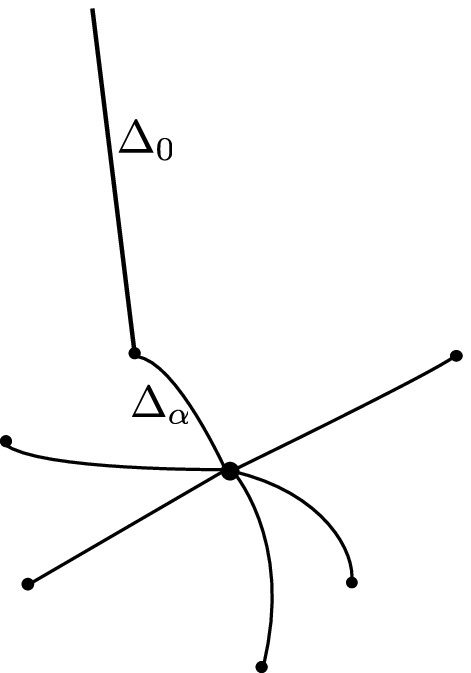}}
\noindent{\ninepoint \baselineskip=2pt {\bf Fig. 14.} {The D6 brane corresponds to a semi-infinite line, $\Delta_{0}$, in the W-plane.  In the mirror this corresponds to a dual sheaf, $E_{0}=L_{\alpha}$.}}
\bigskip
%

\subsec{${\bf P}^1\times {\bf P}^1$ example}

Consider local ${\bf P^1}\times{\bf P}^1$. In this case, $Q_t = (1,1,0,0,-2)$, $Q_s =(0,0,1,1,-2)$, 
$$
|z_1|^2 + |z_2|^2 = 2 |z_0|^2 + r_t, \qquad
|z_3|^2 + |z_4|^2 = 2 |z_0|^2 + r_s, 
$$
and the corresponding curve classes $C_t$ and $C_s$ generate $H_2(X)$. 
The two dual divisor classes, $D_t$ and $D_s$, with the property that $C_t\circ D_t = 1 =  C_s\circ D_s$, and  $C_t\circ D_s = 0 = C_s\circ D_t$, can be taken to correspond to divisors $D_t =D_1$ and $D_s=D_3$ on the figure. 
\bigskip
\centerline{\epsfxsize 2.0truein\epsfbox{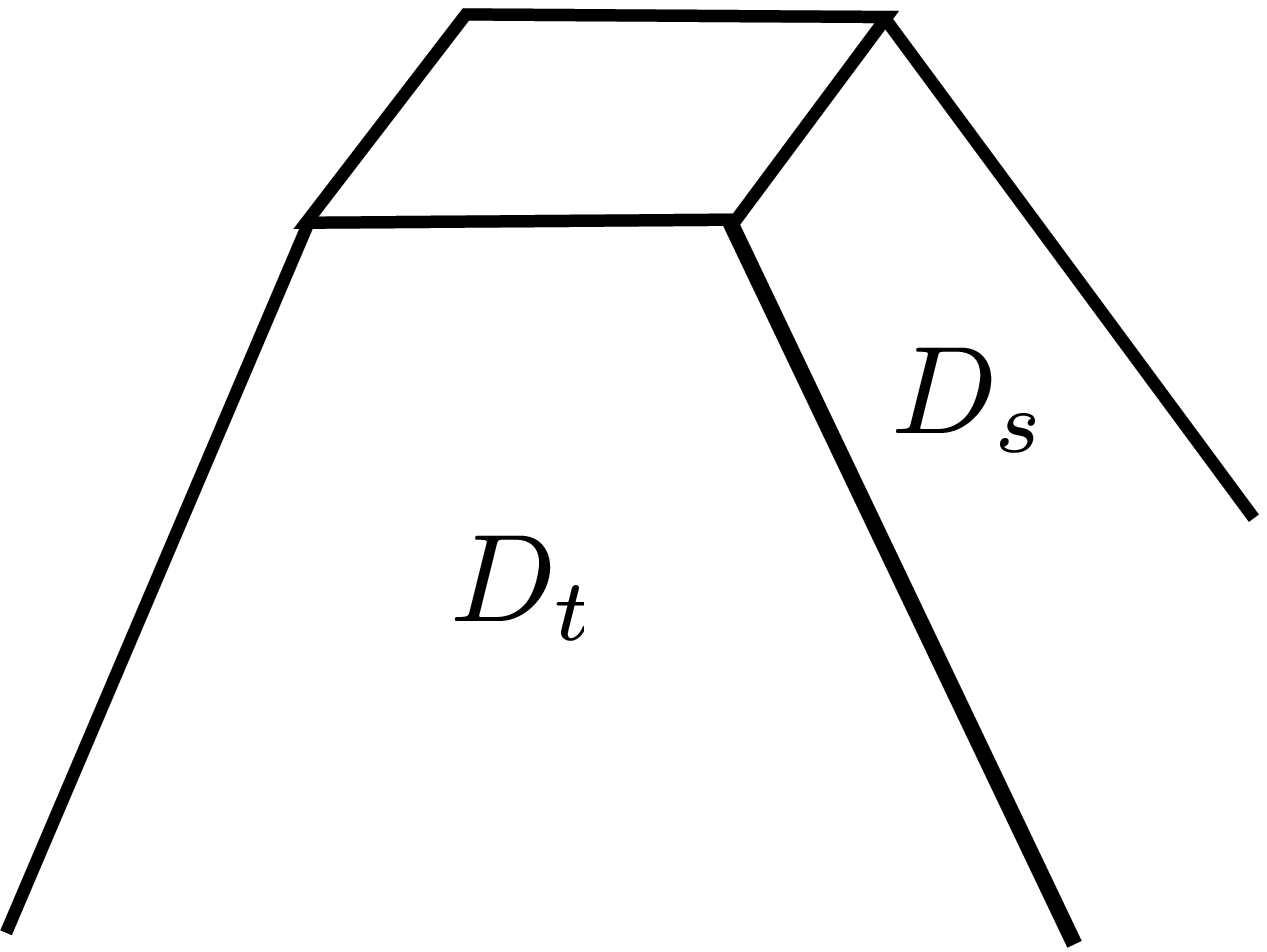}}
\noindent{\ninepoint \baselineskip=2pt {\bf Fig. 15.} {The toric base of local ${\bf P^1} \times {\bf P^1}$, with the non-compact divisors labeled by $D_{s}$ and $D_{t}$.}}
\bigskip
There is one compact divisor $D_S = D_0$ with 
$$
D_S = -2 D_s -2 D_t
$$
The divisor $D_S$ restricts to the surface $S={\bf P^1} \times {\bf P}^1$.

The four three-cycles $\Delta_{\alpha}$ on $Y$ map to four exceptional sheaves supported on $X$, whose charges span the compact homology of $X$. The D4, D2 and the D0 branes correspond to a  collection of exceptional sheaves %
$$E_3={\cal O}_{S}(-2,-2), \; E_2 ={\cal O}_{S}(-1,-2)[-1], \;E_4 ={\cal O}_{S}(-2,-1)[-1], \; E_1=  {\cal O}_{S}(-1,-1)[-2],$$
 supported on $S$. To compute the charges, we use the fact that a D4-brane with a ${\cal O}_{S}(m,n)$ line bundle has Chern class
$$
 D_S e^{n D_t+m  D_s-{1\over 2}K_S}(1+{c_2(S)\over 24}),
$$
which equals\foot{We are using that the canonical class $K_S$ of the surface $S$ 
equals the class of the divisor $D_S$ restricting to it in a Calabi-Yau, $K_S= D_S$ and
$D_S \cdot D_s = C_t$, $D_S \cdot D_t = C_s,$
$D_t \cdot D_t \cdot D_t = D_s \cdot D_s \cdot D_s = {1\over 4}$, $D_t \cdot D_t \cdot D_s=
D_t \cdot D_s \cdot D_s = -{1\over 4}.$
Moreover $D_S c_2(S)$ is just the Euler characteristic of $S$, which is $4$ in this case.}
$$(D_S + (m+1)C_t+(n+1)C_S +(m+1)(n+1) {\rm pt})(1+{c_2(S) \over 24}).$$
 
This gives the charges of fractional D-branes are
\eqn\QC{{\eqalign{
\Delta_3 \; &=(D_S - C_t - C_s + {\rm pt})\Big(1+ {c_2(S)\over 24}\Big)
\cr  \Delta_2\; &  = (-D_S + C_t)\Big(1+ {c_2(S)\over 24}\Big),\cr 
\Delta_4  \;& =( - D_S +C_s )\Big(1+ {c_2(S)\over 24}\Big),\cr 
\Delta_1\; &=D_S\Big(1+ {c_2(S)\over 24}\Big).
}}
}
The non-vanishing intersection numbers can be derived from \GU\  and found to be
$$n_{32} = n_{34} = n_{21}=n_{41} = 2, \qquad n_{13}= 4,$$
in agreement with mirror symmetry and section 3.

Now we add a D6 brane corresponding to the simplest choice,  
$$E_0={\cal O}_X[-1],$$
a trivial line bundle on the Calabi-Yau, with charge
$$\Delta_0\;= -X(1+{1\over 24} c_2(X)).
$$
Clearly, $Ext^{k}_{X}(E_0,E_0)=0$, except for $k=0,3$. We can compute the spectrum of bifundamentals using \foot{Recall, see footnote 17, that the sheaf $E_{\alpha}$ is really $i_* E_{\alpha}$, inherited from $S$ by the embedding $i:S\rightarrow X$. 
Therefore, what we call $Ext^k_X({\cal O}_X, E_{\alpha})$ is $Ext^{k}_{X}({\cal O}_{X}, i_{*}E_{\alpha}) = Ext^{k}_{S}(i^{*}{\cal O}_{X},E_{\alpha}) = Ext^{k}_{S}({\cal O}_{S},E_{\alpha}) = H^{k}(S,E_{\alpha})$.  We thank E. Sharpe for an explanation of this point.}
$$
Ext^k_{X}({\cal O}_X, E_{\alpha}) = H^{k}(S,E_{\alpha}),
$$
$$Ext^k_{X}(E_\alpha, E_{\beta}) = Ext^{k-p+q}_{X}(E_{\alpha}[p], E_{\beta}[q]),$$
and ${\rm dim}H^0({\bf P}^1, {\cal O}(m)) =m+1$, ${\rm dim}H^1({\bf P}^1, {\cal O}(-m)) =-m-1,$ we 
 find that 
$$Ext^1_{X}(E_0, E_{3}) = {\bf C},$$
while the rest $Ext^{0,1}$'s vanish in this sector, so there is precisely one chiral bifundamental multiplet between $E_0$ and $E_3$.\foot{The intersection numbers in the D4-D2-D0 quiver follow similarly.  Using that $Ext_S^k(V_{\alpha}, V_{\beta}) = H^{k}(S,V_{\alpha}^* \otimes V_{\beta})$, since $V_{\alpha}$'s are bundles on $S={\bf P^1}\times{\bf P}^1$. 
This
implies that, for example $Ext^1_{X}(E_{3}, E_2) = H^0(S,{\cal O}(1,0)) = {\bf C}^2$, and $Ext^2_{X}(E_{3}, E_1) = H^2(S,{\cal O}(1,1)) = {\bf C}^4=
Ext^1_{X}(E_1,E_3)$, as claimed.}

$$
n_{03} = 1.
$$

\newsec{Monodromy and $B$-field}

One particularly simple example of monodromy in the moduli space corresponds to changing $B$ by integer values.
Shifts of the NS-NS B-field by a two form $D \in H^{2}(X,{\bf Z})$
\eqn\Bshift{
B \rightarrow B + D, \qquad  D \in H^{2}(X,{\bf Z})
}
are a symmetry of string theory. The closed string theory is invariant under this, essentially per definition since we come back to the same point in the moduli space. 
This does not mean that the states are invariant - in particular, the D-branes are not invariant.
Turning on a B-field is the same as turning on lower dimensional brane charges. The induced charges are given by
\eqn\shift{
\Delta \rightarrow \Delta e^D,
}
for any sheaf with chern class $\Delta \in H^*(X, {\bf Z})$, as shifting the $B$-field is the same as shifting the field strength $F$ on the brane by $D$. 
The state thus does not come back to itself. 
Since we come back to the same point in the moduli space, we expect the partition function to change as
\eqn\bc{
Z(q) = \sum_{\Delta} \Omega(\Delta) q^{\Delta} \rightarrow Z'(q) = \sum_{\Delta} \Omega(\Delta e^{D}) q^{\Delta}
}
simply corresponding to the fact that $\Delta$ is mapped to a different state $\Delta e^{D}$ at the same point in the moduli space.
$Z$ and $Z'$ are of course equivalent, up to a change of variables. 
This looping around the moduli space can be represented in terms of crossing a sequence of walls of the ``second kind" and of  Seiberg dualities. 

In the present context, there is a subtlety in the realization of \bc, due to the fact that the Calabi-Yau is non-compact,  and that the quiver does not describe all the possible D-branes on $X$. Start with an extended quiver describing bound states of a D6 brane, say corresponding to ${\cal O}_X$,
with compact D4-D2-D0 branes. Shifting B-field as in \Bshift\ corresponds
to adding non-compact D4 brane charge to the D6 brane, if the divisor corresponding to $D$ is non-compact. In this case, we get \bc\ only after summing the contributions of different quivers in which the $\Delta_0$ node corresponds to ${\cal O}_X(D)$, for different $D \in H^{2}(X,{\bf Z})$. 
If however, we consider shifts by two forms with compact support,
$$D \in H^{2}_{cmpct}(X, {\bf Z}),$$
then this is a symmetry of the fixed quiver gauge theory as well. This is because the D6-D4-D2-D0 quiver describes
all the BPS states corresponding to a D6 brane bound to compact D-branes on $X$. 
Thus, the different D6-D4-D2-D0 quiver gauge theories we can get on $X$, at the same point in the moduli space, are classified  by
$$ 
D \in H^{2}(X, {\bf Z})/H^{2}_{cmpct}(X, {\bf Z}).
$$
As an aside, note that if we consider only the D4-D2-D0 quiver, as in section 3,
having no non-compact charge to begin with, any shift of B-field \Bshift\ is a symmetry.

\subsec{The ${\bf P}^1 \times {\bf P}^1$ example}

Consider the effect of changing the B-field on the quiver $Q$ in Figure 5.
$$
B \rightarrow B + n_t D_t + n_s D_s, \qquad n_s,n_t \in {\bf Z}.
$$
Since $H^{2}_{cmpct}(X, {\bf Z})$ is generated by the divisor $D_S$
$$D_S = -2 D_t - 2 D_s.
$$
shifts by
$(n_s,n_t) = (2n , 2n)$ for $n$ an integer, should be a symmetry both of the quiver and of the spectrum. Thus, the different quiver gauge theories one can get by shifting B-field are classified by $(n_t,n_s)$ modulo shifts by $(2,2)$.

Consider increasing the $B$-field by $(n_t,n_s) = (1,0)$. In the language of IIB on $Y$, shifting $B$ by $D_t$ corresponds to shifting
$$z_t, z_s \rightarrow z_t e^{2\pi i} ,z_s.
$$
This deforms the 3-cycles $\Delta_{\alpha}$ as in the Figure 16 to $\Delta_{\alpha}'$, related to the good basis we started out with 
by
\eqn\chbb{\eqalign{
\Delta_3 &\;\rightarrow \; 
\Delta_3' = - \Delta_4  \cr
\Delta_4 &\;\rightarrow \;  \Delta_4' = \Delta_3+2\Delta_4 \cr
\Delta_2 &\;\rightarrow \;  \Delta_2'=- \Delta_1 \cr
\Delta_1 &\;\rightarrow \; \Delta_1' = \Delta_2+2 \Delta_1 
}
}
The flip of orientation of $\Delta_1$ and $\Delta_4$ is necessary for their orientations to agree with $\Delta_{2}'$ and $\Delta_{3}'$. 

\bigskip
\centerline{\epsfxsize 3.2truein\epsfbox{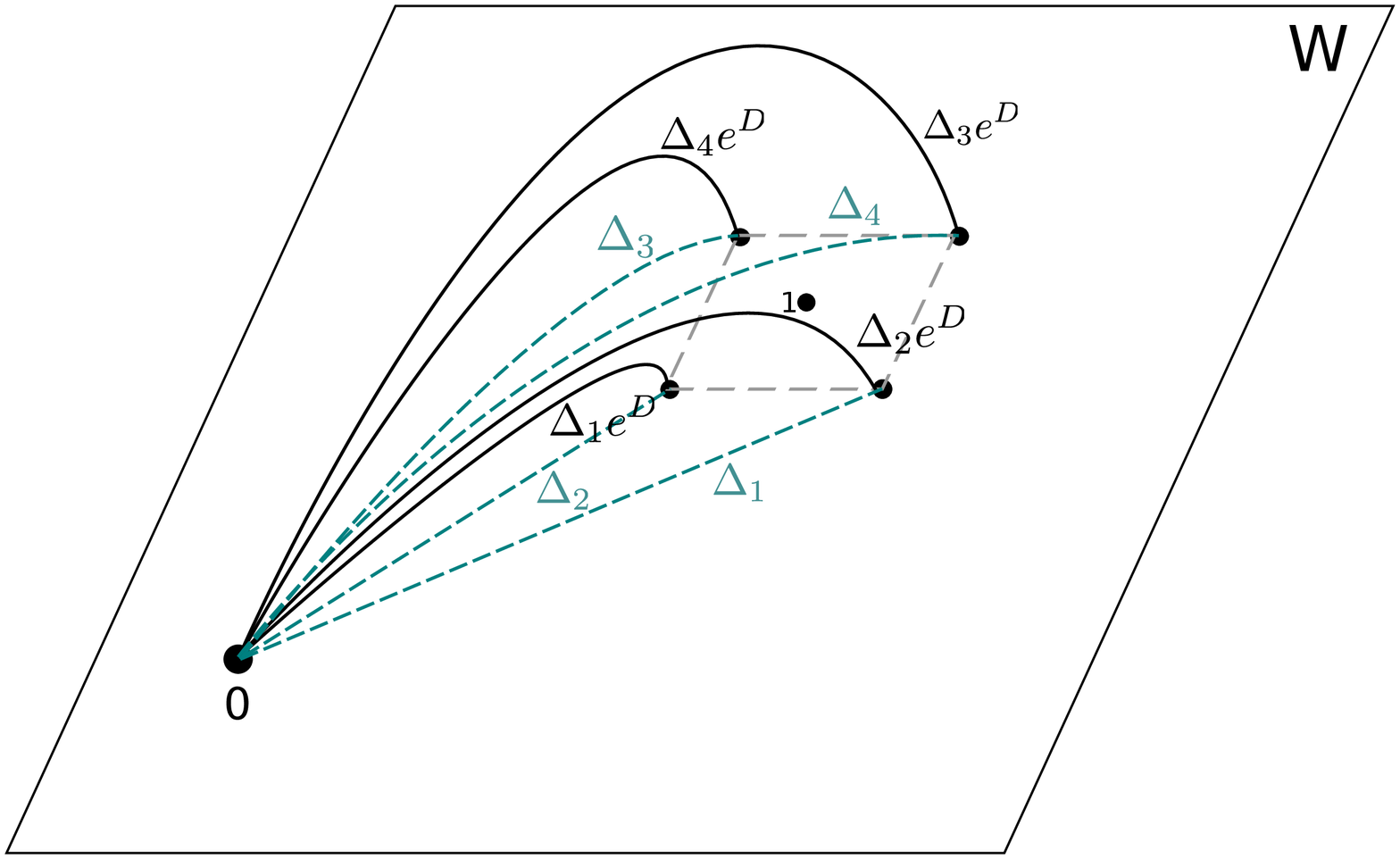}}
\noindent{\ninepoint \baselineskip=2pt {\bf Fig. 16.} {The W-Plane after adding one unit of B-field, (1,0), for local ${\bf P^1}\times {\bf P^1}$.  The deformed cycles $\{\Delta_{i}e^{D}\}$ are shown in black. while the original cycles are shown in blue.}}
\bigskip

Viewed from $X$, $\Delta$ are sheaves on $X$. 
The shift of B-field corresponds to tensoring with a line bundle of first chern class $D_t.$ The chern classes change by
$$
\Delta \rightarrow \Delta' = \Delta e^{D_t},
$$
where above describes how components of $\Delta$ change in a fixed basis for $H_{*}(X)$.  To begin with, for example, $\Delta_3$, and $\Delta_2$ correspond to ${\cal O}_{S}(-2,-2)$ and ${\cal O}_{S}(-1,-2)[-1]$ respectively. After we change the B-field, they pick up charge, and get mapped to ${\cal O}_{S}(-2,-1)$ and ${\cal O}_{D_0}(-1,-1)[-1]$, which correspond to $-\Delta_4$ and $-\Delta_1$, respectively, in agreement with \chbb. Similarly, $\Delta_4$, corresponding to ${\cal O}_{S}(-2,-1)[-1]$ gets mapped to ${\cal O}_S(-2,0)[-1]$.
The chern class of this, $\Delta_4'$ equals     
$$
\Delta_4' = -(D_S +C_t-C_s - \rm{pt})(1-{c_2(S)\over 24})= 
 \Delta_3+ 2\Delta_4,
$$     
in agreement with \chbb.

In the quiver, the shift of the $B$-field by $D_t$ corresponds to a sequence of two Seiberg dualities, 
where we dualize first the node $\Delta_2$, (as we did in the last section) and then the node $\Delta_3$ (Figure 17).
It is easy to see from this that the D4-D2-D0 quiver is invariant under the shift of the B-field, although the D-branes on the nodes get replaced by a linear combinations of the ones 
that we had started out with, given by \chbb. 

\bigskip
\centerline{\epsfxsize 5.0truein\epsfbox{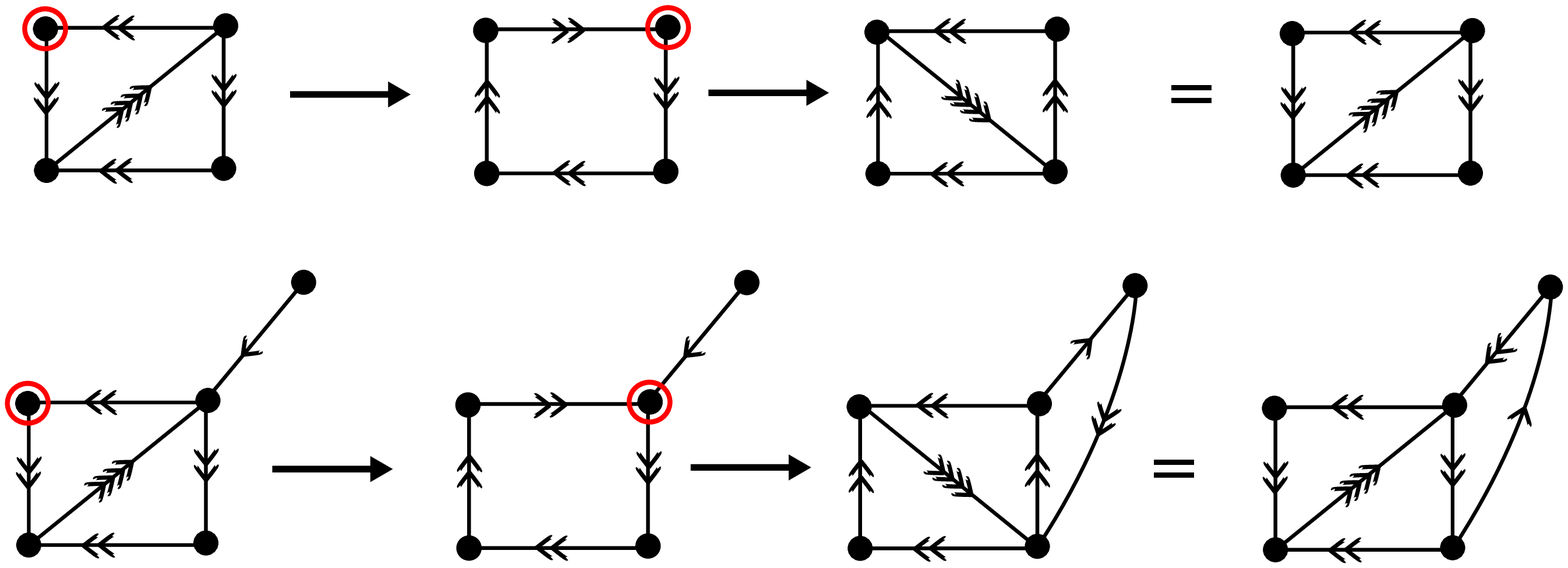}}
\noindent{\ninepoint \baselineskip=2pt {\bf Fig. 17.} {Seiberg Duality corresponding to a shift in the B-field (1,0) for local ${\bf P^1} \times {\bf P^1}$.  The unframed quiver on top is invariant up to a permutation of the nodes, while the framed quiver develops new framing arrows.}}
\bigskip

Now consider the theory with the D6 brane, by including the node $\Delta_0$, which has the bundle ${\cal O}_X[-1]$. The shift of the B-field acts on all the 
D-branes, so it acts on the D6 brane as well, mapping it to $E_0'= {\cal O}_X(D_t)[-1]$.
This implies that the D6 brane picks up D4 brane charge corresponding to the non-compact divisor $D_t$.
Since no other nodes carry this charge,  $E_0'$ becomes the new framing node, with charge 
$\Delta_0'$ induced by the B-field
$$
\Delta_0  = -X \rightarrow \Delta_0' = - Xe^{D_t} 
$$ 
The extended quiver, with the $\Delta_0$ node included also transforms by two Seiberg dualities, Figure 17. The quiver in this case is {\it not}  the same as before, since $\Delta_{\alpha}$ have different intersection numbers with $\Delta_0'$ than with $\Delta_0$.
In particular
$$
n'_{03} =  \Delta'_{0} \circ \Delta_3 =2, \qquad n'_{40} = \Delta_4 \circ \Delta'_0 = 1.
$$
This can also be verified directly from sheaf cohomology, showing that for example $Ext^1(E_0', E_3) = H^0_S({\cal O}(-2,-3)) =2$.   The quiver gauge theory also has a new superpotential term
$$
{\cal W} = \sum_{i, j = 1,2} \epsilon^{ij} {\rm Tr} \,p \,A_i \,q_j + \sum_{i,j,a,b=1,2}\epsilon^{ij}\epsilon^{ab}\bigl(  {\rm Tr} A_i B_{a} D_{jb} + {\rm Tr} {\tilde A}_i {\tilde B}_a D_{jb}\bigr).
$$
In the previous section, we showed that dualizing a node $\Delta_{*}$ in the quiver is realized as a geometric transition in the dimer model
counting the BPS degeneracies, where the face corresponding to $\Delta_{*}$ is dualized. We showed that this implies the 
\KS\ wall crossing formula 
\eqn\dwn{A_{Q'}^{(1)} = A_{-\Delta_{*}}^{-1} A_Q^{(1)} A_{-\Delta_{*}},}
together with a change of variables.
Here, we apply this twice, corresponding to a sequence of two Seiberg dualities:
$$A_{Q'}^{(1)} =  A_{-\Delta_{3}}^{-1}A_{-\Delta_{2}}^{-1} A_Q^{(1)} A_{-\Delta_{2}} A_{-\Delta_{3}}.$$ 
%
%
In the dimer model, this is a change of the boundary conditions at infinity, replacing the top of the cone by an edge two sites long (see Figure 18 and 19).
It is easy to see that this is consistent with the crystal one would have derived from the quiver 
$Q'$ by localization.\foot{The relation of Seiberg duality, wall crossing and crystals was first noted in \JW . That work was one inspiration for this research.} 
\bigskip
\centerline{\epsfxsize 4.0truein\epsfbox{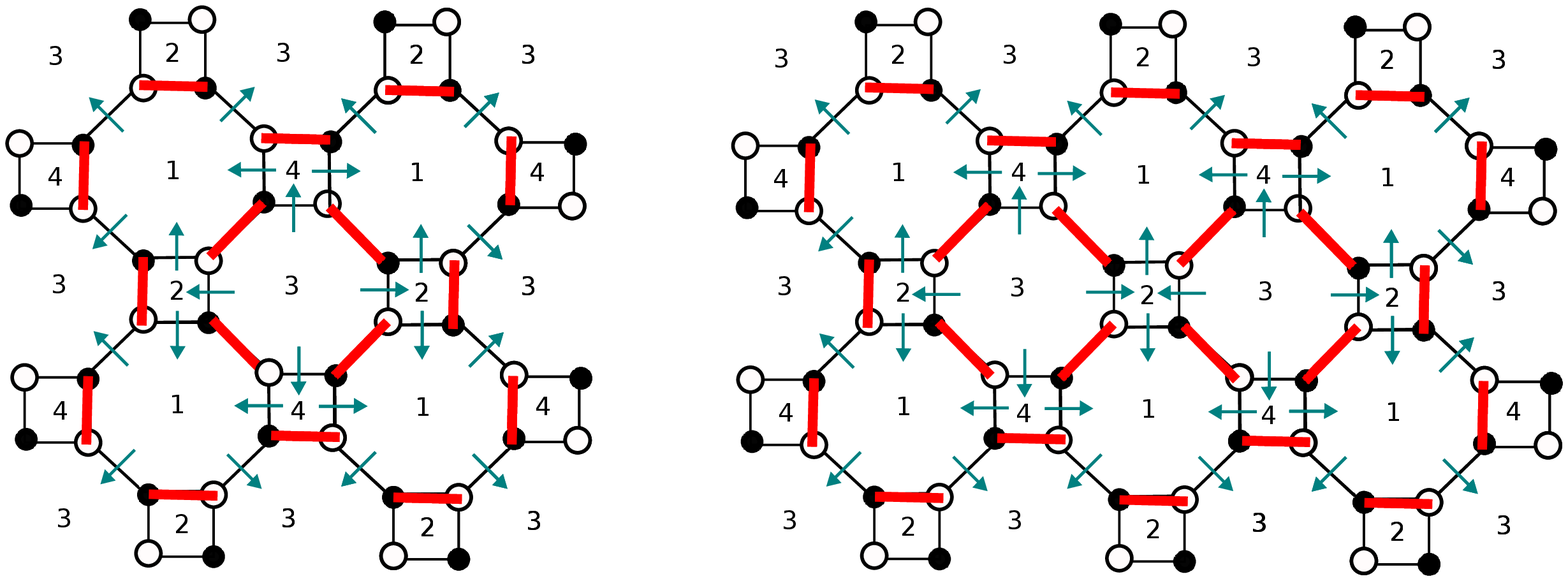}}
\noindent{\ninepoint \baselineskip=2pt {\bf Fig. 18.} {The vacuum dimer configuration for ${\bf P^1} \times {\bf P^1}$ grows an edge corresponding to a shift in the B-field by (1,0).}}
\bigskip

\bigskip
\centerline{\epsfxsize 3.0truein\epsfbox{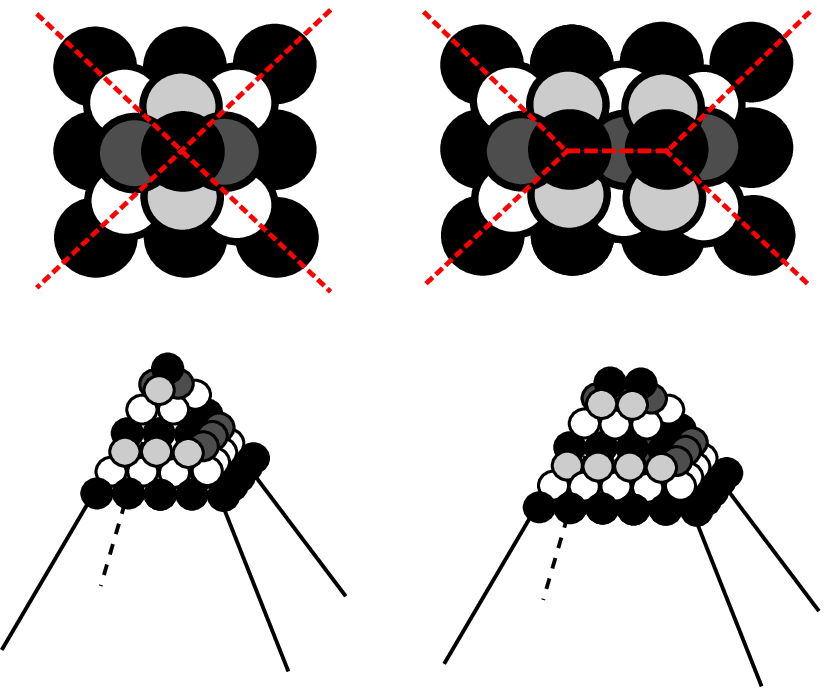}}
\noindent{\ninepoint \baselineskip=2pt {\bf Fig. 19.} {The apex of the ${\bf P^1} \times {\bf P^1}$ crystal grows an edge corresponding to a shift in the B-field by (1,0).}}
\bigskip

Repeating this $m$ times, shifting 
$B$ to $B+ mD_t$
we end up with a quiver $Q_m$ with  $n_{03} =m+1$, and  $n_{40}= m$.
%
%
This corresponds to growing a ridge in the crystal, $m+1$ atoms long as in \refs{\sz,\JW}.  In the next section, we will give a physical interpretation to this observation.

More generally, all the possible inequivalent quivers one can obtain in this way correspond to shifts of B-field by the inequivalent choices in  $H^2(X, {\bf Z})/H^2_{cmpct}(X, {\bf Z})$.
%
%
%
%
%
For example, it is easy to show (either by Seiberg duality, or direct computation) 
that shifting the B-field by $- D_t$, replaces the node
$E_0$ by $ {\cal O}_X(- D_t)[-1]$,  keeping everything else the same. This means that only $n_{04}=1$ is nonzero for arrows beginning or ending on $\Delta_0$. Similarly, replacing $E_0$ by $ {\cal O}_X(- D_s)[-1]$, we get only $n_{02}=1$. Finally, we get only $n_{01}=1$, by
taking  $E_0={\cal O}_X(- D_t - D_s)[-1]$, see figure 20.

\bigskip
\centerline{\epsfxsize 5.0truein\epsfbox{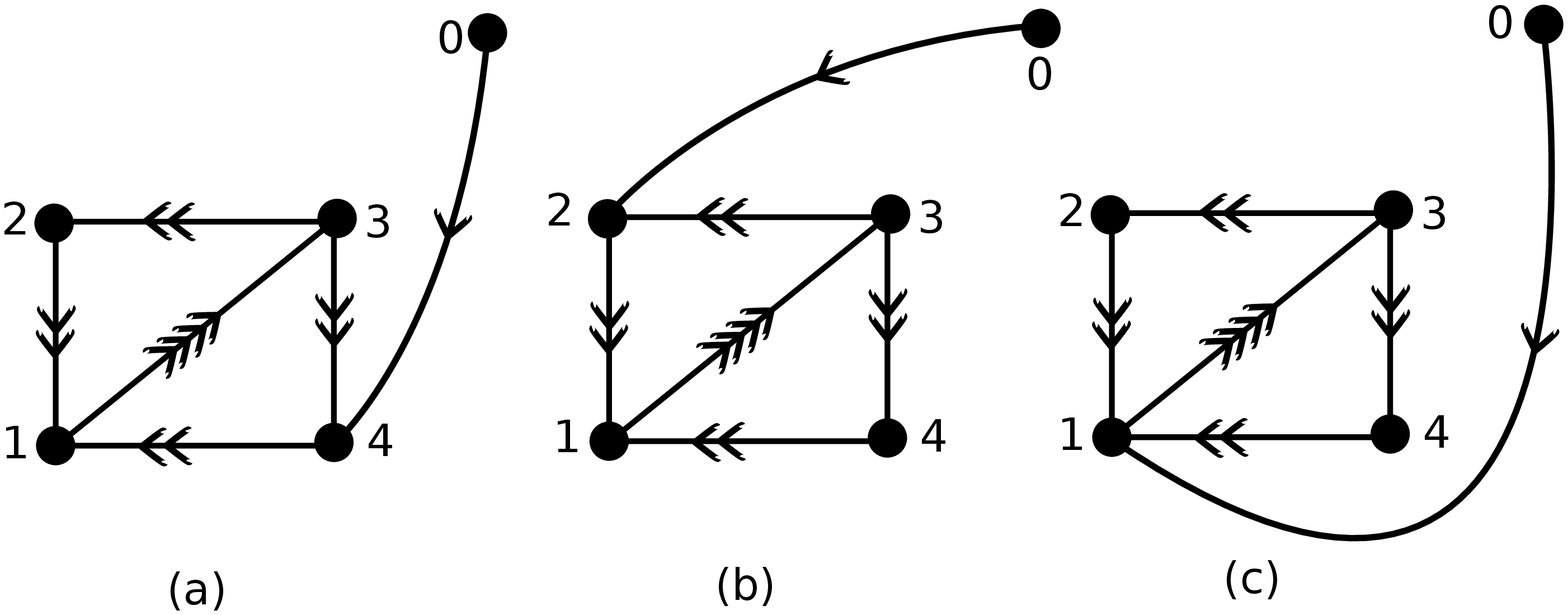}}
\noindent{\ninepoint \baselineskip=2pt {\bf Fig. 20.} {Framings for local ${\bf P^1} \times {\bf P^1}$, corresponding to negative shifts of the B-field that replace ${\cal O}_{X}[-1]$ by (a) ${\cal O}_{X}(-D_{t})[1]$, (b) ${\cal O}_{X}(-D_{s})[1]$, and (c) ${\cal O}_{X}(-D_{t}-D_{s})[1]$.}}
\bigskip

Now consider shifts of the B-field by $D=- D_S \in H^2_{cmpct}(X, {\bf Z})$,
$$
B \rightarrow B-  D_S.
$$
This is a symmetry of the quiver. It can be implemented by changing the B-field by $D_t$, $D_s$, $D_t$ and $D_s$,  corresponding to a sequence of four Seiberg dualities on the quiver, which in the end leaves the quiver invariant, except that the charges of the nodes of the quiver change.\foot{The order in which we cross the walls matters, only in so much that the description is simplest for one particular order, and for that ordering we can describe this by Seiberg duality. Otherwise, this does not have a simple interpretation in the gauge theory. The end result, however, is independent of the order.} This can also be seen from the perspective of the Kontsevich-Soibelman algebra.
This corresponds to conjugation of $A_{Q}^{(1)}$ by
$ A_{-\Delta_{2}} A_{-\Delta_{3}}$, $A_{-\Delta'_{4}} A_{-\Delta'_{3}}$, $A_{-\Delta_{2}''} A_{-\Delta_{3}''}$, and $A_{-\Delta_{4}'''} A_{-\Delta_{3}'''},$
respectively, 
where the four $\Delta_3$'s, for example, correspond to the four $\Delta_3$ nodes in the four quivers we get along the way.
This can be rewritten as conjugation by 
\eqn\conj{M =M_{\Delta_3} M_{\Delta_3'} M_{\Delta_3''} M_{\Delta_3'''},}
in terms of the operator
$$ M_{\Delta} =  A_{-\Delta} A_{\Delta}.
$$
$M_{\Delta}$ is a monodromy operator, implementing  the shift of the charge\foot{Since
$
M_{\Delta}^{-1}  e_{\Delta'} M_{\Delta}= (1-e_{-\Delta})^{-\Delta \cdot \Delta'}  (1-e_{\Delta})^{\Delta \cdot \Delta'} e_{\Delta'}.
$}
$$
M_{\Delta} :\;\; e_{\Delta'}\rightarrow (-1)^{\Delta \cdot \Delta'}  e_{(\Delta' - ({\Delta' \cdot \Delta})\Delta)}.
$$
If we denote by
$$
M: \;\;e_{\Delta} \rightarrow e_{f(\Delta)},
$$
the net effect on the partition function is that
$$
M: \;\; \sum_{\Delta} \Omega_{Q}(\Delta) q^{\Delta} \rightarrow  \sum_{\Delta} \Omega_{Q}(\Delta) q^{f(\Delta)} = 
\sum_{\Delta} \Omega_{Q}(f^{-1}(\Delta)) q^{\Delta} ,$$
where we changed variables in the last line. It can be shown that $f^{-1}(\Delta)$ is exactly what is needed for this to correspond to turning on the B-field, $-D_S$,
namely, $f(\Delta) = \Delta e^{D_S}$.

\newsec{Geometry of the D6 brane bound states}

In this section, we begin by considering a fixed state, a D6 brane bound to D4 branes, D2 branes and D0 branes of charge $\Delta$, and ask how its degeneracies change as we shift the B-field by $D\in H^2(X, {\bf Z})$. Turning on a B-field is the same as turning on lower dimensional brane charges on the D6 brane. The induced charges are given by
\eqn\shift{
\Delta \rightarrow \Delta e^D,
}
as shifting the $B$-field is the same as shifting the field strength $F$ on the brane by $D$. When we come back to the same point in the moduli space, the states have been re-shuffled by \shift . So the degeneracies of any one state will change (at the same time, the spectrum as a whole is invariant, as we emphasized in the previous section).
The degeneracies, and how they change, can be found from the quivers and crystals describing the brane. A state of arbitrary charge $\Delta$, in general, corresponds to a complicated configuration of the melting crystal ${\cal C}$. 

It turns out that the vacuum of the crystal ${\cal C}$, and certain subset of its states have a beautifully simple geometric description, closely related to the geometry of the Calabi-Yau.  We will spend some time explaining this, and then use this result to show that, in the limit of the large B-field, 
$$
D \rightarrow \infty
$$
counting of quiver degeneracies reduces to the Donaldson-Thomas theory as formulated in \refs{\QF,\MNOPI,\MNOPII}, in terms of a counting ideal sheaves on $X$.  The latter was discovered by trying to give a physical interpretation to the combinatorics of the topological vertex \TV\  and Gromov-Witten theory. In this way, we will be able to derive the famous Donaldson-Thomas/Gromov-Witten correspondence directly from considering quiver representations in this limit. The proposal that the DT/GW correspondence holds in the large B-field limit was put forward in \DM\ using split attractors
and verified in \AOVY\ for Calabi-Yau manifolds without compact 4-cycles, using M-theory.  

\subsec{Geometry of D6 branes}

To begin with, consider the D6 brane itself, the sheaf ${\cal O}_X[-1]$, which corresponds to the entire crystal, and 
$$
\Delta  =  -X = \Delta_0.
$$
The crystal ${\cal C}$  is a cone in the lattice $\Lambda = {\bf Z}_{\geq 0}^3$, whose geometry is closely related to the geometry of the Calabi-Yau. Namely, consider the intersection of ${\cal C}$ with a sublattice $\Lambda_0 \subset \Lambda$ corresponding to points of color $\alpha$, where
$\alpha$ is the framing node. The subset of lattice points ${\cal C}_0 = {\cal C}\cap \Lambda_0$ correspond to holomorphic functions on $X$. 

This can be seen as follows \refs{\HVH,\Hananythree,\Kennaway}. Recall that the cone ${\cal C}$ is generated by the ${\bf T}^3$ weights of the paths in $A_0$, starting at the framing node. The subcone ${\cal C}_0$ corresponds to paths ending on node ${\alpha_0}$. Because $n_{0\alpha}=1$, such paths can be viewed as starting and ending at the node of color $\alpha$. These correspond to single trace operators ${\rm Tr}\, O$ in the quiver, where we take the ranks to infinity, so that there are no relations between the traces. The chiral operators $O$ are generated by a set of monomials $M_i$ corresponding to the shortest loops in the quiver, beginning and ending on node $\alpha$. The monomials $M_{i}$ all commute, since two paths in the quiver of the same $R$-charge and the same endpoints are equivalent in the path algebra. Thus, we can simply write $O$ as $O =M_{1}^{n_1} M_{2}^{n_2} \ldots M_{k}^{n_k}$. Since the order is irrelevant, the space of operators ${\rm Tr}\, O$ is the same as considering all gauge invariant chiral operators in the abelian quiver theory corresponding to a single D0 brane on $X$, described in the quiver by taking all ranks to be $1$.  Since the moduli space of a D0 brane is $X$, the later is the space of holomorphic functions on $X$. 

The space of holomorphic functions on $X$ is also a lattice, closely related to the geometry of $X$. As we reviewed in section 5, $X$ comes with a set of coordinates $z_i,$
satisfying \cons\
\eqn\consa{
\sum_i Q_i^a |z_i|^2 = r_a,
}
modulo gauge transformations \gauge . We can view $X$ as a fibration over the toric base obtained by forgetting the phases of $z_i$  and using $|z_i|^2$ as 
coordinates.
Consider the integral points in the base, where 
$$
N_i = |z_i|^2, \qquad N_i \in {\bf Z}_{\geq 0}.
$$
For each such point, we get a monomial in $z_i$
$$
\prod_i z_i^{N_i}
$$
which is a function on $X$, since
\eqn\intersect{
\sum_i Q_i^a N_i=0.
}
We have set $r_a=0$ in \cons , as we are starting out with singular $X$. This is because the quiver we started out with has just one framing node, and so ${\cal C}$ was a cone with an apex at the origin. Since the Calabi-Yau is three dimensional, the space of all such monomials is a three dimensional lattice, the singular cone ${\cal C}_0$.

Now consider what happens to the D6 brane as we increase the $B$-field 
$$B \rightarrow B+ D.$$
Increasing the B-field is the same as turning on flux on the D6 brane,  tensoring ${\cal O}_X[-1]$ with a line bundle of first Chern class $D$. The D6 brane becomes ${\cal O}_X(D)[-1]$, and the charge of the state becomes 
$$
\Delta= -Xe^{D}.
$$
Here, we assume
\eqn\div{
D = \sum_{a} n_a D_a, \qquad n_a\geq 0
}
where $D_a$ generate the Kahler cone. We claim that the state $\Delta$ corresponds to the deformed crystal ${\cal C}(D)$ whose lattice sites ${\cal C}_0 (D)= {\cal C}(D)\cap \Lambda_0$ are
holomorphic sections of an $O(D)$ bundle over $X$ instead. This corresponding to solving
\eqn\bnd{
\sum_i Q_i^a N_i=n_a.
}
As ${\cal C}_0$ is a discretized version of the base of Calabi-Yau, modifying ${\cal C}_0$ as in \bnd\ modifies the moduli. 

\bigskip
\centerline{\epsfxsize 2.0truein\epsfbox{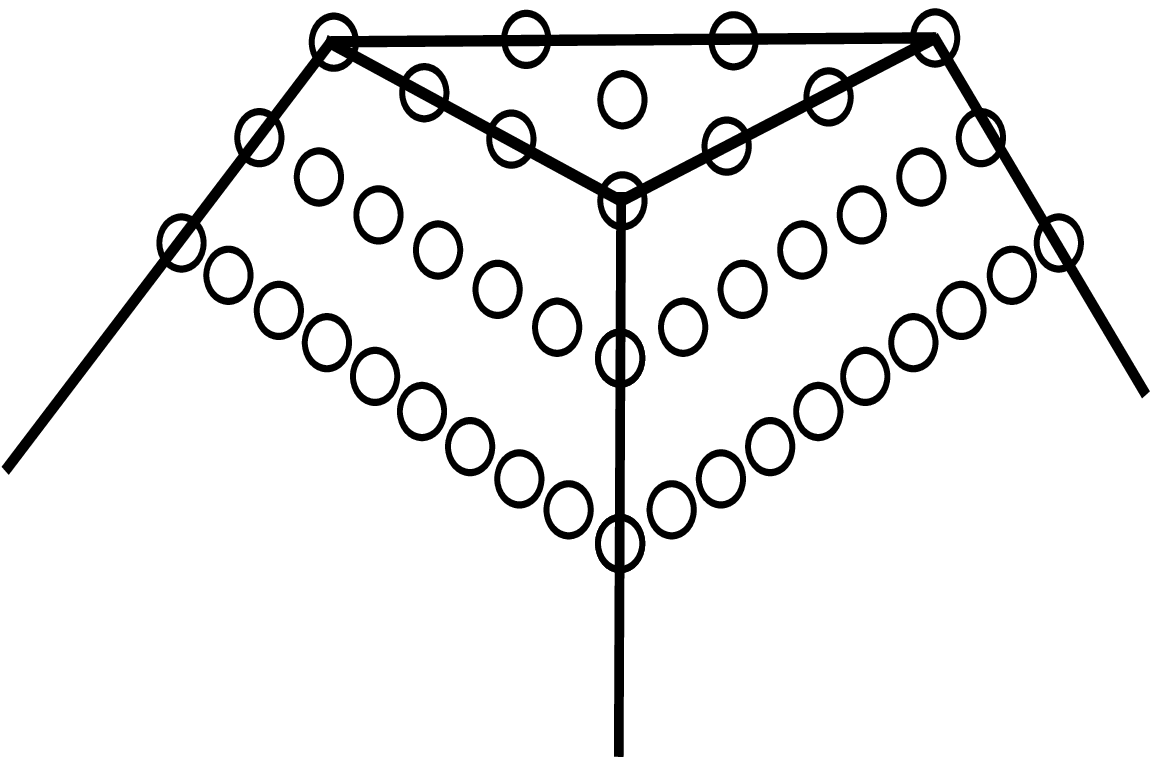}}
\noindent{\ninepoint \baselineskip=2pt {\bf Fig. 21.} {Integral points in the toric base of local ${\bf P}^{2}$ correspond to holomorphic sections of ${\cal O}(D)$.}}
\bigskip
%

The cases $D \in H^{2}_{cmpct}(X, {\bf Z})$ and $D\in H^2(X, {\bf Z})/H^2_{cmpct}(X, {\bf Z})$, should be discussed separately, since they differ in character. 
When  
\eqn\ncp{D=D_0\in H^2(X, {\bf Z})/H^2_{cmpct}(X, {\bf Z}),}
the quiver changes since the D6 brane node becomes ${\cal O}_X(D)[-1]$, so
$$\Delta = (-X)e^{D_0}=\Delta_0$$ 
corresponds to the ``vacuum" configuration of the crystal ${\cal C}(D_0)\subset \Lambda$ associated with this new quiver. We are growing the edges of the crystal, corresponding to increasing the lengths of curves, 
but no faces open up. In other words, shifts of the B-field by non-compact divisors \ncp\ correspond to purely non-normalizable deformations of the crystal and the Calabi-Yau. 
Next, consider shifting the B-field by $D$, such that 
\eqn\cp{D-D_0\in H^2_{cmpct}(X, {\bf Z}),}
This $D- D_0 \neq 0$ corresponds to shifting the normalizable moduli of the Calabi-Yau, opening up faces in the toric base, and deforming ${\cal C}_0$ without changing its asymptotics.
This adds compact D4 brane charge to the D6 brane, and this can always be described in the quiver we had started with. To verify that the configuration of the crystal corresponds to ${\cal O}_X(D)[-1],$
we need to express its Chern character
$$
\Delta = -Xe^{D}
$$
in terms of the Chern characters of the nodes of the quiver 
$$
\Delta = \Delta_0 + \sum_{\alpha} N_{\alpha}   \Delta_{\alpha},
$$
where $\Delta_0$ corresponds to the D6 brane node,
$$
\Delta_0= -X e^{D_0}.
$$
Note that ranks $N_{\alpha}$ depend both on $D$ and $D_0.$ The set of ${N}_{\alpha}$'s we obtain in this way corresponds to how many atoms of color $\alpha$ we need to remove to 
describe ${\cal O}_X(D)[-1]$. In specific examples, this is completely straight forward, the only subtlety being that, to get integers, one has to keep careful track of the curvature contributions to the charges, which we have been mostly suppressing so far.  It would be nice to find a general proof of this. 

\subsec{An example}

Consider the local ${\bf P^1}\times{\bf P}^1$. In this case, $Q^1 = (1,1,0,0,-2)$, $Q^{2} =(0,0,1,1,-2)$ 
so the ``pure" D6 brane ${\cal O}_X[-1]$ on $X$ corresponds to the set of points 
$$N_1+N_2 = 2 N_0, \qquad\;\; N_3+N_4 = 2 N_0,
$$
in ${\Lambda_0}.$ The cone direction here is parameterized by $N_0$. At fixed $N_0$, we get a square with $(2N_0+1)\times(2N_0+1)$ integral points. It is easy to see that this agrees with we get by considering either of the two quivers in section 3, and a subset of points corresponding to node $\alpha_0 = 3$, in this case. Of course, the finer structure of the crystal 
${\cal C}$ is different in the two cases, but the geometry of ${\cal C}_0$ is the same.

\bigskip
\centerline{\epsfxsize 4.0truein\epsfbox{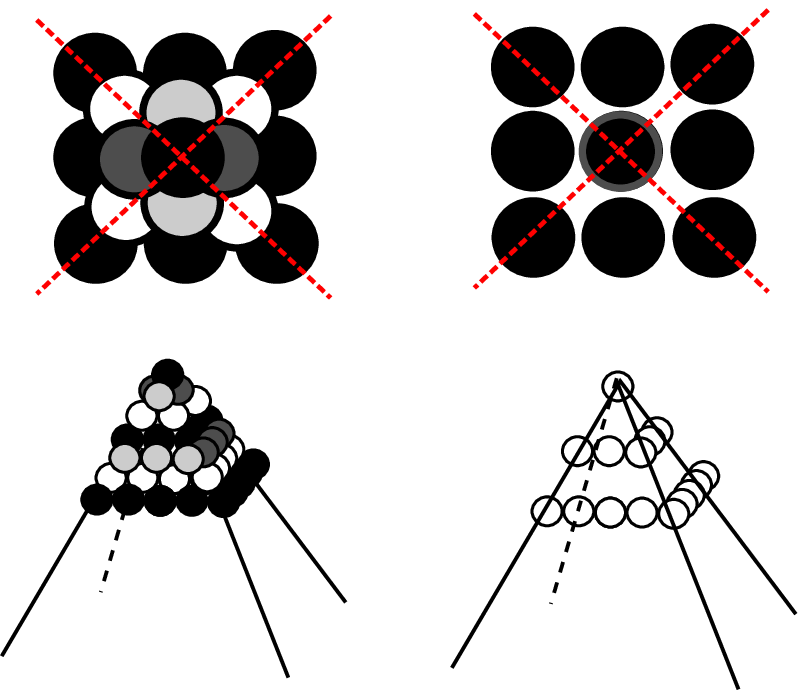}}
\noindent{\ninepoint \baselineskip=2pt {\bf Fig. 22} {The crystal for local ${\bf P^1}\times{\bf P}^1$.}}
\bigskip

Take now ${\cal O}_X(m D_t)[-1]$. This is the D6 brane node of the quiver $Q_m$ in the previous section. 
The corresponding crystal ${\cal C}$ gets deformed to ${\cal C}(m D_t)$ by replacing the apex of the cone with an edge $m+1$ sites long. The crystal sites of ${\cal C}_0(mD_t)$ are given by 
\eqn\edge{N_1+N_2 = 2 N_0+m, \qquad\;\; N_3+N_4 = 2 N_0,
}
which corresponds to giving the curve $C_t$ length $m+1$.
Add to this $-n$ units of the compact D4 brane charge, by taking $D = (m+2n)D_t + 2n D_s$ instead. From what we had said above, we expect a face of $(m+2n +1) \times (2n+1)$ nodes of color $\Delta_3$  to open up at the appex
\eqn\face{N_1+N_2 = 2 N_0+m+2n, \qquad\;\; N_3+N_4 = 2 N_0+2n.
}
\bigskip
\centerline{\epsfxsize 4.0truein\epsfbox{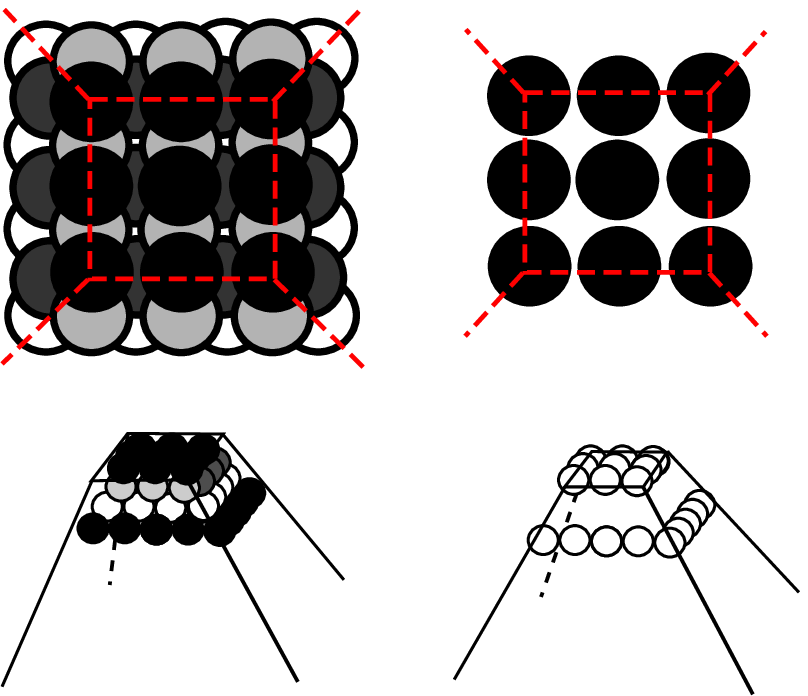}}
\noindent{\ninepoint \baselineskip=2pt {\bf Fig. 23.} {The crystal for local ${\bf P^1}\times{\bf P}^1$ with compact B-field.}}
\bigskip

Now we will show that this precisely describes the state  ${\cal O}_X(D)[-1]$, whose charge is
$$\Delta = -X(1+{c_2(X)\over 24}) e^{(m+2n)D_t+2n D_s}.
$$
in terms of the crystal associated with the quiver $Q_m$. The quiver has the D6 brane node ${\cal O}_X(m D_t)[-1]$
of charge 
$${\Delta_0} = -X(1+{c_2(X)\over 24}) e^{mD_t}.
$$
The difference of these charges  
\eqn\diff{\Delta - \Delta_0 = n D_0 + (n+m)n C_s+n^2 C_t-({4\over 3}n^3 + {1\over 2} m n^2 - {1\over 6} n) {\rm pt},
}
has to be carried by the D4-D2-D0 nodes of the quiver, and moreover, should correspond  to the nodes 
we needed to remove to go from \edge\ to \face.  The number of sites we remove
from the crystal ${\cal C}$ corresponding to the quiver $Q_m$ is
\eqn\pyr{\eqalign{
\sum_{i=0}^{n-1}& (2i+1+m)(2i+1) \Delta_3 + (2i+1)(2i+2 +m) \Delta_2\cr 
+& (2i+2)(m+2 i +1)\Delta_4 +(2i+2)(2i+2+m)\Delta_1}.
}
Recall that \QC\ 
$$\eqalign{
\Delta_3 &= (D_S - C_t - C_s + {\rm pt})(1+{1\over 24} c_2(S))\cr
\Delta_4 &= (-D_S + C_s)(1+{1\over 24} c_2(S))\cr
\Delta_2 &= (-D_S + C_t)(1+{1\over 24} c_2(S))\cr
\Delta_1 &= D_S (1+{1\over 24} c_2(S))}.
$$
Adding up \pyr\ and using that, for a divisor $D_S$ which restricts to a surface $S$ in the Calabi-Yau
\eqn\cc{
c_2(X) D_S = (c_2(S) - c_1(S)^2) D_S
}
and
$$
c_2(S) D_S = \chi(S) {\rm pt}, \qquad  c_1(S)^2 D_0 = (12 - \chi(S)){\rm pt},
$$
we recover \diff . 
Above $\chi(S)$ is the euler characteristic of $S$. Here, $S={\bf P^1}\times {\bf P}^1$, and $\chi(S)=4$.

While we did the explicit computation for the quiver $Q_m$, we could have just as well used the dual quiver $Q_m'$,
obtained by dualizing node 2. This changes the microscopics of the crystal, so ${\cal C}(D)$ changes, however the shape of the crystal and ${\cal C}_0(D)$ stays the same. 
This had better be the case, as ${\cal C}_0$ does not know about the full quiver, but only about a subset of its nodes that are untouched by the quiver mutation. Explicitly, the charges of the nodes of $Q_m$ and $Q_m'$ are related by
$$\eqalign{
\Delta_2' &= \Delta_1+2\Delta_2\cr
\Delta_1' &= -\Delta_2\cr
\Delta_3' &= \Delta_3\cr
\Delta_4' &= \Delta_4
}
$$
while $\Delta_0=\Delta_0'$. Using this to rewrite \pyr, we get
$${\eqalign{
\sum_{i=0}^{n-1} &(2i+1+m)(2i+1) \Delta_3' + 
(2i+2)(2 i +1+m)\Delta_4'\cr
+& (2i+2)(2i+2+m)\Delta_2'+ (2i+3)(2i+2 +m) \Delta_1'. }}
$$
It is easy to see that this counts the nodes in the cut-off top of the crystal based on the quiver $Q'_m$.

In the next subsection, we will consider meltings of ${\cal C}(D)$. This counts a subset of states of the original crystal ${\cal C}$, which naturally should correspond to bound states of D6 brane with an ${\cal O}_X(D)$ bundle on it, with lower dimensional branes.
As an aside, note that, while this problem can be phrased in terms of counting a subset of states of the original quiver, we can try to define it more directly as well, in terms of a D4-D2-D0 quiver, with a framing node corresponding to ${\cal O}_X(D)[-1]$. However, such quivers appear to contain bifundamentals in $Ext^{0,3}(E_{\alpha}, E_{\beta})$, which correspond to ghosts, not chiral multiplets \refs{\Wij,\HeckV}.

\subsec{Geometric interpretation of the D6 brane bound states}

Take a D6 brane $-X$, with any number of
{\it i.} D4 branes wrapping surfaces $S$ in $H_4(X, {\bf Z})$, {\it ii.}
D2 branes wrapping curves $C$ in $H_2(X, {\bf Z})$ and {\it iii.} D0 branes where we require $D_S$, $C$ to be positive\foot{More precisely, we require C to be in the Mori Cone of X, and we require S to be a very ample divisor so that $S\cdot D>0$ for any C, S satisfying these conditions \msw.}, and the number of D0 branes to be non-negative. 

Changing the B-field
by $D$ \shift , gives this state a simple geometric interpretation in terms of removing {\it i.} faces corresponding to toric divisors $D_S$, restricting to $S$, {\it ii.} edges corresponding to $C$ and {\it iii.} vertices of ${\cal C}_0(D)$. This holds for any $D$ large enough that the ``edges'', ``faces" and ``vertices" of ${\cal C}_0(D)$ have an unambiguous meaning., and the positivity constraint comes from the fact that we can only remove, but not add sites along the edges, faces and vertices.\foot{To be completely clear, arbitrary meltings of the crystal correspond to a larger set of charges that do not obey this, but the other states do not have such an intuitive description.}

\bigskip
\centerline{\epsfxsize 4.5truein\epsfbox{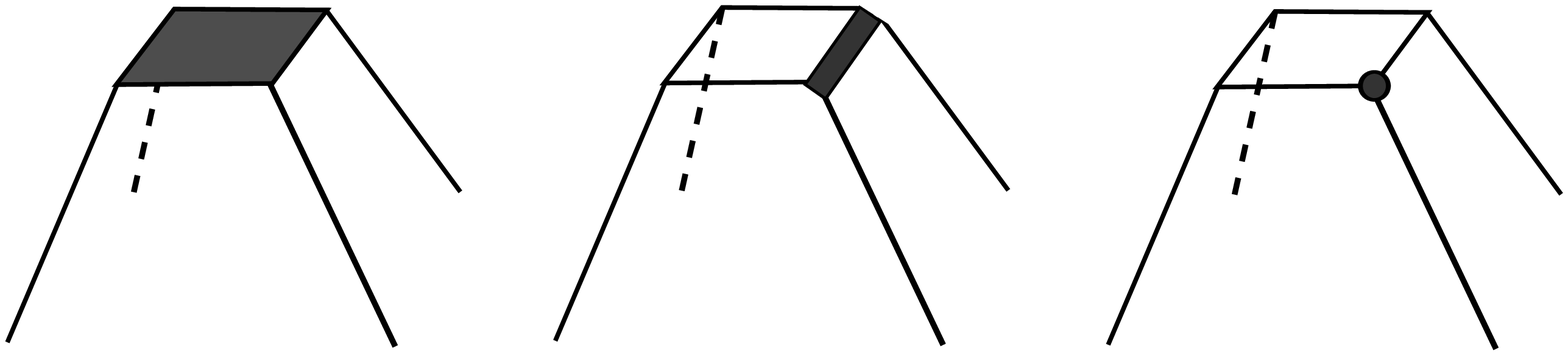}}
\noindent{\ninepoint \baselineskip=2pt {\bf Fig. 24} {Melting a face (a) corresponds to adding compact D4 brane charge, melting an edge (b) adds D2 brane charge, and melting a node in ${\cal C}_{0}(D)$ adds D0 brane charge.}}
\bigskip

As we discussed above, the D6 brane ${\cal O}_X(D)[-1]$ in the background of B-field $D$ is described by the crystal ${\cal C}(D)$ such that ${\cal C}_0(D)$ are 
integral points in the Calabi-Yau with Kahler class $D$. Adding to this a D4 brane on a divisor $D_S$ corresponds to changing the bundle on the D6 brane to ${\cal O}_X(D-D_S)[-1]$, 
and hence changing ${\cal C}(D)$ to ${\cal C}(D-D_S)$. This simply changes the Kahler class of the Calabi-Yau base by $D_S$. From the perspective of the crystal ${\cal C}(D),$
${\cal C}(D-D_S)$ is obtained by removing sites along the face corresponding to $D_S$. If $D_S$ is a compact divisor, we remove a finite number of sites. Consistency requires that the charge carried by these be the charge of the
D4 brane on $D_S$ in this background. In other words, if $S$ is the surface the divisor $D_S$ restricts to, the charge should be that of ${\cal O}_S(D)$, obtained from the pure D4 brane ${\cal O}_S$ on $D_S$ by
shifting the B-field by $D$.

Explicitly, the crystals ${\cal C}(D)$ and ${\cal C}(D-D_S)$ carry the charges of the D6 branes corresponding to ${\cal O}_X(D)[-1]$ and ${\cal O}_X(D-D_S)[-1],$
$$
-Xe^{D}(1+{1\over 24} c_2(X))
$$
and 
$$
-Xe^{D-D_S}(1+{1\over 24} c_2(X)).
$$
Thus, the sites that we remove in going from one crystal to the other must carry the difference of the charges. This is
$$
D_S(1-{1\over 2!} D_S + {1\over 3!} D_S^2 + {1\over 24} c_2(X)) e^{D} = D_S(1 + {1\over 24} c_2(S))
 e^{D-{1\over 2}K_S}$$
which equals the charge of a D4 brane on $S$ in the background B-field ${\cal O}_S(D)$. We used here the fact that one can think of the sheaf supported on a surface $S$ as a bundle on $S$ twisted by $-1/2$ of the canonical line bundle $K_S$ of the 
surface, and that $K_S = D_S,$ on a Calabi-Yau. 

Similarly, a D2 brane wrapping a curve $C$ in $X$ corresponding to ${\cal O}_C(D)$ is obtained by 
removing an edge in the crystal along $C$. To see this, suppose that $C$ lies on the intersection of two divisors
$C=D_S D_T$. Then, analogously to what we had done for the D6 branes and the D4 branes, we can express the D2 brane as a difference of the D4 branes on $D_S$ when we change the background from $D$ to $D-D_T$. The corresponding 
charges are 
\eqn\co{
D_S e^{D-{1\over 2} K_S}(1+{1\over 24} c_2(S))
}
and
\eqn\ct{
D_S e^{D-D_T-{1\over 2} K_S}(1+{1\over 24} c_2(S)).
}
The difference of \co\ and \ct\ is 
$$
C \Big(1 - {1\over 2} (D_T+D_S)\Big)e^D =C\big(1+{1\over 2}c_1(C)\big)e^D,
$$
which is the Chern class of the sheaf ${\cal O}_C(D)$ supported on $C$ in the background B-field $D$.  

We could also remove $n$ edges along $C$ in the crystal, but then we have the choice of melting additional edges along the face or down the side of the crystal.  The difference in charges corresponds exactly to $n{\cal O}_{C}(D) + k[pt]$ for some k determined by the crystal structure, where the additional D0 charge arises from the fact that in a general crystal, the lengths of multiple melted edges will not all be the same.

\subsec{An Example}

In our local ${\bf P^1} \times {\bf P}^1$ example, consider $X$ with the B-field shifted by $D= (m+2n)D_t + 2n D_s$.
Removing a face worth of sites from ${\cal C}_0$ (and the relevant sites from its refinement ${\cal C}$),
\eqn\pyrtwo{\eqalign{
&(2n+1+m)(2n+1) \Delta_3 + (2n+1)(2n+2 +m) \Delta_2\cr
 + &(2n+2)(m+2 n +1)\Delta_4 +(2n+2)(2n+2+m)\Delta_1}
}
and adding up the charges, we get
$$
D_S+ (m+2n+1)C_s+ (2n+1)C_t + \Big((2n+1+m)(2n+1) +{1\over 6}\Big){\rm pt} 
$$
which is the same as
$$
D_S(1+{1\over 24}c_2(S)) e^{D-{1\over 2} K_S},
$$
the chern character of ${\cal O}_S(D)$, where $S={\bf P}^1\times {\bf P}^1$, and $K_S = -2 D_s-2D_t = D_S$.

Similarly, removing an edge, corresponding to $C_t$ say, we remove sites of net charge
$$
(2n+1+m) \Delta_3 + (2n+2 +m) \Delta_2 + (m+2 n +1)\Delta_4 +(2n+2+m)\Delta_1
$$
Adding this up, the net charge is
$$
C_t + (2n+1+m){\rm pt} = C_t(1+{1\over 2} c_1(C_t))e^{D}
$$
corresponding to ${\cal O}_{C_t}(D)$.

To summarize, the states that have a simple geometric description
are those we obtain from ``pure" D6, D4, D2 (and of course D0 branes) ${\cal O}_X$, ${\cal O}_S$ and ${\cal O}_C$ by turning on a B-field $B\rightarrow B+D,$ in other words  ${\cal O}_X(D)$, ${\cal O}_S(D)$ and ${\cal O}_C(D)$.

\subsec{The large $B$-field limit and DT/GW correspondence}

Now consider ${\cal C}(D)$ in the limit of large $D\rightarrow \infty$, or more precisely
\eqn\inft{n_a \rightarrow \infty}
where
\eqn\div{
D = \sum_{a} n_a D_a, \qquad n_a\geq 0.
}
This is the large radius limit of $X$, where the lengths of all edges in ${\cal C}(D)$ go to infinity. When we consider meltings of this crystal,
we need to specify what we keep fixed in this limit. Without doing anything, working in terms of the fixed basis of $H_*(X, {\bf Z})$, corresponding to $X$, $D_a$, $C_a$ and ${\rm pt}$ the states corresponding to melting of ${\cal C}(D)$ would all have infinite charges. Moreover, because the edges of the crystal are infinitely long, the counting problem itself is not well defined.

\bigskip
\centerline{\epsfxsize 3.0truein\epsfbox{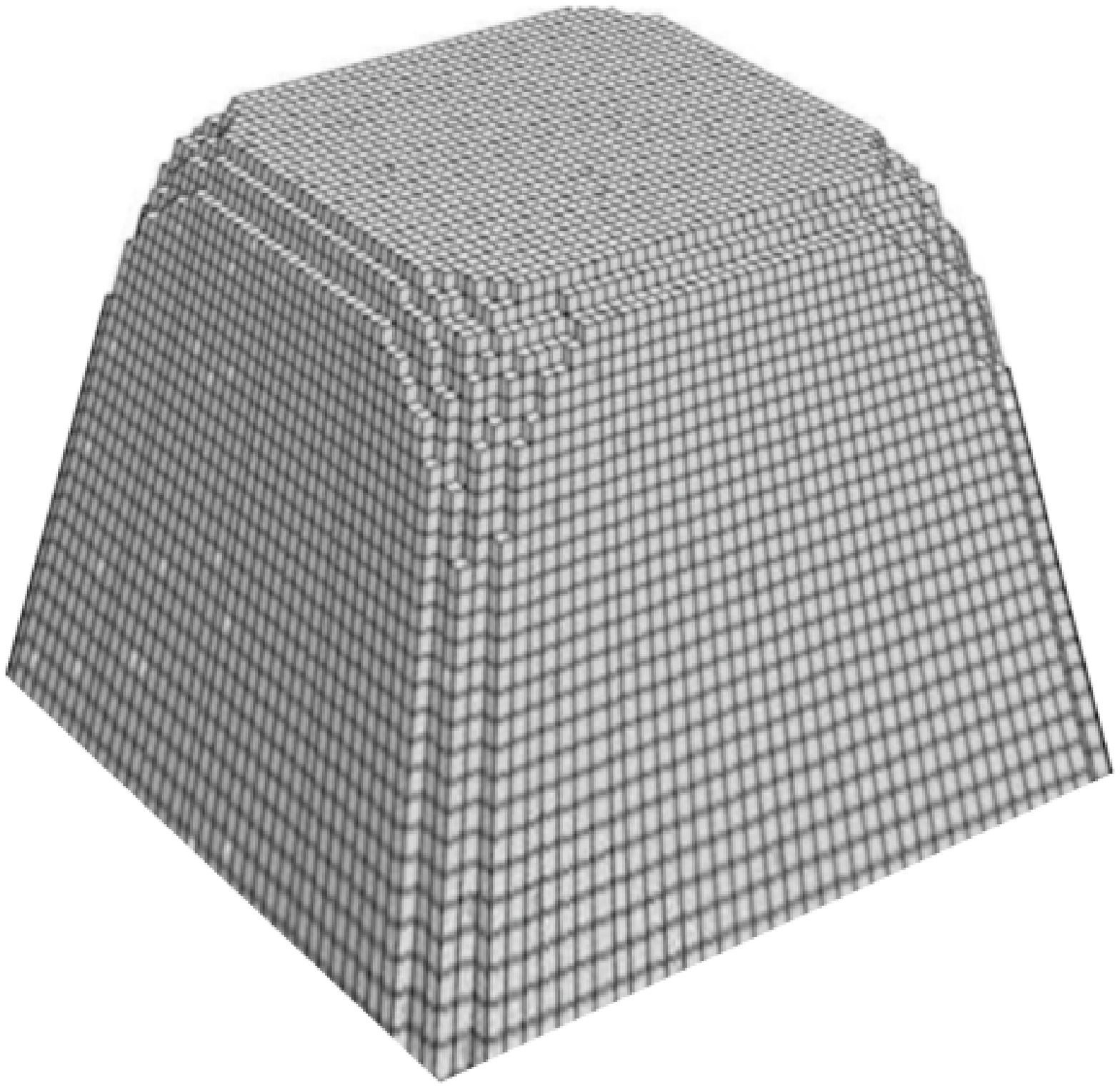}}
\noindent{\ninepoint \baselineskip=2pt {\bf Fig. 25} {Melting the local ${\bf P^1}\times{\bf P}^1$ crystal with a large compact B-field turned on.}}
\bigskip

To remedy this, consider changing the basis to
$$
X' = X e^D, \qquad D_a'=D_a e^D, \qquad C_a' = C_a e^D,
$$ 
and counting states whose charges in terms $X'$, $D_a'$, $C_a'$ and ${\rm pt}$ are finite, i.e. states of the form
\eqn\charge{
X'+ k_a D_a' + \ell_a C'_a +  m \;{\rm pt}
}
where $k,\ell$ and $m$ are all finite. In terms of the quiver we started with,
this is implemented by taking a limit in the weight space, where the weights 
of all but one node go to zero, 
\eqn\limt{
q_{\alpha} \rightarrow 0, \qquad\qquad{\alpha=1, \ldots k-1}
}
keeping the product of all the weights fixed,
$$\prod_{\alpha = 1}^k  q_{\alpha} = q.
$$
Since the sum of the charges of the nodes is one unit of D0 brane charge, 
states weighted by $q^m$ carry $m$ units of D0 brane charge. 

Taking the limits \inft\ and  \limt\ together, makes the state counting well defined. In the limit, the only local excitations of the crystal ${\cal C}(D)$ that survive are those where equal numbers of nodes of all colors are excited -- corresponding to removing pure D0 branes, weighted by $q^m$. One can think of this as a kind of a phase transition in the crystal  ${\cal C}(D)$ where excitations of individual nodes freeze out, and one can only remove atoms in groups weighted by $q$. In this limit, ${\cal C}(D)$ {\it becomes} ${\cal C}_0(D)$.
Moreover, the crystal can melt only from the vertices, and near each vertex \QF, the crystal ${\cal C}_0(D)$ looks like a copy of the ${\bf C}^3$ crystal in \ORV . In ${\bf C^3}$ we only have D0 branes to begin with, so the crystals ${\cal C}$ and ${\cal C}_0$ for this coincide. Note that, while we were originally counting signed partitions, with signs 
$(-1)^{d(\Delta)}$ 
from \dsign, in the large $D$ limit the signs become simply $(-1)^{m^2} = (-1)^m$, up to an overall sign, since we are restricting to configurations where all ranks are equal.

But, in addition, some large excitations survive as well. There are edge excitations which carry finite number of D2 brane charge $\ell_a$ in \charge, and also
excitations along faces carrying D4 brane charges $k_a$. We have shown in the previous section that, even at finite $D$, excitations
along an edge $\sum_a \ell_a C_a$ in $H_2(X,{\bf Z})$ 
carry charge 
$$\sum_a \ell_a C_a'=(\sum_a \ell_a C_a)e^D.$$
Similarly, removing the face in ${\cal  C}_0(D)$ in the class $\sum_k D_a$
corresponds to adding charge 
$$\sum_a k_a D_a' = (\sum_a k_a D_a)e^D$$
to the D6 brane $-X'$. 
We also showed that adding the D4 brane is the same as shifting $D$ by $\sum_{a} k_{a} D_{a}$. This clearly does not affect the degeneracies of D0-D2-D6 branes in the limit where we take $D$ to infinity --  it only shifts what we mean by $D$, in agreement with \DVV\DM . 

The relation of topological string amplitudes on $X$ with certain melting crystals was observed in \ORV.  The combinatorics of the topological vertex \TV\ and the A-model topological string on $X$ is the same \ORV\ as the combinatorics of ${\bf C}^3$ crystals glued together over the edges between ${\bf C}^3$ patches in $X$.
In \QF\ a physical explanation of this was proposed, by relating the crystals to D6 brane bound states.
Starting with the crystal ${\cal C}_0(D)$ corresponding to integral points in the base of a toric Calabi-Yau $X$, \QF\ showed that in the limit where one takes the Kahler class $D$ of $X$ to infinity, the crystal degenerates to ${\bf C}^3$ crystals glued together over long legs, exactly as in \ORV . On the other hand, it was  shown that in the same limit, the crystal counts bound states of a D6 brane on $X$, with D2 and D0 branes (in the language of sheaves, these are ideal sheaves on $X$). The count in \QF\
was formulated in terms of the maximally supersymmetric SYM on $X$, topologically twisted, and non-commutative. The conjecture of \QF\ relating the topological string on $X$ to counting bound states of a single D6 brane on $X$ with D0 and D2 branes, for any Calabi-Yau $X$, is known as the Gromov-Witten/Donaldson-Thomas correspondence \refs{\MNOPI,\MNOPII}. 
Recently, it was proven for toric threefolds by \MOOP.

We have shown that the bound states of a D6 brane on $X$, with D4, D2 and D0 branes described by a quiver $Q$ and in the background of B-field $D$, are counted by crystals ${\cal C}(D)$, at any $D$. The crystal roughly corresponds to integral points in the base of the Calabi-
Yau with Kahler class $D$, though the precise microscopic details depend on $Q$.  In the limit of infinite $D$, the microscopic structure is lost, and ${\cal C}(D)$ becomes the same as the crystal ${\cal C}_0(D)$ -- and hence the same as the crystal in \ORV\QF. Thus, the count of the D6 brane bound states from the six dimensional perspective of \QF\ and the $0+1$ dimensional quiver quantum mechanics of D6 branes bound to D4-D2 and D0 branes, agree -- but only in this limit.\foot{This is in accord with \DM, which pointed out that the correspondence of  \QF\ can hold only in the limit of infinite B-field. One should be able to understand this a consequence of essentially infinite non-commutativity turned on in \QF.} We have thus re-derived the Gromov-Witten/Donaldson-Thomas correspondence of \refs{\QF, \MNOPI, \MNOPII} from the quiver perspective. 

More than that, we provided an answer to the question raised in \Okoun : what is the crystal ${\cal C}_0(D)$ in \QF\ is counting at finite $D$? The crystal ${\cal C}_0(D)$, or more precisely its refinement ${\cal C}(D)$,  is counting Donaldson-Thomas invariants defined as the Witten indices of the quiver quantum mechanics describing one D6 brane on $X$, bound to D4, D2 and D0 branes in the background $B$ field $D$.
\newsec{Acknowledgments}

We would like to thank T. Dimofte, Y. Nakayama,  N. Reshetikhin, E. Sharpe,  C. Vafa and M. Yamazaki for very helpful discussions. We are especially grateful to D. Jafferis, A. Hanany, R. Kenyon, H. Ooguri and Y. Soibelman for explanations of their work. This work was supported by the Berkeley Center for Theoretical Physics, by the National Science Foundation (award number 0855653), by the Institute for the Physics and Mathematics of the Universe, and by the US Department of Energy under Contract DE-AC02-05CH11231.

\appendix{A}{The Conifold}

We will show how we can use Seiberg duality and dimer mapping explained in section 4, to reproduce the results of \refs{\sz, \benyoung}\ on the partition function of the conifold.  Recall that the conifold has only two nodes, which makes this model especially simple.  The conifold quiver is shown for chamber n (with n framing nodes) in Figure 26 while the conifold dimer is shown in Figure 27.
\bigskip
\centerline{\epsfxsize 2.0truein\epsfbox{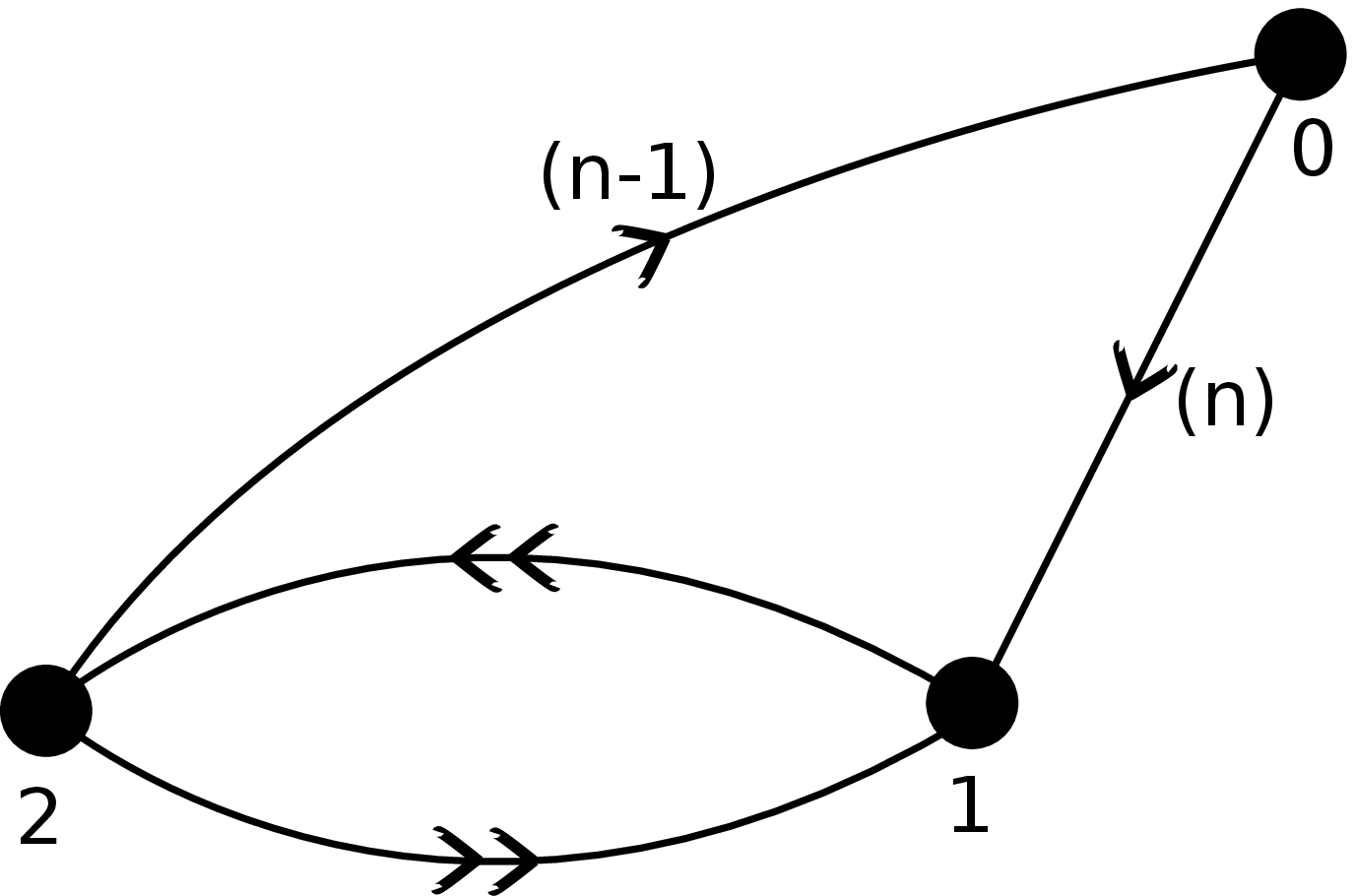}}
\noindent{\ninepoint \baselineskip=2pt {\bf Fig. 26.} {The conifold quiver for chamber n.}}
\bigskip

Starting from the configuration shown in Figure 27.a, we can perform Seiberg Duality on Node 1.  Seiberg Duality results in a dual face for 1, as explained in section 4.  However, the resulting brane tiling has two-valent vertices, which correspond to mass terms in the superpotential.  Integrating out results in the quiver shown in Figure 27.c.  This transformation takes the quiver from chamber $n$ to chamber $n+1$.  As an explicit example of the techniques in section 4, we will derive the exact wall crossing formula, along with the change of variables from the dimer model.

The exact partition function for the conifold is known \refs{\sz, \benyoung}.  In chamber n, it is given by,
$$
Z(n,q_{1},q_{2}) = M(1,-q_{1}q_{2})^{2}\prod_{k\geq 1}\bigl(1+q_{1}^{k}(-q_{2})^{k-1}\bigr)^{k+n-1}\prod_{k\geq n}\bigl(1+q_{1}^{k}(-q_{2})^{k+1}\bigr)^{k-n+1}
$$
where $M(x,q)$ is the MacMahon function,
$$
M(x,q) \equiv \prod_{m=1}^{\infty}\Bigl({1\over 1- xq^{m}}\Bigr)^{m}
$$
\bigskip
\centerline{\epsfxsize 5.0truein\epsfbox{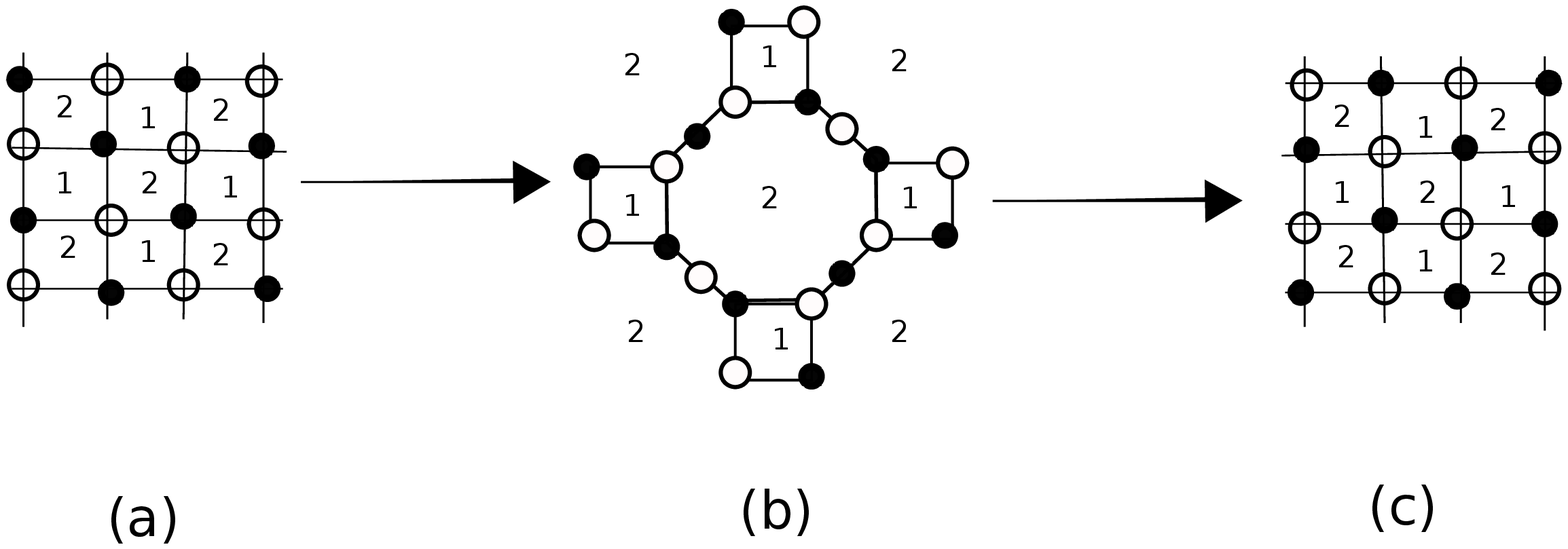}}
\noindent{\ninepoint \baselineskip=2pt {\bf Fig. 27.} {The effect of Seiberg Duality on the conifold dimer.  In the second step, we have integrated out the fields with a mass term in the superpotential coming from the 2-valent vertices.}}
\bigskip

Now consider crossing the wall from chamber n to chamber n+1.  This gives,
$$
Z(n+1) = Z(n,q_{1},q_{2})(1+q_{1})^{\langle q_{1},q_{\Delta}\rangle} = Z(n,q_{1},q_{2})(1+q_{1})^{-n}
$$
since for the conifold, every melting configuration must have the same intersection number, $\Delta_{1} \circ \Delta=-n$. \foot{Note that in Section 4 we only considered dualizing on nodes with $n_{0*}=1$ or $0$ framing arrows.  However, the conifold is simple enough to explicitly check that the dimer intersection number equals the quiver intersection number for arbitrary $n_{0*}$.}  By a simple change of variables \foot{This change of variables includes additional minus signs relative to \wcb.  This is because in this infinite product form, we have implicitly absorbed signs in the $\{q_{i}\}$.  These signs flip (from the $(-1)^{d(\Delta)}$ factors) when we cross the wall from chamber $n$ to chamber $n+1$.}
$${\eqalign{
q_{1} &= -Q_{2}^{-1} \cr
q_{2} &= -Q_{1}Q_{2}^{2} }}
$$
we find,
$$
Z(n+1) = M(1,-Q_{1}Q_{2})^{2}\prod_{k\geq 1}\bigl(1+Q_{1}^{k}(-Q_{2})^{k-1}\bigr)^{k+n+1}\prod_{k\geq n+1}\bigl(1+Q_{1}^{k}(-Q_{2})^{k+1}\bigr)^{k-n}
$$
which agrees with the general formula for the conifold partition function in chamber (n+1).

\appendix{B}{The local ${\bf P}^2$ example}

As another example, consider the toric quiver corresponding to $X = {\cal O}(-3) \rightarrow {\bf P}^2$. The D4-D2-D0 quiver has three nodes, corresponding to 
${\tilde E}_3 ={\cal O}_S(-3)$, ${\tilde E}_1= {\cal O}_S(-2)[-1]$, ${\tilde E}_2 = {\cal O}_S(-1)[-2]$. The charge of ${\cal O}_S(n)$
is
\eqn\co{
D_S e^{nD_t -{1\over 2} K_S}(1+{1\over 24} c_2(S))
}
where $D_S$ is the divisor corresponding to the surface, and $D_t$ generates the Kahler class, so 
$D_S = -3 D_t$. Also, $D_t D_S = C_t$ and per definition $D_t C_t =1.$ Thus, \co\ can be rewritten as
\eqn\cot{
D_S + (n+{3\over 2}) C_t +{1\over 2} (n+1)(n+2)\,{\rm pt} -{1\over 8}\,{\rm pt}
}
We can write
$$\eqalign{
{\tilde \Delta}_3 &= D_S + {\rm pt} + (-{3 \over 2} C_t +{1\over 4} {\rm pt})\cr
{\tilde \Delta}_2 &= D_S + 2 C_t  + (-{3 \over 2} C_t +{1\over 4} {\rm pt})\cr
{\tilde \Delta}_1 &= -D_S -C_t + ({3 \over 2} C_t -{1\over 4} {\rm pt}).}
$$ 
This has $n_{31}=n_{12}=3$, $n_{23}=6$, where $n_{ij}$ is the number of arrows from node $i$ to node $j$.  Add to this the D6 brane, corresponding to ${\cal O}_X[-1]$,
whose charge is
$$
{\tilde\Delta}_0 =-X(1+{c_2(X)\over 24}),
$$
Since $c_2(X) \cdot D_S= (c_2(S) - c_1^2(S))D_S =-6$, it is easy to see that $n_{03}=1$ and $n_{02}=n_{01}=0$.  The intersection numbers can be computed by setting up the usual $Ext$ machinery, and then reducing this to cohomology calculations on $S={\bf P}^{2}$.  For example, $Ext^{1}_{X}({\cal O}_{X}[-1],{\cal O}_{S}(-3)) = Ext^{2}_{X}({\cal O}_{X},{\cal O}_{S}(-3)) = H^{2}({{\bf P}^{2}},{\cal O}_{S}(-3)) = {\bf C}$, while all the other $Ext$ groups vanish in this sector.\foot{The last step follows from ${\rm dim} H^{0}({\bf P^k},{\cal O}(m)) = {m+k \choose k}$, ${\rm dim} H^{k}({\bf P^k},{\cal O}(m)) = {-m-1 \choose -k-m-1}$, and ${\rm dim H^n}({\bf P^k},{\cal O}(m))=0$, for $n\neq 0,k$, see e.g. \Asp.}

This is not a toric quiver yet, we need to dualize node ${\tilde \Delta}_{1}$. We get,
$$\eqalign{
{\Delta}_3& ={\tilde \Delta}_3= D_S + {\rm pt} + (-{3 \over 2} C_t +{1\over 4} {\rm pt})\cr
{\Delta_2}& = {\tilde \Delta}_2 + 3 {\tilde \Delta_3}= -2 D_S - C_t  +2 ({3 \over 2} C_t -{1\over 4} {\rm pt})\cr
{\Delta}_1 &=-{\tilde \Delta}_1 =  D_S +C_t - ({3 \over 2} C_t -{1\over 4} {\rm pt}).}
$$ 
and $\Delta_0 = {\tilde \Delta}_0$, unchanged.  The new bundles are given by, $E_{3}={\cal O}_{S}(-3)$, $E_{1}={\cal O}_{S}(-2)[-2]$, $E_{2}={\widetilde{\cal O}}_{S}$.
This has $n_{13}=n_{21}=n_{32} = 3$, $n_{03}=1$,
and the superpotential
$$
W = \sum_{i,j,k=1}^4 \epsilon_{ijk} {\rm Tr}A_i B_j C_k
$$ where $n_{ij}$ is the number of arrows from node $i$ to node $j$. The resulting quiver is in figure 28.

\bigskip
\centerline{\epsfxsize 2.0truein\epsfbox{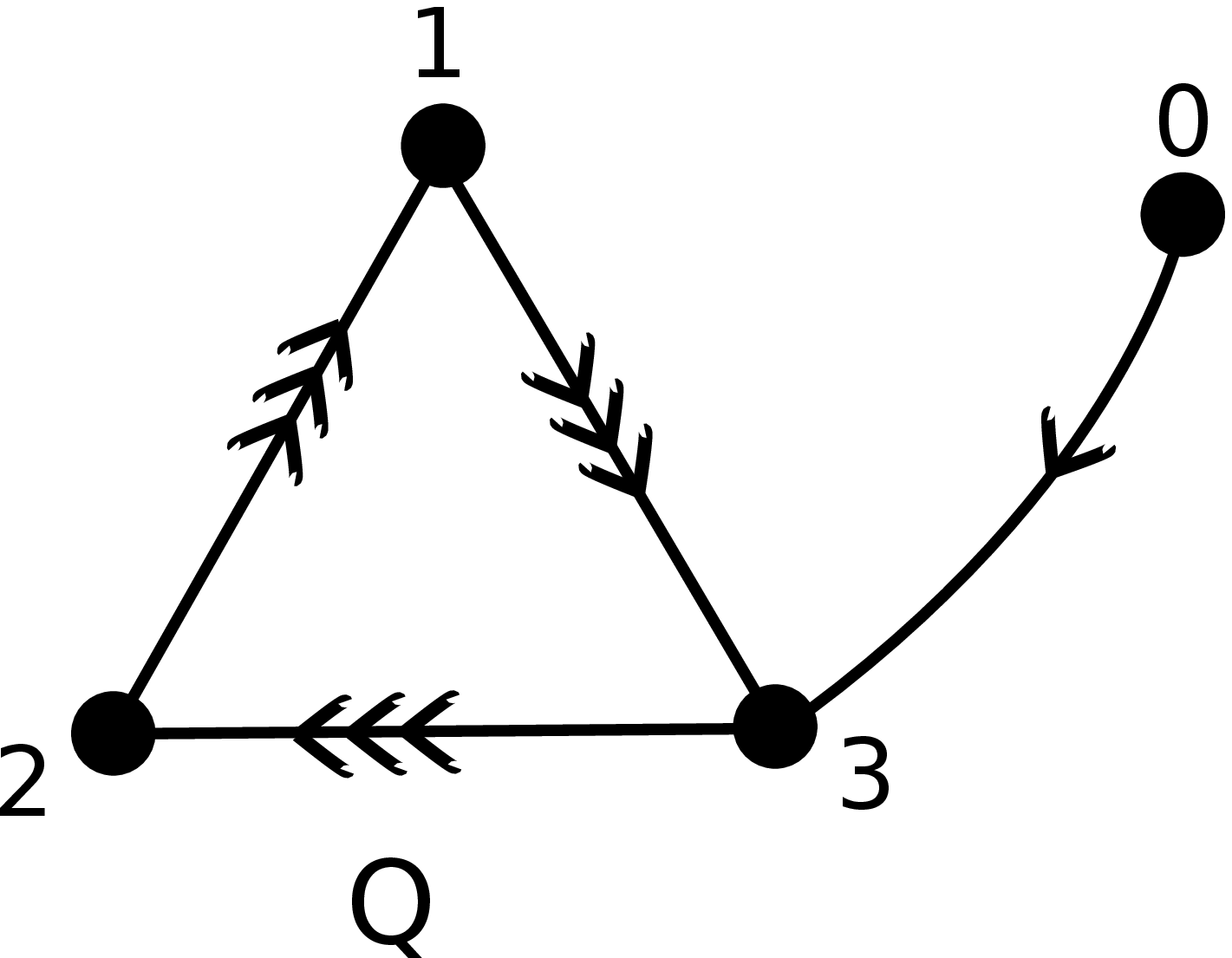}}
\noindent{\ninepoint \baselineskip=2pt {\bf Fig. 28} {The quiver for the orbifold phase of local ${\bf P}^{2}$.}}
\bigskip

The crystal ${\cal C}$ corresponding to this quiver is on the left in the figure 29. The corresponding crystal ${\cal C}_0$ is on the right.
This corresponds to a set of points
$$N_1+N_2 +N_3 = 3N_0, 
$$
in agreement with the fact that for local ${\bf P}^2$,
$Q = (1,1,1,-3)$.

\bigskip
\centerline{\epsfxsize 3.5truein\epsfbox{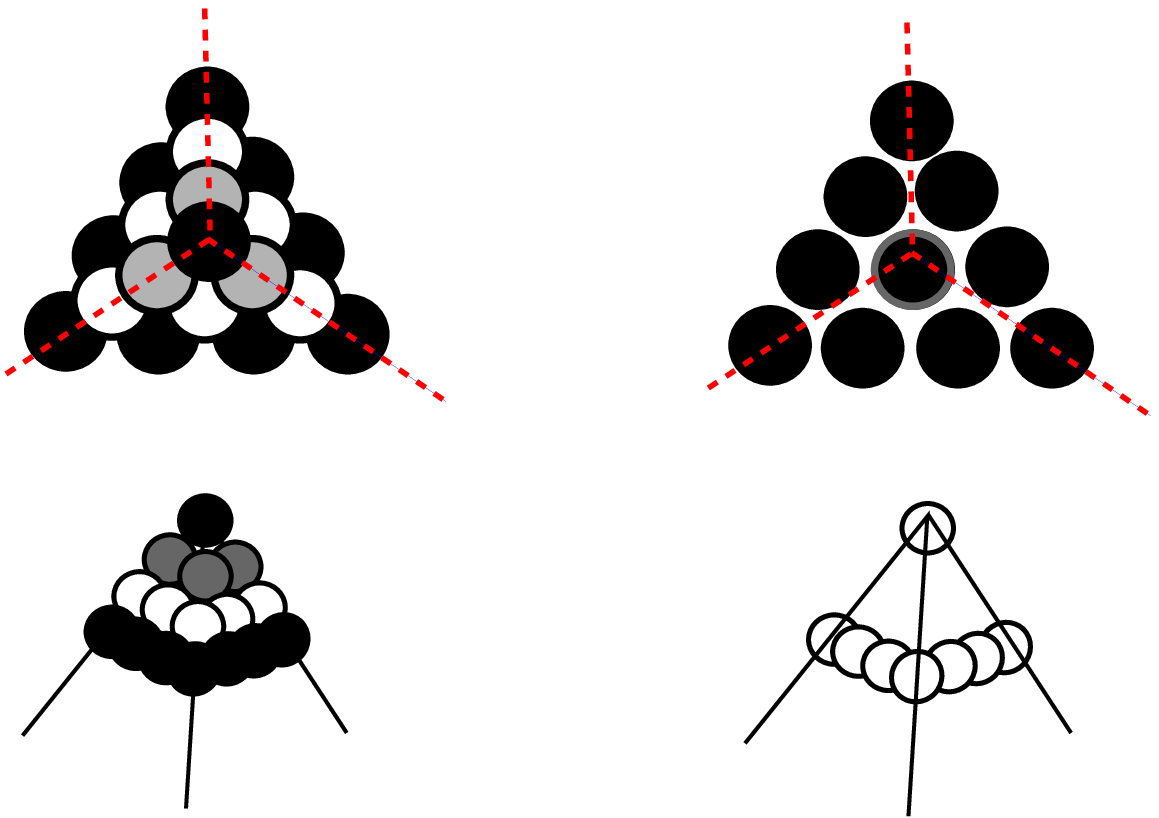}}
\noindent{\ninepoint \baselineskip=2pt {\bf Fig. 29.} {The full crystal, ${\cal C}$ for local ${\bf P}^{2}$ is shown on the left, while the subcrystal, ${\cal C}_{0}$ corresponding to holomorphic functions is shown on the right.     }}
\bigskip

Changing the B-field to $D = - n D_S = 3n D_t$, the lattice ${\cal C}_0(D)$ becomes (see figure 30)
$$N_1+N_2 +N_3 = 3N_0 + 3n.
$$
Getting ${\cal C}(D)$ from ${\cal C}$ corresponds to removing sites 
$$
\sum_{i=0}^{n-1}{(3i+1)(3i+2)\over 2} \Delta_3 +{(3i+2)(3i+3)\over 2} \Delta_2+ {(3i+3)(3i+4)\over 2} \Delta_1 
$$
\bigskip
\centerline{\epsfxsize 5.0truein\epsfbox{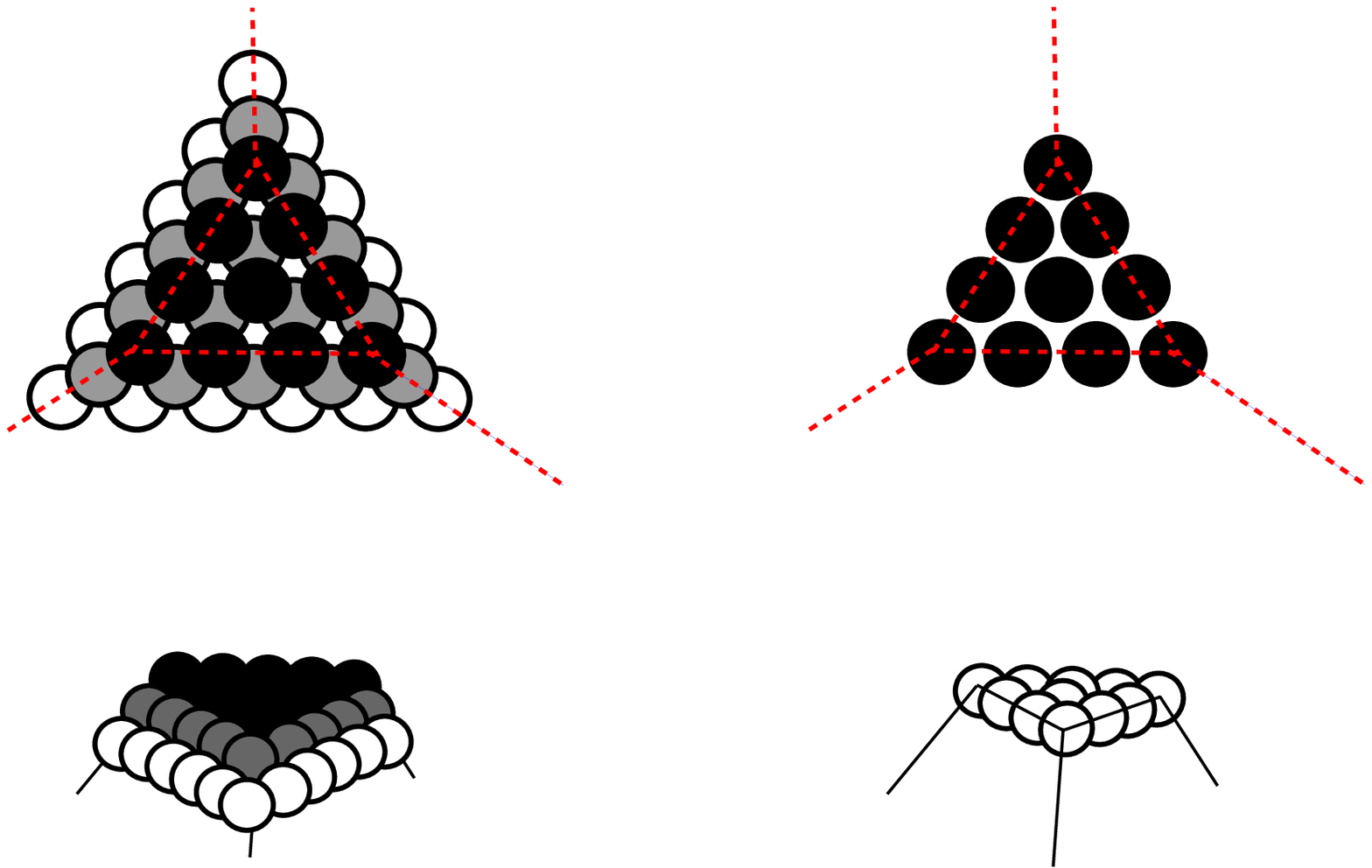}}
\noindent{\ninepoint \baselineskip=2pt {\bf Fig. 30.} {The full crystal, ${\cal C}(D)$, and the subcrystal, ${\cal C}_{0}(D)$ for local ${\bf P}^{2}$ after turning on a large B-field.}}
\bigskip

Summing up the charges, we find
$$
n D_S +{3\over 2} n^2 C_t+( {3\over 2}n^3 - {1\over 4}n ){\rm pt}
$$
This accounts for the charge the D6 brane picks up by putting it in the background B-field, which takes it to ${\cal O}_X(D)[-1]$, with charge
$$
\Delta =-Xe^{D}(1+{c_2(X)\over 24}),
$$
Namely, it is easy to see that this agrees with the difference 
$$\Delta-\Delta_0 = nD_S  - {n^2 \over 2}D_S D_S + {n^3 \over 3!} D_SD_SD_S + {n\over 24} C_2(X) D_S
$$
using 
$$
D_S D_S  =-3 C_t, \qquad D_SD_SD_S = 9, 
$$ 
Similarly, the face of the crystal carries charge
$$
{(3n+1)(3n+2)\over 2} \Delta_3 +{(3n+2)(3n+3)\over 2} \Delta_2+ {(3n+3)(3n+4)\over 2} \Delta_1 
$$
which equals
$$
D_S + (3 n +{3\over 2})C_t + ({(3n+1)(3n+2)\over 2}+ {1\over 4}){\rm pt}.
$$
From above, we see that this is the charge of 
$${\cal O}_S(D),$$
as we claimed in the text. Similarly, the charge of an edge is
$$
(3n+1)\Delta_3+(3n+2)\Delta_2+(3n+3)\Delta_3.
$$
This equals
$$
C_t + (3n+1){\rm pt},
$$
the charge of
$$
{\cal O}_{C_t}(D).
$$

\listrefs

\end